%% file: main.tex
\documentclass[fleqn,10pt]{article}
\usepackage{latexsym, graphicx, epsfig, amsmath, amssymb,amsfonts}
\usepackage{natbib,amsthm,version}
\usepackage{amsbsy,bm,multirow,enumerate}
\usepackage[titletoc,page]{appendix}
\usepackage[mathscr]{eucal}
\usepackage{mathtools}
\usepackage{color}
\usepackage[utf8]{inputenc}
\usepackage[english]{babel}
\usepackage{amsthm}

\newtheorem{theorem}{Theorem}[section]

\newtheorem{remark}{Remark}

\theoremstyle{definition}
\newtheorem{definition}{Definition}[section]

\title{
Multiphase Flows of $N$ Immiscible Incompressible Fluids:
A Reduction-Consistent and Thermodynamically-Consistent Formulation
and Associated Algorithm
} 
\author{
  S. Dong\thanks{ Email: sdong@purdue.edu} \\
  Center for Computational and Applied Mathematics \\
  Department of Mathematics \\
  Purdue University \\
  West Lafayette, Indiana, USA
 } 

\date{}
\begin{document}
\maketitle



\input Abstract


\vspace{0.05cm}

\input Introduction

\input Method

\input Algorithm

\input Tests

\input Summary

\section*{Acknowledgement}
This work was partially supported by
NSF (DMS-1318820, DMS-1522537). 

\input Appendix

\input AppendB

\input AppendC

\input AppendD

\input AppendE

\bibliographystyle{plain}
\bibliography{nphase,obc,mypub,nse,sem,contact_line,interface,multiphase}

\end{document}

%% file: Abstract.tex
\begin{abstract}

  We present a reduction-consistent and thermodynamically
  consistent formulation and an associated numerical algorithm
  for simulating the dynamics of an isothermal mixture
  consisting of $N$ ($N\geqslant 2$) immiscible incompressible
  fluids with different physical properties (densities,
  viscosities, and pair-wise surface tensions).
  By reduction consistency we refer to the property that
  if only a set of $M$ ($1\leqslant M\leqslant N-1$) fluids
  are present in the system then
  the N-phase governing equations and boundary conditions will
  exactly reduce to those for the corresponding  $M$-phase system.
  By theromdynamic consistency we refer to the property
  that the formulation honors the thermodynamic
  principles.
  Our N-phase formulation is developed
  based on a more general method that allows for the systematic
  construction of reduction-consistent formulations,
  and the method suggests the existence of many possible forms of 
  reduction-consistent and thermodynamically consistent N-phase
  formulations.
  Extensive numerical experiments have been presented
  for flow problems involving multiple fluid components and
  large density ratios and large viscosity ratios, and
  the simulation results are compared with the physical
  theories or the available physical
  solutions. The comparisons demonstrate that our  method
  produces physically accurate results for this class of problems.

\end{abstract}

%% file: Introduction.tex
\section{Introduction}
\label{sec:intro}

%

This paper concerns the formulation and simulation of isothermal
multiphase flows consisting of $N$ ($N\geqslant 2$) immiscible
incompressible fluids with possibly very different physical properties
(e.g.~densities, dynamic viscosities, and pair-wise surface tensions).
Following our previous works~\cite{Dong2014,Dong2015,Dong2017}
and with a slight abuse of notation,
we will refer to such problems as N-phase flows, where $N$ denotes
the number of different fluid components in the system, not necessarily
the number of material phases.
Our primary concern is the reduction consistency and thermodynamic
consistency in the formulation of such problems.
By thermodynamic consistency we refer to the property that
the formulation should honor the thermodynamic principles
(e.g.~mass conservation, momentum conservation, second law of thermodynamics,
Galilean invariance). Reduction consistency
is rooted in the following simple observation about N-phase systems:
\begin{itemize}

\item
  Given an N-phase system, if some fluid components are absent
  from the system such that only $M$ ($1\leqslant M\leqslant N-1$)
  fluids are present, then this N-phase system is equivalent
  to the smaller M-phase system consisting of the fluids that are present.

\end{itemize}
We insist that the mathematical formulation for the N-phase system
should correspondingly satisfy the same property, namely,
\begin{enumerate}[($\mathscr{C}0$):]

\item
  If only $M$ ($1\leqslant M\leqslant N-1$) fluids are present in
  the N-phase system (while the other fluids are absent),
  then the N-phase formulation should reduce to the corresponding
  M-phase formulation.

\end{enumerate}
We refer to this property as the reduction consistency.


The overall approach taken in this work falls into
the phase field (or diffuse interface) framework,
and we are primarily interested in the simulation of N-phase systems
with three or more fluid components (i.e.~$N\geqslant 3$).
For two-phase flows we refer to
several comprehensive reviews
(see e.g.~\cite{AndersonMW1998,SethianS2003,ScardovelliZ1999,Tryggvasonetal2001}
and the references therein) of
this and related approaches.
Multiphase problems involving three or more fluid
components have attracted a growing interest,
and a number of researchers have contributed
to the advance of this field; see e.g.~\cite{KimL2005,BoyerL2006,Kim2009,BoyerLMPQ2010,Kim2012,HeidaMR2012,Dong2014,BoyanovaN2014,Dong2015,BrannickLQS2015,ZhangW2016,BanasN2017,WuX2017}, among others.
Among the past studies, a handful of phase field models
(e.g.~\cite{KimL2005,HeidaMR2012,Dong2014}) have
been developed that take into account the conservation laws
and the constitutive relations dictated by thermodynamic
principles.
Reduction consistency issues are investigated
for a three-phase and a multi-phase Cahn-Hilliard
model (without hydrodynamic interactions) \cite{BoyerL2006,BoyerM2014},
and these studies have signified the importance in the form of the free energy
density function.
While  the current work focuses on the hydrodynamic interactions of
multiple fluids,
certain consistency issues encountered here can be
analogous to those facing the materials community
for multi-component
materials (see e.g.\cite{Steinbachetal1996,FolchP2005,BolladaJM2012,TothPG2015}).


In a previous work \cite{Dong2014} we have proposed
a general phase field model for formulating
 an isothermal system of $N$ ($N\geqslant 2$)
immiscible incompressible fluids. The model is derived
based on and honors the mass conservation of the $N$
individual fluid components, the momentum conservation,
the second law of thermodynamics,
and the Galilean invariance principle.
In such a sense it is a thermodynamically consistent model.
This model is formulated based on a volume-averaged
mixture velocity, which can be rigorously shown to be divergence
free \cite{Dong2014}. It is fundamentally
different from those of \cite{KimL2005,HeidaMR2012},
which are based on a mass-averaged velocity (not divergence free).
%
This N-phase model is generalized in \cite{Dong2015},
and a class of general order parameters has been introduced
to formulate the N-phase system.

While the model of \cite{Dong2014,Dong2015} is thermodynamically
consistent, it nonetheless falls short with respect to the
reduction consistency. Motivated by this inadequacy and inspired
by the discussions of consistency issues 
in \cite{BoyerM2014}, we have very recently in \cite{Dong2017}
combined a modified thermodynamically consistent N-phase
model and the reduction-consistency considerations, and
developed a method for simulating wall-bounded N-phase
flows and N-phase contact angles.
We have specifically considered the following set of reduction-consistency
conditions on the N-phase formulation~\cite{Dong2017}: \\
{\it
If only a set of $M$ ($1\leqslant M\leqslant N-1$) fluids
are present in the N-phase system, then
}
\begin{enumerate}[($\mathscr{C}1$):]
{\it
\item
  the N-phase free energy density function should reduce to the
  corresponding M-phase free enegy density function;

\item
  the set of N-phase governing equations should reduce to that for
  the corresponding M-phase system, together with a set of identities
  corresponding to the absent fluids;

\item
  the set of boundary conditions for the N-phase system should reduce
  to that for the corresponding M-phase system, together with a set of
  identities corresponding to the absent fluids.
}
\end{enumerate}
Note that the consistency conditions ($\mathscr{C}2$) and
($\mathscr{C}3$) are imposed for both the momentum equations
and the phase field equations.


By assuming a constant mobility matrix in the formulation,
we have explored in \cite{Dong2017} the implications of the above consistency
conditions on the N-phase governing equations and boundary conditions.
It is found  that to satisfy
the reduction-consistency conditions the mobility matrix should
take a particular form (specific form given in \cite{Dong2017})
and that the ``multi-well'' potential free energy density function
needs to satisfy a set of properties as given in \cite{Dong2017}.
The reduction-consistency problem is thus  boiled down
to the following:
\begin{itemize}

\item
  Given an arbitrary set of pairwise surface tension values, how does
  one construct the multi-well potential free energy density function
  to satisfy the properties given in \cite{Dong2017}?
  
\end{itemize}
If one could construct such a multiwell potential energy density function,
a fully reduction-consistent N-phase formulation could
be obtained.
This construction problem is unfortunately highly non-trivial and challenging, and
it so far remains an open question. 
It is noted that in \cite{BoyerM2014}
the consistency of a Cahn-Hilliard model (no hydrodynamic
interactions) is studied under a set of weaker consistency
conditions. The resultant property on the potential energy density
function from \cite{BoyerM2014} is weaker (it is
a subset of the required properties), and
does not ensure the reduction consistency of the momentum
equations. 
%
Due to the lack of an appropriate potential free energy
density function to ensure full reduction consistency,
in \cite{Dong2017} a particular potential free energy
density form has been adopted to arrive at a specific
N-phase formulation, which ensures only a partial
reduction  consistency of the  governing equations 
(between $N$ phases and
two phases only).
In \cite{Dong2017} we have also developed a set of
reduction-consistent N-phase contact-angle
boundary conditions based on the consistency property
($\mathscr{C}3$).

%

Recognizing the enormous challenge with the approach of \cite{Dong2017}
to fulfill the reduction consistency,
we present in this paper a different approach 
to achieve full reduction consistency and thermodynamic
consistency for the N-phase formulations.
The key distinction lies in dropping the assumption that
the mobility matrix be constant.
This allows us to devise the mobility matrix and
the free energy density function individually in an untangled
fashion, which can satisfy certain appropriate
reduction properties separately.
Full reduction consistency for the set of N-phase governing
equations can then be guaranteed based on these
individual reduction properties.


More specifically, we present developments in the
the following aspects in this work:
\begin{itemize}

\item
  We present a thermodynamically consistent phase field model,
  which 
  is different from those of \cite{Dong2014,Dong2017},
  for the
  hydrodynamic interactions of the N-fluid mixture. The development
  process of this model mirrors that of \cite{Dong2014},
  but it leads to a different model due to a different representation
  of the mass balances of individual fluid components and
  different constitutive relations to satisfy the second law
  of thermodynamics. This model serves as the starting point for
  reduction consistency considerations in this work, and it is critical to
  the success in achieving full reduction consistency in
  the N-phase formulation.

\item
  We introduce the concepts of reduction compatibility and reduction consistency 
  for a set of variables, functions, and equations,
  and look into some useful properties of reduction-consistent
  and reduction-compatible functions/variables.

\item
  We provide a method (Theorem \ref{thm:thm_1}) that 
  guarantees the reduction consistency of the set of N-phase governing
  equations from the aforementioned phase field model.
  The method consists of a set of sufficient conditions
  with regard to the reduction consistency and reduction compatibility
  of the mobility matrix and terms involving the
  free energy density function.
  The method is quite general, and suggests many ways to construct
  reduction-consistent and thermodynamically consistent
  N-phase formulations.

\item
  We suggest a specific form for the mobility matrix and
  the free energy density function that satisfy the reduction
  properties dictated 
   by the method. This leads to a specific reduction-consistent
  and thermodynamically consistent N-phase formulation.

\item
  We present a numerical algorithm for solving the governing equations
  of this N-phase formulation, together with a set of reduction-consistent
  boundary conditions. In particular, we look into how to
  algorithmically deal with
  the variable mobility matrix involved therein.  
  
\end{itemize}


The novelties of this paper lie in several aspects:
(i) the method (Theorem \ref{thm:thm_1}) to systematically
construct reduction-consistent N-phase
governing equations;
(ii) the specific reduction-consistent and thermodynamically
consistent N-phase formulation; and
(iii) the numerical algorithm for solving the set of
reduction-consistent and thermodynamically consistent
N-phase phase field equations with a variable mobility matrix.
To the best of the author's knowledge, the N-phase formulation
presented herein is the first fully reduction-consistent
and thermodynamically consistent mathematical formulation 
for the hydrodynamic interactions of incompressible N-phase flows.


The rest of this paper is structured as follows.
In Section \ref{sec:formulation} we introduce
the ideas of reduction compatibility and
reduction consistency for a set of functions and equations,
and present a method that allows for the systematic construction
of reduction-consistent N-phase governing equations.
We also present a specific
reduction-consistent and thermodynamically consistent
N-phase formulation based on this method.
In Section \ref{sec:alg} we present an efficient numerical
algorithm for solving the N-phase phase field equations
with a variable mobility matrix. This,
together with the algorithm for the N-phase momentum
equations summarized in Appendix E,
provides an effective method for simulating
incompressible N-phase flows with the reduction-consistent
and thermodynamically consistent formulation.
In Section \ref{sec:test} we provide extensive numerical
experiments to test the method developed herein, and
the simulation results are compared with physical
theories and exact physical solutions
for problems involving multiple fluid components
and large contrasts in densities and viscosities.
Section \ref{sec:summary} then concludes the paper
with some closing remarks.
In Appendix A we outline the development of the thermodynamically
consistent phase field model for an isothermal
mixture of $N$ ($N\geqslant 2$) immiscible incompressible fluids,
on which the current work is based.
We summarize the key steps in the derivation of the model
based on the mass conservation, momentum conservation, and
the second law of thermodynamics.
In Appendix B we provide proofs for several useful properties
about reduction-consistent and reduction compatible
functions listed in Section \ref{sec:formulation}.
Appendix C and Appendix D provide proofs for
the Theorems \ref{thm:thm_1} and \ref{thm:thm_2}
given in the main body of the text.
Appendix E summarizes a numerical algorithm for solving
the N-phase momentum equations.

%% file: Method.tex
\section{Reduction-Consistent and Thermodynamically Consistent N-phase Formulation}
\label{sec:formulation}

%
%


Consider an isothermal mixture of $N$ ($N\geqslant 2$)
immiscible incompressible fluids contained in some flow domain.
In this section we present a reduction-consistent and thermodynamically-consistent
formulation for this system.
The formulation is developed based on a thermodynamically consistent
phase field model for the N-fluid mixture.
In the Appendix A we have outlined the derivation of
this  N-phase model based on
the mass conservation, momentum conservation,
and the second law of thermodynamics.
The development process for this model mirrors that
for the model of \cite{Dong2014}. However, owing to
a different way to represent the mass balances
and to specify the constitutive relations to ensure
the second law of thermodynamics, here we arrive at a model
that is different from those of \cite{Dong2014,Dong2017}.
In this model the dynamics of the N-phase mixture
is described by the equations
\eqref{equ:nse_original}--\eqref{equ:CH_original}.
The model honors the mass conservation of the $N$
individual fluid components, the momentum conservation
and the second law of thermodynamics, and it is
Galilean invariant. Therefore
it is said to be thermodynamically consistent. 
We refer the reader to Appendix A for the details in
the development of the model.
This model serves as the starting point, and 
in subsequent developments
we concentrate on how to fulfill reduction consistency 
with this model.

Given a mixture of $N$ ($N\geqslant 2$) immiscible incompressible fluids,
let $\tilde{\rho}_i$ ($1\leqslant i\leqslant N$) and
$\tilde{\mu}_i$ ($1\leqslant i\leqslant N$) denote the constant
density and constant dynamic viscosity of pure fluid
$i$ (before mixing), respectively.
Let $\sigma_{ij}$ ($1\leqslant i\neq j\leqslant N$) denote
the constant surface tension associated with the interface formed between
fluid $i$ and fluid $j$, satisfying the following property
\begin{equation}
  \left\{
  \begin{split}
    &
    \sigma_{ij}=\sigma_{ji}, \quad 1\leqslant i,j\leqslant N, \\
    &
    \sigma_{ij} > 0, \quad 1\leqslant i<j\leqslant N, \\
    &
    \sigma_{ii} = 0, \quad 1\leqslant i\leqslant N.
  \end{split}
  \right.
  \label{equ:sigma_cond}
\end{equation}
Let $c_i(\mathbf{x},t)$ ($1\leqslant i\leqslant N$) and
$\rho_i(\mathbf{x},t)$ ($1\leqslant i\leqslant N$)
denote the volume fraction and density
of fluid $i$ {\em within the mixture}, which are field functions of
space $\mathbf{x}$ and time $t$.
Let $\vec{c}=(c_1, c_2, \dots,c_N)$, and
$\rho(\vec{c})$ and $\mu(\vec{c})$ denote
the mixture density and mixture dynamic viscosity.
These and other related variables are defined
in more detail in the Appendix A.

\subsection{Reduction Compatibility and Reduction Consistency}
\label{sec:def_consis}

As mentioned in the Introduction section, certain equivalence relations
exist between an N-phase system and smaller M-phase systems ($1\leqslant M\leqslant N-1$).
If some fluid components are absent, then the N-phase system
is physically equivalent to a smaller multiphase system 
consisting of the fluids that are present.
We next explore these equivalence relations and introduce
the concepts of reduction consistency and reduction compatibility
for a set of variables/functions and equations.
This provides the basis for the study of reduction consistency
of N-phase governing equations.

To make the idea more concrete, let us first consider the case
in which one fluid component is absent from the N-phase
system. Suppose the $k$-th fluid ($1\leqslant k\leqslant N$)
is absent from the N-phase system, i.e.~the system is characterized
by
\begin{equation}
  c_k^{(N)} \equiv 0, \quad
  \rho_k^{(N)}\equiv 0 \quad
  \text{for some} \ 1\leqslant k\leqslant N,
  \label{equ:special_nphase_system}
\end{equation}
where the superscript in $(\cdot)^{(N)}$ accentuates
the point that the variable is with respect to the
N-phase system.
We will use this convention about the superscript throughout this paper
and, if possible, will omit this superscript for
brevity where no confusion arises.

We assume that the ordering of the fluids in the resultant
$(N-1)$-phase system follows that of the original N-phase
system (excluding fluid $k$). In other words, the following
correspondence relations for the volume
fractions $c_i$ ($1\leqslant i\leqslant N$) hold:
\begin{equation}
  c_i^{(N)} = \left\{
  \begin{array}{ll}
    c_i^{(N-1)}, & 1\leqslant i\leqslant k-1, \\
    0, & i=k, \\
    c_{i-1}^{(N-1)}, & k+1\leqslant i\leqslant N,
  \end{array}
  \right.
  \quad
  \text{or} \ \
  \left\{
  \begin{split}
    &
  c_i^{(N-1)} = 
  \left\{
  \begin{array}{ll}
    c_i^{(N)}, & 1\leqslant i\leqslant k-1, \\
    c_{i+1}^{(N)}, & k\leqslant i\leqslant N-1,
  \end{array}
  \right. \\
  &
  c_k^{(N)}=0.
  \end{split}
  \right.
  \label{equ:volfrac_correspondence}
\end{equation}
The density of fluid $i$ {\em within the mixture}, $\rho_i$ ($1\leqslant i\leqslant N$),
has a correspondence relation analogous to the above
between the original N-phase and the resultant
$(N-1)$-phase systems.
On the other hand, the constant density of pure fluid $i$,
$\tilde{\rho}_i$ ($1\leqslant i\leqslant N$),
has a similar correspondence relation but with
some difference:
\begin{equation}
  \tilde{\rho}_i^{(N)} = \left\{
  \begin{array}{ll}
    \tilde{\rho}_i^{(N-1)}, & 1\leqslant i\leqslant k-1, \\
    \tilde{\rho}_{i-1}^{(N-1)}, & k+1\leqslant i\leqslant N,
  \end{array}
  \right.
  \quad
  \text{or} \ \
  \tilde{\rho}_i^{(N-1)} = 
  \left\{
  \begin{array}{ll}
    \tilde{\rho}_i^{(N)}, & 1\leqslant i\leqslant k-1, \\
    \tilde{\rho}_{i+1}^{(N)}, & k\leqslant i\leqslant N-1.
  \end{array}
  \right. 
  \label{equ:const_density_correspondence}
\end{equation}
The critical difference lies in that even though fluid $k$ is absent from
the system ($c_k^{(N)}= 0$, $\rho_k^{(N)}= 0$), the 
density of pure fluid $k$ remains the same
non-zero constant ($\tilde{\rho}_k^{(N)}\neq 0$).   
The constant dynamic viscosities $\tilde{\mu}_i$ ($1\leqslant i\leqslant N$)
have a correspondence relation analogous
to \eqref{equ:const_density_correspondence}.

The correspondence relations \eqref{equ:volfrac_correspondence} and \eqref{equ:const_density_correspondence}
characterize two different types of variables.
The distinction between them
lies in that in the latter type there is no constraint
on the $k$-th variable of the original N-phase system
if fluid $k$ is absent from the system.
It is important to distinguish these two
types of correspondence relations and the two types of variables.
Equations \eqref{equ:volfrac_correspondence} and
\eqref{equ:const_density_correspondence}
describe how the variables $c_i$ ($1\leqslant i\leqslant N$)
and $\tilde{\rho}_i$ ($1\leqslant i\leqslant N$)
transform, respectively, if any  fluid $k$ ($1\leqslant k\leqslant N$)
is absent from the N-phase system.

Let us now look into how a given set of functions of
$\vec{c}=(c_1,\cdots,c_N)$ 
defined on the N-phase system transforms if
any one fluid $k$ ($1\leqslant k\leqslant N$)
is absent from the system.
Intuitively, if these functions transform in a way similar to $c_i$,
we say that they are reduction consistent.
If they transform in a way similar to $\tilde{\rho}_i$,
we say that they are reduction compatible.

%
More specifically, we consider the set of variables
$v_i^{(N)}(\vec{c}^{(N)})$ ($1\leqslant i\leqslant N$)
defined on the N-phase system for all $N=1, 2, 3,\dots$,
and investigate their transformations if any fluid is absent
from the N-phase system for $N\geqslant 2$.
Similarly, we also study the transformations of the sets of variables
$v_{ij}^{(N)}(\vec{c}^{(N)})$ ($1\leqslant i,j\leqslant N$; $N\geqslant 1$) and
$v^{(N)}(\vec{c}^{(N)})$ ($N\geqslant 1$) if any fluid is
absent from the system.
We define the reduction compatibility and
reduction consistency
of these sets of variables (or functions) as follows.
\begin{definition}
  \label{def:def_vi_compatible}
  A set of variables $v_i^{(N)}(\vec{c}^{(N)})$ ($1\leqslant i\leqslant N$;
  $N=1,2,3\dots$) is
  said to be reduction-compatible if for any $N\geqslant 2$, this set
  transforms as follows
  when any fluid $k$ ($1\leqslant k\leqslant N$) is absent from
  the N-phase system:
  \begin{equation}
  v_i^{(N)} = \left\{
  \begin{array}{ll}
    v_i^{(N-1)}, & 1\leqslant i\leqslant k-1, \\
    v_{i-1}^{(N-1)}, & k+1\leqslant i\leqslant N,
  \end{array}
  \right.
  \label{equ:def_vi_compatible}
  \end{equation}
  where $v_i^{(N)}=v_i^{(N)}(\vec{c}^{(N)})$
  and $v_i^{(N-1)}=v_i^{(N-1)}(\vec{c}^{(N-1)})$,
  and $\vec{c}^{(N)}$ and $\vec{c}^{(N-1)}$ are
  connected by the correspondence relation \eqref{equ:volfrac_correspondence}.
\end{definition}
\begin{definition}
  \label{def:def_vi_consis}
  A set of variables $v_i^{(N)}(\vec{c}^{(N)})$ ($1\leqslant i\leqslant N$; $N=1,2,3\dots$)
  is said to be reduction-consistent if (i)
  this set is reduction-compatible,
  and (ii) for any $N\geqslant 2$, this set
  satisfies the following additional property
  when any fluid $k$ ($1\leqslant k\leqslant N$) is absent from the
  N-phase system:
  \begin{equation}
    v_k^{(N)} = 0.
    \label{equ:def_vi_consis}
  \end{equation}
  where $v_k^{(N)} = v_k^{(N)}(\vec{c}^{(N)})$.
  
\end{definition}
\begin{definition}
  \label{def:def_vij_compatible}
  A set of variables $v_{ij}^{(N)}(\vec{c}^{(N)})$ ($1\leqslant i,j\leqslant N$;
  $N=1,2,3\dots$)
  is said to be reduction-compatible if for any $N\geqslant 2$, this set
  transforms as follows when any fluid $k$ ($1\leqslant k\leqslant N$)
  is absent from the N-phase system:
  \begin{equation}
    v_{ij}^{(N)} = \left\{
    \begin{array}{ll}
      v_{ij}^{(N-1)}, & 1\leqslant i\leqslant k-1, \ 1\leqslant j\leqslant k-1, \\
      v_{ij-1}^{(N-1)}, & 1\leqslant i\leqslant k-1, \ k+1\leqslant j\leqslant N, \\
      v_{i-1j}^{(N-1)}, & k+1\leqslant i\leqslant N, \ 1\leqslant j\leqslant k-1, \\
      v_{i-1j-1}^{(N-1)}, & k+1\leqslant i\leqslant N, \ k+1\leqslant j\leqslant N,
    \end{array}
    \right.
    \label{equ:def_vij_compatible_1}
  \end{equation}
  where $v_{ij}^{(N)} = v_{ij}^{(N)}(\vec{c}^{(N)})$ and
  $v_{ij}^{(N-1)} = v_{ij}^{(N-1)}(\vec{c}^{(N-1)})$,
  and $\vec{c}^{(N)}$ and $\vec{c}^{(N-1)}$ are
  connected by the correspondence relation \eqref{equ:volfrac_correspondence}.
\end{definition}
\begin{definition}
  \label{def:def_vij_consis}
  A set of variables $v_{ij}^{(N)}(\vec{c}^{(N)})$ ($1\leqslant i,j\leqslant N$;
  $N=1,2,3\dots$)
  is said to be
  reduction-consistent if (i) this set
  is reduction-compatible, and (ii) for any $N\geqslant 2$,
  this set satisfies the following additional property when
  any fluid $k$ ($1\leqslant k\leqslant N$) is absent from
  the N-phase system:
  \begin{equation}
    v_{ki}^{(N)} = v_{ik}^{(N)} = 0, \quad 1\leqslant i\leqslant N
    \label{equ:def_vij_consis}
  \end{equation}
  where $v_{ik}^{(N)} = v_{ik}^{(N)}(\vec{c}^{(N)})$
  and $v_{ki}^{(N)}=v_{ki}^{(N)}(\vec{c}^{(N)})$.
  
\end{definition}
\begin{definition}
  \label{def:def_v_consis}
  A set of variables $v^{(N)}(\vec{c}^{(N)})$ ($N=1,2,3\dots$) is said
  to be reduction consistent if for any $N\geqslant 2$,
   it transforms as follows when any fluid $k$ ($1\leqslant k\leqslant N$)
  is absent from the N-phase system:
  \begin{equation}
    v^{(N)}(\vec{c}^{(N)}) = v^{(N-1)}(\vec{c}^{(N-1)})
  \end{equation}
  where $\vec{c}^{(N)}$ and $\vec{c}^{(N-1)}$ are
  connected by the correspondence relation \eqref{equ:volfrac_correspondence}.

\end{definition}

\begin{remark}
  For a reduction-consistent set of functions $v_i^{(N)}$ ($1\leqslant i\leqslant N$)
  defined on the N-phase system,
if any fluid $k$ ($1\leqslant k\leqslant N$)
is absent from the N-phase system,
then the $k$-th function in this set will vanish while
the other ($N-1$) functions will reduce to
the corresponding functions $v_i^{(N-1)}$ ($1\leqslant i\leqslant N-1$) for
the smaller ($N-1$)-phase system.
For a  reduction-consistent function $v^{(N)}$ defined on the N-phase
system, if any fluid $k$ ($1\leqslant k\leqslant N$)
is absent from the system,
then this function will reduce to the corresponding
function $v^{(N-1)}$ for the smaller ($N-1$)-phase system.
For example, according to the definitions
$\nabla^2 c_i$ ($1\leqslant i\leqslant N$)
is a reduction-consistent set of variables,
and $\rho(\vec{c})=\sum_{i=1}^N\tilde{\rho}_i c_i$
is a reduction-consistent function.
\end{remark}

Based on the transformation properties of variables,
 we can  look into how a given set of
equations of the N-phase system transforms if
any fluid $k$ ($1\leqslant k\leqslant N$) is absent from the system.
Specifically,
we define the reduction consistency of
a set of equations as follows.
\begin{definition}
  \label{def:def_vi_equ_consis}
  A set of equations 
  \begin{equation*}
    v_i^{(N)}(\vec{c}^{(N)}) = 0, \quad 1\leqslant i\leqslant N; \quad N=1,2,3\dots
  \end{equation*}
  is said to be reduction-consistent if the set
  of variables $v_i^{(N)}(\vec{c}^{(N)})$ ($1\leqslant i\leqslant N$; $N=1,2,3\dots$)
  is reduction-consistent.
  
\end{definition}
\begin{definition}
  \label{def:def_v_equ_consis}
  A set of equations
  \begin{equation}
    v^{(N)}(\vec{c}^{(N)}) = 0, \quad N=1,2,3\dots
  \end{equation}
  is said to be reduction-consistent if the set of variables
  $v^{(N)}(\vec{c}^{(N)})$ ($N=1,2,3\dots$) is reduction-consistent.
  
\end{definition}


\begin{remark}
  For a reduction-consistent set of equations $v_i^{(N)}=0$ ($1\leqslant i\leqslant N$)
  defined on the N-phase system,
if any fluid $k$ ($1\leqslant k\leqslant N$)
is absent,
then the $k$-th equation in this set will reduce to
an identity and the other ($N-1$) equations
will reduce to the corresponding equations $v_i^{(N-1)}=0$ ($1\leqslant i\leqslant N-1$)
for the smaller ($N-1$)-phase system.
For a  reduction-consistent equation $v^{(N)}=0$
defined on  the N-phase system, if any fluid $k$ ($1\leqslant k\leqslant N$)
is absent, then this equation
will reduce to the corresponding equation $v^{(N-1)}=0$
for the smaller ($N-1$)-phase system.
\end{remark}

%

We next write down some useful properties about the
reduction-compatible and reduction-consistent variables/functions
of the N-phase system. 
It is straightforward to verify these properties
based on the definitions. In the following we omit the superscript ${(N)}$ 
and assume that the variables are defined for all N-phase
systems ($N=1,2,3\dots$). For example,
the statement below ``$v_i(\vec{c})$ ($1\leqslant i\leqslant N$) is
a reduction-consistent set of functions'' refers to
``$v_i^{(N)}(\vec{c}^{(N)})$ ($1\leqslant i\leqslant N$; $N=1,2,3\dots$) is
a reduction-consistent set of functions'' to be exact, and
the statement below ``$v(\vec{c})$ is a reduction-consistent function''
refers to ``$v^{(N)}(\vec{c}^{(N)})$ ($N=1,2,3\dots$) is a reduction-consistent
set of functions'', etc.
\begin{enumerate}[($\mathscr{T}$1):]

\item
  Reduction consistency implies reduction compatibility. 
  The reverse is not true.
  
\item
  If $v_i(\vec{c})$ ($1\leqslant i\leqslant N$) and
  $w_i(\vec{c})$ ($1\leqslant i\leqslant N$) are two reduction-consistent
  sets of functions, then
  $a v_i(\vec{c})+b w_i(\vec{c})$ ($1\leqslant i\leqslant N$)
  form a reduction-consistent
  set of functions, where $a$ and $b$ are constants.
  The same property holds for reduction consistent sets of
  functions $v_{ij}(\vec{c})$ ($1\leqslant i,j\leqslant N$) and $w_{ij}(\vec{c})$
  ($1\leqslant i,j\leqslant N$).

\item
  If $v_i(\vec{c})$ ($1\leqslant i\leqslant N$) and
  $w_i(\vec{c})$ ($1\leqslant i\leqslant N$) are two reduction-compatible
  sets of functions, then
  $a v_i(\vec{c})+b w_i(\vec{c})$ ($1\leqslant i\leqslant N$)
  form a reduction-compatible
  set of functions with constants $a$ and $b$.
  The same property holds for two reduction-compatible sets of
  functions $v_{ij}(\vec{c})$ ($1\leqslant i,j\leqslant N$) and $w_{ij}(\vec{c})$
  ($1\leqslant i,j\leqslant N$).

\item
  If $v_i(\vec{c})$ ($1\leqslant i\leqslant N$) are a reduction-consistent
  set of functions and $w_i(\vec{c})$ ($1\leqslant i\leqslant N$)
  are a reduction-compatible
  set of functions, then $v_i(\vec{c})w_i(\vec{c})$ ($1\leqslant i\leqslant N$)
  form a reduction-consistent set of functions.

\item
  If $v(\vec{c})$ and $w(\vec{c})$ are two reduction-consistent
  functions, then $a v(\vec{c})+b w(\vec{c})$ is a reduction
  consistent function for constants $a$ and $b$,
  and $v(\vec{c})w(\vec{c})$ is also a reduction-consistent function.
  
\item
  If $v_i(\vec{c})$ ($1\leqslant i\leqslant N$) are a reduction-consistent
  set of functions and $w(\vec{c})$ is a reduction-consistent function,
  then $v_i(\vec{c})w(\vec{c})$ ($1\leqslant i\leqslant N$)
  form a reduction-consistent set of functions.

\item
  If $v_{ij}(\vec{c})$ ($1\leqslant i,j\leqslant N$) are
  a reduction-consistent set of functions, and
  $w_i(\vec{c})$ ($1\leqslant i\leqslant N$) are a reduction-compatible set of
  functions, then $\sum_{j=1}^N v_{ij}w_j$ ($1\leqslant i\leqslant N$)
  and $\sum_{i=1}^N v_{ij}w_i$ ($1\leqslant j\leqslant N$)
  form two reduction-consistent sets of functions.

\item
  If $v_{ij}(\vec{c})$ ($1\leqslant i,j\leqslant N$) are a
  reduction-compatible set of functions, and
  $w_i(\vec{c})$ ($1\leqslant i\leqslant N$) are a reduction-consistent
  set of functions, then $\sum_{j=1}^N v_{ij}w_j$ ($1\leqslant i\leqslant N$)
  and $\sum_{i=1}^N v_{ij}w_i$ ($1\leqslant j\leqslant N$)
  form two reduction-compatible sets of functions.

\item
  If $v_i(\vec{c})$ ($1\leqslant i\leqslant N$) are a
  reduction-consistent set of functions, then
  $\sum_{i=1}^N v_i(\vec{c})$ is a reduction-consistent
  function.

\item
  If $v_{ij}(\vec{c})$ ($1\leqslant i,j\leqslant N$)
  are a reduction-consistent set of functions,
  then $\sum_{j=1}^{N}v_{ij(\vec{c})}$ ($1\leqslant i\leqslant N$)
  $\sum_{i=1}^N v_{ij}(\vec{c})$ ($1\leqslant j\leqslant N$)
  are two reduction-consistent sets of functions,
  and $\sum_{i,j=1}^Nv_{ij}(\vec{c})$ is a reduction-consistent function.

\item
  If $v_{ij}(\vec{c})$ ($1\leqslant i,j\leqslant N$) are
  a reduction-compatible set of functions
  and $w_i(\vec{c})$ ($1\leqslant i\leqslant N$) are
  a reduction-consistent set of functions,
  then $\sum_{i,j=1}^N v_{ij}w_iw_j$ is a reduction-consistent function.

\item
  If $v_{ij}(\vec{c})$ ($1\leqslant i,j\leqslant N$) are
  a reduction-consistent set of functions
  and $w_i(\vec{c})$ ($1\leqslant i\leqslant N$) are
  a reduction-compatible set of functions,
  then $\sum_{i,j=1}^N v_{ij}w_iw_j$ is a reduction-consistent function.
  
\item
  If $v(\vec{c})$, $v_i(\vec{c})$ ($1\leqslant i\leqslant N$),
  and $v_{ij}(\vec{c})$ ($1\leqslant i,j\leqslant N$)
  are reduction-consistent (resp.~reduction-compatible) sets of
  functions, then $\mathscr{P}v$, $\mathscr{P}v_i$ ($1\leqslant i\leqslant N$),
  and $\mathscr{P}v_{ij}$ ($1\leqslant i,j\leqslant N$)
  are also reduction-consistent (resp.~reduction-compatible)
  sets of functions, where $\mathscr{P}$ stands for
  one of the operators $\frac{\partial}{\partial t}$,
  $\nabla$, or $\nabla^2$. If $v$, $v_i$ and $v_{ij}$
  are vector functions, then $\mathscr{P}$ can also
  be divergence and curl operators.

\item
  If a function $v$ is independent of $\vec{c}$, then it is
  a reduction consistent function.

\item
  $c_i$ ($1\leqslant i\leqslant N$) and
  $\rho_i(\vec{c})$ ($1\leqslant i\leqslant N$)
  are two reduction-consistent sets of variables;
  $\rho(\vec{c})$ and $\mu(\vec{c})$ (given by \eqref{equ:rho_expr} and \eqref{equ:mu_expr})
  are each a reduction-consistent
  function.

\item
  $\tilde{\rho}_i$ ($1\leqslant i\leqslant N$) and
  $\tilde{\mu}_i$ ($1\leqslant i\leqslant N$)
  are two reduction-compatible sets of variables.
  $\sigma_{ij}$ ($1\leqslant i,j\leqslant N$)
  are a reduction-compatible set of variables.
  
\end{enumerate}
A proof of the properties ($\mathscr{T}7$)--($\mathscr{T}12$)
is provided in the Appendix B.



Let us now consider how  reduction-consistent functions
transform if more than one fluid components are
absent from the N-phase system.
If any one fluid is absent from the system,
then a reduction-consistent function $v^{(N)}$
will reduce to the function $v^{(N-1)}$ for
the corresponding ($N-1$)-phase system.
By repeatedly applying this property, we conclude
that if $K$ ($1\leqslant K\leqslant N-1$) fluid
components are absent from the N-phase system
then a reduction-consistent function $v^{(N)}$
will reduce to the function $v^{(N-K)}$ for the corresponding ($N-K$)-phase
system.

Similarly, given a reduction-consistent set of functions
$v_i^{(N)}$ ($1\leqslant i\leqslant N$) for the N-phase system,
if any one fluid is absent,
then the function in this set with the index corresponding
to the absent fluid will vanish identically while the other ($N-1$)
functions will reduce to the functions
$v_i^{(N-1)}$ ($1\leqslant i\leqslant N-1$) for the corresponding
($N-1$)-phase system. By repeatedly applying this property
to the resultant ($N-1$)-phase, ($N-2$)-phase, \dots \ systems,
we conclude that if $K$ ($1\leqslant K\leqslant N-1$) fluid
components are absent from the N-phase system, then
those $K$ functions in this set with indices corresponding to
the absent fluids will vanish identically while the other ($N-K$)
functions will reduce to the functions $v_i^{(N-K)}$ ($1\leqslant i\leqslant N-K$) for
the corresponding ($N-K$)-phase system.

It then follows from the above discussions that,
given a reduction-consistent set of equations for the N-phase
system, if $K$ ($1\leqslant K\leqslant N-1$) fluid components
are absent from the system,
then those $K$ equations in this set with indices corresponding
to the absent fluids will each reduce to an identity,
while the other ($N-K$) equations will reduce to
the corresponding equations for the smaller ($N-K$)-phase system.


\subsection{Reduction Consistency of N-Phase Governing Equations}

%

Let us now look into the reduction consistency of the
governing equations given by \eqref{equ:nse_original}--\eqref{equ:CH_original}
for the N-phase system ($N=1,2,3\dots$), that is, how these equations
transform if any fluid component is absent from the system.
Define
\begin{subequations}
  \begin{equation}
    \mathscr{M}(\vec{c}) = \rho(\vec{c})\left(\frac{\partial\mathbf{u}}{\partial t} +
    \mathbf{u}\cdot\nabla\mathbf{u} \right)
    + \tilde{\mathbf{J}}\cdot\nabla\mathbf{u}
     +\nabla p
    - \nabla\cdot\left[\mu(\vec{c})\mathbf{D}(\mathbf{u}) \right]
    + \sum_{i=1}^N \nabla\cdot\left[
    \nabla c_i \otimes \frac{\partial W}{\partial(\nabla c_i)}
    \right],
    \label{equ:def_nse}
  \end{equation}
  \begin{equation}
    \mathscr{N}(\vec{c}) = \nabla\cdot\mathbf{u},
    \label{equ:def_continuity}
  \end{equation}
  \begin{equation}
    \mathscr{F}_i(\vec{c}) = \frac{\partial c_i}{\partial t} + \mathbf{u}\cdot\nabla c_i
    - \sum_{j=1}^N \nabla\cdot\left[
      m_{ij}(\vec{c})\nabla\left(
      \frac{\partial W}{\partial c_j}
      - \nabla\cdot\frac{\partial W}{\partial\nabla c_j}
      \right)
      \right],
    \quad 1\leqslant i\leqslant N,
    \label{equ:def_CH}
  \end{equation}
  \begin{equation}
    \mathscr{H}_i(\vec{c}) = \frac{\partial W}{\partial c_i},
    \quad 1\leqslant i\leqslant N,
    \label{equ:def_dWdci}
  \end{equation}
  \begin{equation}
    \bm{\mathscr{G}}_i(\vec{c}) = \frac{\partial W}{\partial(\nabla c_i)},
    \quad 1\leqslant i\leqslant N,
    \label{equ:def_dWdgradc_orig}
  \end{equation}
  \begin{equation}
    \mathscr{I}_i(\vec{c}) = \nabla\cdot\frac{\partial W}{\partial(\nabla c_i)},
    \quad 1\leqslant i\leqslant N.
    \label{equ:def_dWdgradc}
  \end{equation}
\end{subequations}
We have the following result.
\begin{theorem}
  \label{thm:thm_1}
  If
  \begin{itemize}
  \item
    $m_{ij}(\vec{c})$ ($1\leqslant i,j\leqslant N$) are a reduction-consistent
    set of variables,

  \item
    $\mathscr{H}_i(\vec{c})$ ($1\leqslant i\leqslant N$) are
    a reduction-compatible set of variables, and

  \item
    $\bm{\mathscr{G}}_i(\vec{c})$ ($1\leqslant i\leqslant N$) are
    a reduction-compatible set of variables,
    
  \end{itemize}
  then
  \begin{itemize}
  \item
    Equation \eqref{equ:nse_original} is a reduction-consistent equation;

  \item
    Equation \eqref{equ:continuity_original} is a reduction-consistent equation;

  \item
    The $N$ equations in \eqref{equ:CH_original} are a reduction-consistent
    set of equations.
    
  \end{itemize}
  
\end{theorem}
\noindent A proof of this theorem is provided in  Appendix C.


\begin{remark}
This theorem provides a set of sufficient conditions for
the reduction consistency of the N-phase governing equations.
If one can choose a reduction-consistent
set of functions for the coefficients $m_{ij}(\vec{c})$,
and choose a free energy density function $W(\vec{c},\nabla\vec{c})$
such that $\mathscr{H}_i(\vec{c})$ and $\bm{\mathscr{G}}_i(\vec{c})$
each forms a reduction-compatible set of variables,
then the theorem guarantees that the
resultant N-phase governing equations are reduction consistent.
In other words, if $K$ fluids ($1\leqslant K\leqslant N-1$)
are absent from the N-phase system,
then the N-phase governing equations will exactly reduce to
the ($N-K$)-phase governing equations that correspond to
the ($N-K$)-phase system formed by those ($N-K$) fluids that are present,
together with $K$ additional identities,
identically satisfied by the zero volume-fraction fields
corresponding to the absent fluids.

\end{remark}

\begin{remark}

  Suppose the free energy density function takes the following
  form
  \begin{equation*}
    W(\vec{c},\nabla\vec{c}) =
    \sum_{i,j=1}^{N} \frac{\lambda_{ij}}{2}\nabla c_i\cdot\nabla c_j +
    \text{(multiwell potential term)}
  \end{equation*}
  where the constants $\lambda_{ij}$ are called the mixing energy density coefficients (symmetric),
  and can be
  related to other physical parameters such as the pairwise surface tensions
  by invoking the consistency condition ($\mathscr{C}1$) (see
  e.g.~\cite{Dong2015,Dong2017}).
  The multiwell potential term is assumed to be independent of $\nabla\vec{c}$.
  Then
  $
  \bm{\mathscr{G}}_i(\vec{c}) = \sum_{j=1}^N \lambda_{ij}\nabla c_j
  $
  ($1\leqslant i\leqslant N$).
  Therefore, if $\lambda_{ij}$ ($1\leqslant i,j\leqslant N$) are
  a reduction-compatible set, $\bm{\mathscr{G}}_i(\vec{c})$ ($1\leqslant i\leqslant N$)
  will be a reduction-compatible set of variables according to
  the property ($\mathscr{T}8$) from Section \ref{sec:def_consis}.
  Similarly, if the free energy density function takes the form
  \begin{equation*}
    W(\vec{c},\nabla\vec{c}) =
    \sum_{i=1}^{N} \frac{\lambda_{i}}{2}\left|\nabla c_i\right|^2 + 
    \text{(multiwell potential term)}
  \end{equation*}
  where $\lambda_i$ are constants, 
  then
  $
  \bm{\mathscr{G}}_i(\vec{c}) = \lambda_{i}\nabla c_i
  $
  ($1\leqslant i\leqslant N$).
  If $\lambda_i$ ($1\leqslant i\leqslant N$) is a reduction-compatible
  set, $\bm{\mathscr{G}}_i(\vec{c})$ ($1\leqslant i\leqslant N$) will form
  a reduction-consistent (and thus also reduction-compatible)
  set of variables according to property ($\mathscr{T}4$) from
  Section \ref{sec:def_consis}.
  
\end{remark}


We next suggest a specific form for $m_{ij}(\vec{c})$ and
for $W(\vec{c},\nabla\vec{c})$ that satisfy the conditions
for Theorem \ref{thm:thm_1}.
Let $f(c)$ denote a non-negative continuous function with the property
\begin{equation}
  \left\{
  \begin{split}
    &
    f(c) = 0, \ \ \text{if} \ c\leqslant 0; \\
    &
    f(c) > 0, \ \ \text{if} \ c>0.
  \end{split}
  \right.
  \label{equ:func_f_property}
\end{equation}
In the current work we will use the following  function for $f(c)$,
\begin{equation}
  f(c) = \left\{
  \begin{array}{ll}
    0, & \text{if} \ c<0 \\
    2c, & \text{if} \ c\geqslant 0.
  \end{array}
  \right.
  \label{equ:def_fc}
\end{equation}
Let $\tilde{m}_{ij}$ ($1\leqslant i,j\leqslant N$) denote
a set of non-negative constants with the property 
$\tilde{m}_{ij}=\tilde{m}_{ji}$ ($1\leqslant i,j\leqslant N$)
and $\tilde{m}_{ii}=0$ ($1\leqslant i\leqslant N$),
and that they form 
a reduction-compatible set of variables for $N=1,2,3,\cdots$.
Some specific examples
for such a set of constants are
$
\tilde{m}_{ij} = \tilde{\rho}_i\tilde{\rho}_j(1-\delta_{ij}),
$
$
\tilde{\mu}_i\tilde{\mu}_j(1-\delta_{ij}),
$
or $\sigma_{ij}$,
where $\delta_{ij}$ denotes the Kronecker delta.
In this work we use the following $\tilde{m}_{ij}$,
\begin{equation}
  \tilde{m}_{ij} = m_0(1-\delta_{ij}) = \left\{
  \begin{array}{ll}
  m_0, & 1\leqslant i\neq j\leqslant N, \\
  0, & 1\leqslant i=j\leqslant N
  \end{array}
  \right.
  \label{equ:def_mij_tilde}
\end{equation}
where $m_0>0$ is a positive constant.
These $\tilde{m}_{ij}$ can be shown to form a reduction-compatible set
in a straightforward fashion based on the definition.
We then define $m_{ij}(\vec{c})$ as follows,
\begin{equation}
  \left\{
  \begin{split}
    &
    m_{ij}(\vec{c}) = -\tilde{m}_{ij}f(c_i)f(c_j),
    \quad 1\leqslant i\neq j\leqslant N \\
    &
    m_{ii}(\vec{c}) = -\sum_{\substack{j=1\\ j\neq i}}^N m_{ij}(\vec{c})
    = f(c_i)\sum_{\substack{j=1\\ j\neq i}}^N\tilde{m}_{ij}f(c_j),
    \quad 1\leqslant i\leqslant N.
  \end{split}
  \right.
  \label{equ:def_mij}
\end{equation}
Note that $m_{ij}(\vec{c})$ ($1\leqslant i,j\leqslant N$)
as defined above
satisfy the conditions \eqref{equ:mij_cond}
and \eqref{equ:mij_cond_1}.
So the matrix $\mathbf{m}$ (see equation \eqref{equ:mob_mat_def})
formed by these $m_{ij}(\vec{c})$ is symmetric
positive semi-definite.

For the free energy density function we consider
the following form
\begin{equation}
  W(\vec{c},\nabla\vec{c}) =
  \sum_{i,j=1}^N \frac{\lambda_{ij}}{2}\nabla c_i\cdot\nabla c_j
  + \beta \sum_{i,j=1}^N \frac{\sigma_{ij}}{2}\left[
    g(c_i) + g(c_j) - g(c_i+c_j)
    \right]
  \label{equ:free_energy}
\end{equation}
where
\begin{equation}
  \left\{
  \begin{split}
    &
    \lambda_{ij} = -\frac{3}{\sqrt{2}}\eta\sigma_{ij}, \quad
    1\leqslant i,j\leqslant N, \\
    &
    \beta = \frac{3}{\sqrt{2}}\frac{1}{\eta}, \\
    &
    g(c) = c^2(1-c)^2
  \end{split}
  \right.
  \label{equ:free_energy_param}
\end{equation}
and $\eta$ is the scale of characteristic interfacial thickness
of the diffuse interfaces.
We assume that the values for the
pairwise surface tensions $\sigma_{ij}$ among the $N$ fluids
are such that the $N\times N$ symmetric matrix formed by
$\lambda_{ij}$ is positive semi-definite in order to ensure the
non-negativity of the first term on the right hand side
of \eqref{equ:free_energy}.
This free energy density function is equivalent to
a form originally suggested in \cite{BoyerM2014}.

With the $m_{ij}(\vec{c})$ and $W(\vec{c},\nabla\vec{c})$ defined above,
we have the following result:
\begin{theorem} 
  \label{thm:thm_2}
  (a)  The functions $m_{ij}(\vec{c})$ ($1\leqslant i,j\leqslant N$)
  as defined by \eqref{equ:def_mij} are a  reduction-consistent set of functions.
  (b) The free energy density function $W(\vec{c},\nabla\vec{c})$
  defined in \eqref{equ:free_energy} is a reduction-consistent function.
  (c) With $W(\vec{c},\nabla\vec{c})$ given by \eqref{equ:free_energy},
  the functions $\mathscr{H}_i(\vec{c})$ ($1\leqslant i\leqslant N$) as defined
  by \eqref{equ:def_dWdci} and the functions
  $\bm{\mathscr{G}}_i(\vec{c})$ ($1\leqslant i\leqslant N$) as
  defined by \eqref{equ:def_dWdgradc_orig}
  are each a reduction-compatible set of functions.

\end{theorem}
\noindent A proof of this theorem is provided in Appendix D.


We conclude based on Theorems \ref{thm:thm_1} and \ref{thm:thm_2}
that, with $m_{ij}(\vec{c})$ given by \eqref{equ:def_mij}
and $W(\vec{c},\nabla\vec{c})$ given by \eqref{equ:free_energy},
the N-phase governing equations \eqref{equ:nse_original}--\eqref{equ:CH_original}
are fully reduction-consistent.
With these forms for $m_{ij}(\vec{c})$ and $W(\vec{c},\nabla\vec{c})$,
equations \eqref{equ:nse_original} and \eqref{equ:CH_original}
are transformed into
\begin{equation}
    \rho(\vec{c})\left(\frac{\partial\mathbf{u}}{\partial t} +
    \mathbf{u}\cdot\nabla\mathbf{u} \right)
    + \tilde{\mathbf{J}}\cdot\nabla\mathbf{u}
    = -\nabla p
    + \nabla\cdot\left[\mu(\vec{c})\mathbf{D}(\mathbf{u}) \right]
    - \sum_{i,j=1}^N \nabla\cdot\left(
    \lambda_{ij}\nabla c_i \otimes \nabla c_j
    \right),
    \label{equ:nse}
\end{equation}
\begin{equation}
    \frac{\partial c_i}{\partial t} + \mathbf{u}\cdot\nabla c_i
    = \sum_{j=1}^N \nabla\cdot\left[
      m_{ij}(\vec{c})\nabla\left(
      -\sum_{k=1}^N \lambda_{jk}\nabla^2 c_k +
      \mathscr{H}_j(\vec{c})
      \right)
      \right],
    \quad 1\leqslant i\leqslant N,
    \label{equ:CH}
\end{equation}
where $\rho(\vec{c})$ and $\mu(\vec{c})$ are
given by \eqref{equ:rho_expr} and \eqref{equ:mu_expr}, respectively,
$\lambda_{ij}$ ($1\leqslant i,j\leqslant N$) are given in
\eqref{equ:free_energy_param}, and
\begin{equation}
  \left\{
  \begin{split}
    &
    \tilde{\mathbf{J}} = -\sum_{i,j=1}^N\tilde{\rho}_im_{ij}(\vec{c})\nabla\left[
    -\sum_{k=1}^{N}\lambda_{jk}\nabla^2 c_k + \mathscr{H}_j(\vec{c})
    \right], \\
    &
    \mathscr{H}_i(\vec{c}) = \frac{\partial W}{\partial c_i}
    = \beta\sum_{j=1}^N \sigma_{ij}\left[
      g^{\prime}(c_i) - g^{\prime}(c_i+c_j)
      \right],
    \quad 1\leqslant i\leqslant N.
  \end{split}
  \right.
  \label{equ:H_expr}
\end{equation}
The N-phase formulation represented by \eqref{equ:nse},
\eqref{equ:continuity_original} and \eqref{equ:CH}
fully satisfies the reduction consistency conditions
($\mathscr{C}1$) and ($\mathscr{C}2$).
This formulation
is reduction-consistent and thermodynamically consistent.



\begin{remark}

  The $m_{ij}(\vec{c})$ and $W(\vec{c},\nabla\vec{c})$ functions suggested
  above are only one way to fulfill the conditions of Theorem \ref{thm:thm_1}.
  We would like to point out that
  it is possible to choose other forms for $f(c)$,
  the constants $\tilde{m}_{ij}$, or the free energy density
  function $W(\vec{c},\nabla\vec{c})$ to satisfy these conditions,
  thus leading to other reduction-consistent and thermodynamically
  consistent N-phase formulations.
  For example, the free energy density form (analogous to
  the one from \cite{Dong2015})
  $
  W(\vec{c},\nabla\vec{c})=\sum_{i,j=1}^N\frac{\Lambda_{ij}}{2}\nabla c_i\cdot\nabla c_j
  + b\sum_{i=1}^N c_i^2(1-c_i)^2,
  $
  where $\Lambda_{ij}\sim -\sigma_{ij}^2$ ($1\leqslant i,j\leqslant N$)
  and $b$ is some constant, also leads to
  $\mathscr{H}_i(\vec{c})$ and $\bm{\mathscr{G}}_i(\vec{c})$
  functions that are reduction compatible.
  One can also employ for example
  \begin{equation*}
    f(c) = \left\{
    \begin{array}{ll}
      0, & c<0 \\
      (2c)^k, & c\geqslant 0
    \end{array}
    \right.
    \ (k\geqslant 2 \ \text{is an integer}),
    \quad \text{or} \
    f(c) = \left\{
    \begin{array}{ll}
      0, & c<0 \\
      1-\cos(\pi c), & 0\leqslant c\leqslant 1 \\
      2, & c>1, 
    \end{array}
    \right.
  \end{equation*}
  which leads to a reduction-consistent set of $m_{ij}(\vec{c})$
  as defined in \eqref{equ:def_mij}.
  
\end{remark}

\begin{remark}
  
  If $m_{ij}(\vec{c})$ ($1\leqslant i,j\leqslant N$) are assumed to be 
  all constants and are not identically zeros, then based on
  Definition \ref{def:def_vij_consis}
  $m_{ij}$ ($1\leqslant i,j\leqslant N$) cannot be a reduction-consistent
  set. Therefore, in this case one has to treat the term
  $
  \sum_{j=1}^Nm_{ij}\nabla\left[\mathscr{H}_i(\vec{c})-\mathscr{I}_i(\vec{c}) \right]
  $
  in $\mathscr{F}_i(\vec{c})$
  as a whole, and try to construct $W(\vec{c},\nabla\vec{c})$ such that
  this expression results in a reduction-consistent set.
  This is essentially the approach taken by \cite{BoyerM2014,Dong2017},
  and it is extremely difficult (if not impossible) to construct such
  a $W(\vec{c},\nabla\vec{c})$ to ensure {\bf full reduction consistency}.
  So far, only a partial reduction consistency (e.g.~between $N$ phases
  and two phases) can be achieved with this approach for an arbitrary
  set of given pairwise surface tension values \cite{Dong2017,BoyerM2014}.

\end{remark}

%% file: Algorithm.tex
\section{Numerical Algorithm and Implementation}
\label{sec:alg}


We now look into how to numerically solve the
reduction-consistent and thermodynamically consistent
N-phase governing equations.
Let $\Omega$ denote the flow domain,
and $\partial\Omega$ denote its boundary.
On $\partial\Omega$ we assume that the velocity
distribution is known,
\begin{equation}
  \mathbf{u} = \mathbf{w}(\mathbf{x},t), \quad
  \text{on} \ \partial\Omega
  \label{equ:bc_vel}
\end{equation}
where $\mathbf{w}$ is the boundary velocity.
We consider the following boundary conditions
for the volume fractions $c_i$,
\begin{subequations}
  \begin{equation}
    \sum_{j=1}^Nm_{ij}(\vec{c})\mathbf{n}\cdot\nabla\left[
      -\sum_{k=1}^N \lambda_{jk}\nabla^2 c_k + \mathscr{H}_j(\vec{c})
      \right] = 0,
    \quad 1\leqslant i\leqslant N,
    \quad \text{on} \ \partial\Omega,
    \label{equ:bc_ci_1}
  \end{equation}
  \begin{equation}
    \mathbf{n}\cdot\nabla c_i = 0, \quad
    1\leqslant i\leqslant N,
    \quad \text{on} \ \partial\Omega,
    \label{equ:bc_ci_2}
  \end{equation}
\end{subequations}
where $\mathbf{n}$ is the outward-pointing unit vector
normal to  $\partial\Omega$.
The boundary conditions \eqref{equ:bc_ci_1}
and \eqref{equ:bc_ci_2} correspond to a
wall with neutral wettability (i.e.~90-degree contact
angle) for all the fluid interfaces.
We further assume that the distributions of
the velocity $\mathbf{u}$ and the volume fractions $c_i$
at $t=0$ are known
\begin{subequations}
  \begin{align}
    &
    \mathbf{u}(\mathbf{x},0) = \mathbf{u}_{in}(\mathbf{x}),
    \label{equ:ic_vel} \\
    &
    c_i(\mathbf{x},0) = c_i^{in}(\mathbf{x}), \ \ 1\leqslant i\leqslant N
    \label{equ:ic_phi}
  \end{align}
\end{subequations}
where $\mathbf{u}_{in}$ and $c_i^{in}$ are the initial velocity
and volume fractions.

One notes that
the boundary condition \eqref{equ:bc_vel} is
a reduction-consistent equation.
The $N$ equations given in the boundary condition
\eqref{equ:bc_ci_1} form a reduction-consistent
set of equations, and the $N$ equations given in
\eqref{equ:bc_ci_2} also form a
reduction-consistent set.
Therefore, the boundary conditions
\eqref{equ:bc_vel}, \eqref{equ:bc_ci_1} and \eqref{equ:bc_ci_2}
satisfy the reduction consistency property ($\mathscr{C}3$).


The equations \eqref{equ:nse}, \eqref{equ:continuity_original}
and \eqref{equ:CH}, supplemented by
the boundary conditions \eqref{equ:bc_vel}--\eqref{equ:bc_ci_2}
and initial conditions \eqref{equ:ic_vel}--\eqref{equ:ic_phi},
together constitute the system to be
solved in numerical simulations.
Note that among the $N$ phase field equations in
\eqref{equ:CH} only ($N-1$) equations are independent
in light of \eqref{equ:volfrac_relation} and
\eqref{equ:mij_cond}.
Similarly, only ($N-1$) equations in the boundary condition
\eqref{equ:bc_ci_1} and in \eqref{equ:bc_ci_2}
are independent.
We will employ the first ($N-1$) equations in \eqref{equ:CH}
and in \eqref{equ:bc_ci_1}--\eqref{equ:bc_ci_2}
to solve for
the volume fractions $c_i$ ($1\leqslant i\leqslant N-1$),
and then compute $c_N$ using the relation \eqref{equ:volfrac_relation}.

To facilitate subsequent discussions, we re-write
equation \eqref{equ:nse} in an equivalent form
\begin{equation}
  \frac{\partial\mathbf{u}}{\partial t}
  + \mathbf{u}\cdot\nabla\mathbf{u}
  + \frac{1}{\rho}\tilde{\mathbf{J}}\cdot\nabla\mathbf{u}
  = -\frac{1}{\rho}\nabla P
  + \frac{\mu}{\rho}\nabla^2\mathbf{u}
  + \frac{1}{\rho}\nabla\mu\cdot\mathbf{D}(\mathbf{u})
  - \frac{1}{\rho}\sum_{i,j=1}^N\lambda_{ij}\nabla^2 c_j\nabla c_i
  + \frac{1}{\rho}\mathbf{f}(\mathbf{x},t),
  \label{equ:nse_trans_1}
\end{equation}
where $P = p + \sum_{i,j=1}^N \frac{\lambda_{ij}}{2}\nabla c_i\cdot\nabla c_j$
is an auxiliary pressure, which hereafter will also be loosely referred
to as the pressure, and we have
added an external body force $\mathbf{f}(\mathbf{x},t)$.
We re-write the first ($N-1$) equations in \eqref{equ:CH} as
\begin{equation}
\frac{\partial c_i}{\partial t} + \mathbf{u}\cdot\nabla c_i
    = \sum_{j=1}^N \nabla\cdot\left[
      m_{ij}(\vec{c})\nabla\left(
      -\sum_{k=1}^N \lambda_{jk}\nabla^2 c_k +
      \mathscr{H}_j(\vec{c})
      \right)
      \right] + d_i(\mathbf{x},t),
    \quad 1\leqslant i\leqslant N-1,
    \label{equ:CH_trans_1}
\end{equation}
where we have added in each equation a source
term $d_i(\mathbf{x},t)$ ($1\leqslant i\leqslant N-1$),
which is a prescribed function for the purpose of numerical
testing only and will be set to $d_i=0$ in actual simulations.
We re-write the boundary conditions (first ($N-1$) equations)
\eqref{equ:bc_ci_1}--\eqref{equ:bc_ci_2} as follows:
\begin{subequations}
  \begin{equation}
    \sum_{j=1}^Nm_{ij}(\vec{c})\mathbf{n}\cdot\nabla\left[
      -\sum_{k=1}^N \lambda_{jk}\nabla^2 c_k + \mathscr{H}_j(\vec{c})
      \right] = d_{ai}(\mathbf{x},t),
    \quad 1\leqslant i\leqslant N-1,
    \quad \text{on} \ \partial\Omega,
    \label{equ:bc_ci_1_trans}
  \end{equation}
  \begin{equation}
    \mathbf{n}\cdot\nabla c_i = d_{bi}(\mathbf{x},t), \quad
    1\leqslant i\leqslant N-1,
    \quad \text{on} \ \partial\Omega,
    \label{equ:bc_ci_2_trans}
  \end{equation}
\end{subequations}
where $d_{ai}(\mathbf{x},t)$ ($1\leqslant i\leqslant N-1$)
and $d_{bi}(\mathbf{x},t)$ ($1\leqslant i\leqslant N-1$)
are prescribed source terms on $\partial\Omega$
for the purpose of numerical testing only, and
will be set to $d_{ai}=0$ and $d_{bi}=0$ in
actual simulations.
In these equations $m_{ij}(\vec{c})$ are given
by \eqref{equ:def_mij}, in which $\tilde{m}_{ij}$
are given by \eqref{equ:def_mij_tilde} and
$f(c)$ is defined by \eqref{equ:def_fc}.
$\lambda_{ij}$ are given in \eqref{equ:free_energy_param},
and $\tilde{\mathbf{J}}$ and $\mathscr{H}_i(\vec{c})$
are given in \eqref{equ:H_expr}.


The numerical algorithm presented below
is for the equations
\eqref{equ:nse_trans_1}, \eqref{equ:continuity_original}
and \eqref{equ:CH_trans_1}, together
with the boundary conditions \eqref{equ:bc_vel},
\eqref{equ:bc_ci_1_trans} and \eqref{equ:bc_ci_2_trans}.

The momentum equations \eqref{equ:nse_trans_1} and
\eqref{equ:continuity_original} have the same structure
as those encountered in previous works~\cite{Dong2014,Dong2017}.
Therefore they can be solved using the algorithm
we developed in \cite{Dong2014,Dong2017} for the momentum equations.
For the sake of completeness, we provide a summary
of the scheme for the momentum equations in Appendix E. This is
a semi-implicit splitting type algorithm.
The computations for the pressure and velocity
are de-coupled with this scheme, and it involves
only constant and time-independent coefficient matrices
for both the pressure and the velocity
linear algebraic systems after discretization.


We present below an algorithm
for numerically solving the set of phase field equations
\eqref{equ:CH_trans_1}, together with the boundary
conditions \eqref{equ:bc_ci_1_trans} and
\eqref{equ:bc_ci_2_trans}. The variable nature of the coefficients
$m_{ij}(\vec{c})$ ($1\leqslant i,j\leqslant N$) is a new
feature compared with those encountered in \cite{Dong2014,Dong2015,Dong2017},
and must be dealt with in an appropriate way.

Let $n\geqslant 0$ denote the time step index and $\Delta t$
the time step size.
We use $(\cdot)^n$ to represent the variable $(\cdot)$ at
time step $n$. 
Let $J$ ($J=1$ or $2$) denote the temporal order of accuracy
of the algorithm, and 
\begin{equation}
  \mathbf{R}_i(\vec{c}) = \sum_{j=1}^N m_{ij}(\vec{c})\nabla\left[
    -\sum_{k=1}^N \lambda_{jk}\nabla^2 c_k + \mathscr{H}_j(\vec{c})
    \right],
  \quad 1\leqslant i\leqslant N.
  \label{equ:Ri_expr}
\end{equation}
Given ($\mathbf{u}^{n},c_i^n$), we solve for $c_i^{n+1}$ with
the algorithm as follows,
\begin{subequations}
  \begin{equation}
    \begin{split}
    \frac{\gamma_0 c_i^{n+1} - \hat{c}_i}{\Delta t}
    + \mathbf{u}^{*,n+1}\cdot\nabla c_i^{*,n+1}
    =& \mathcal{K}_0\nabla^2\left[
      -\nabla^2(c_i^{n+1}-c_i^{*,n+1})
      + S(c_i^{n+1}-c_i^{*,n+1})
      \right] \\
    &+ \nabla\cdot \mathbf{R}_i(\vec{c}^{*,n+1})
    + d_i^{n+1},
    \quad 1\leqslant i\leqslant N-1,
    \end{split}
    \label{equ:phase_1}
  \end{equation}
  \begin{multline}
    \mathcal{K}_0\mathbf{n}\cdot\nabla\left[
      -\nabla^2(c_i^{n+1}-c_i^{*,n+1})
      + S(c_i^{n+1}-c_i^{*,n+1})
      \right] \\
    + \mathbf{n}\cdot\mathbf{R}_i(\vec{c}^{*,n+1})
    = d_{ai}^{n+1}, 
    \quad 1\leqslant i\leqslant N-1,
    \ \ \text{on} \ \partial\Omega,
    \label{equ:phase_2}
  \end{multline}
  \begin{equation}
    \mathbf{n}\cdot\nabla c_i^{n+1} = d_{bi}^{n+1},
    \quad 1\leqslant i\leqslant N-1,
    \ \ \text{on} \ \partial\Omega.
    \label{equ:phase_3}
  \end{equation}
\end{subequations}
%
%
If $\chi$ denotes a generic variable, then
in the above equations $\frac{1}{\Delta t}(\gamma_0\chi^{n+1}-\hat{\chi})$
represents an approximation
of $\left.\frac{\partial \chi}{\partial t} \right|^{n+1}$
with the $J$-th order backward differentiation formula (BDF),
with $\gamma_0$ and $\hat{\chi}$ given by
\begin{equation}
  \hat{\chi} = \left\{
  \begin{array}{ll}
    \chi^n, & J=1, \\
    2\chi^n-\frac{1}{2}\chi^{n-1},& J=2;
  \end{array}
  \right.
  \qquad
  \gamma_0 = \left\{
  \begin{array}{ll}
    1, & J=1, \\
    3/2, & J=2.
  \end{array}
  \right.
  \label{equ:def_var_hat}
\end{equation}
$\chi^{*,n+1}$ denotes a $J$-th order explicit approximation of $\chi^{n+1}$
given by
\begin{equation}
  \chi^{*,n+1} = \left\{
  \begin{array}{ll}
  \chi^n, & J=1, \\
  2\chi^n - \chi^{n-1}, & J=2.
  \end{array}
  \right.
  \label{equ:def_var_star}
\end{equation}
%
The positive constant
$\mathcal{K}_0$ is given by
\begin{equation}
  \mathcal{K}_0 = Nm_0 \left| \sum_{i,j=1}^N \lambda_{ij} \right|.
  \label{equ:def_K0}
\end{equation}
%
%
$S$ is a chosen constant, which must satisfy a
condition to be specified later.
$\vec{c}^{*,n+1}$ is defined by
$\vec{c}^{*,n+1} = (c_1^{*,n+1},\dots,c_N^{*,n+1})$.


The key construction in the above algorithm lies
in the two extra terms in the semi-discretized
phase-field equations \eqref{equ:phase_1} and in the boundary conditions
\eqref{equ:phase_2},
$\mathcal{K}_0\nabla^2\left[\nabla^2(c_i^{n+1}-c_i^{*,n+1})  \right]$
and $\mathcal{K}_0S\nabla^2(c_i^{n+1}-c_i^{*,n+1})$,
and the explicit treatment of the
$\nabla\cdot\mathbf{R}_i(\vec{c})$ term.
Note that the two extra terms are both equivalent to zeros,
to the $J$-th order accuracy. 
With these treatments the computations for
different volume fractions $c_i$ ($1\leqslant i\leqslant N-1$) are de-coupled.
Moreover, for each $c_i$ the extra terms in the algorithm allow
us to transform the equation of a $4$-th spatial order
into two {\em de-coupled} 2nd-order equations,
which will become clear below.

Equation \eqref{equ:phase_1} can be written as
\begin{equation}
  \begin{split}
  \frac{\gamma_0}{\mathcal{K}_0\Delta t}c_i^{n+1}
  + \nabla^2(\nabla^2 c_i^{n+1}) - S\nabla^2c_i^{n+1}
  = Z_i
  =& Q_i + \nabla^2(\nabla^2 c_i^{*,n+1})
  - S\nabla^2 c_i^{*,n+1} \\
  &+ \frac{1}{\mathcal{K}_0}\nabla\cdot\mathbf{R}_i(\vec{c}^{*,n+1}),
  \quad 1\leqslant i\leqslant N-1,
  \end{split}
  \label{equ:ci_trans_1}
\end{equation}
where
\begin{equation}
Q_i = \frac{1}{\mathcal{K}_0}\left(
d_i^{n+1} + \frac{\hat{c}_i}{\Delta t}
- \mathbf{u}^{*,n+1}\cdot\nabla c_i^{*,n+1}
\right),
\quad 1\leqslant i\leqslant N-1.
\label{equ:def_Qi}
\end{equation}
Each of the above equations has a form similar to that encountered
in two-phase flows (see e.g.~\cite{DongS2012}). Therefore each of
them can be transformed into two de-coupled
Helmholtz-type equations using the same idea as in two-phase
flows~\cite{DongS2012}.
By adding/subtracting a term $\alpha\nabla^2 c_i^{n+1}$
($\alpha$ denoting a constant to be determined)
on the left hand side (LHS), we can transform \eqref{equ:ci_trans_1}
into
\begin{equation}
  \nabla^2\left[
    \nabla^2 c_i^{n+1} + \alpha c_i^{n+1}
    \right]
  - (\alpha+S)\left[
    \nabla^2 c_i^{n+1}
    - \frac{\gamma_0}{(\alpha+S)\mathcal{K}_0\Delta t} c_i^{n+1}
    \right]
  = Z_i,
  \quad 1\leqslant i\leqslant N-1.
  \label{equ:ci_trans_2}
\end{equation}
By requiring that
$
\alpha = -\frac{\gamma_0}{(\alpha+S)\mathcal{K}_0\Delta t},
$
we obtain
\begin{equation}
  \alpha = \frac{1}{2}\left[
    -S + \sqrt{S^2 - \frac{4\gamma_0}{\mathcal{K}_0\Delta t}}
    \right],
  \quad \text{and the condition} \ \
  S \geqslant \sqrt{\frac{4\gamma_0}{\mathcal{K}_0\Delta t}}.
  \label{equ:def_alpha}
\end{equation}
The chosen constant $S$ must satisfy the
above condition.

Therefore, equation \eqref{equ:ci_trans_2} can be
written equivalently as
\begin{subequations}
  \begin{equation}
    \nabla^2\psi_i^{n+1} - (\alpha+S)\psi_i^{n+1}
    = Z_i, \quad 1\leqslant i\leqslant N-1,
    \label{equ:CH_psi}
  \end{equation}
\begin{equation}
  \nabla^2 c_i^{n+1} + \alpha c_i^{n+1} = \psi_i^{n+1},
  \quad 1\leqslant i\leqslant N-1,
  \label{equ:CH_phi}
\end{equation}
\end{subequations}
where $\psi_i^{n+1}$ ($1\leqslant i\leqslant N-1$) are
auxiliary variables and are defined by
equation \eqref{equ:CH_phi}.
The two equations \eqref{equ:CH_psi} and \eqref{equ:CH_phi}
are Helmholtz type equations, and they
can be solved in a de-coupled fashion.
Note that under the condition for $S$ given in
\eqref{equ:def_alpha}, $\alpha<0$ and $\alpha+S>0$.
In order to solve \eqref{equ:ci_trans_2},
one can first solve \eqref{equ:CH_psi} for $\psi_i^{n+1}$
and then solve \eqref{equ:CH_phi} for $c_i^{n+1}$.

In light of equation \eqref{equ:CH_phi},
the boundary condition \eqref{equ:phase_2} can be transformed into
\begin{multline}
\mathbf{n}\cdot\nabla\psi_i^{n+1}
- (\alpha+S)\mathbf{n}\cdot\nabla c_i^{n+1}
= \mathbf{n}\cdot\nabla\left(
\nabla^2 c_i^{*,n+1} - Sc_i^{*,n+1}
\right)  \\
+ \frac{1}{\mathcal{K}_0}\mathbf{n}\cdot\mathbf{R}_i(\vec{c}^{*,n+1}) 
- \frac{1}{\mathcal{K}_0}d_{ai}^{n+1}, 
\quad 1\leqslant i\leqslant N-1.
\label{equ:bc_psi_1}
\end{multline}
By using \eqref{equ:phase_3} we can further transform the above
equation into
\begin{multline}
\mathbf{n}\cdot\nabla\psi_i^{n+1} =
 \mathbf{n}\cdot\nabla\left(
\nabla^2 c_i^{*,n+1} - Sc_i^{*,n+1}
\right)
+ \frac{1}{\mathcal{K}_0}\mathbf{n}\cdot\mathbf{R}_i(\vec{c}^{*,n+1}) \\
+ (\alpha+S) d_{bi}^{n+1}
- \frac{1}{\mathcal{K}_0}d_{ai}^{n+1},
\quad 1\leqslant i\leqslant N-1.
\label{equ:bc_psi_2}
\end{multline}
These are the boundary conditions for the auxiliary variables
$\psi_i^{n+1}$ ($1\leqslant i\leqslant N-1$).


We employ the spectral element method~\cite{SherwinK1995,KarniadakisS2005,ZhengD2011}
for spatial discretizations in this work. Let us now consider
how to implement the above algorithm using 
$C^0$ spectral elements.
We first derive the weak forms for the equations \eqref{equ:CH_psi}
and \eqref{equ:CH_phi}, by assuming that all variables are
in the continuum space. Then we restrict the test and trial
functions to the appropriate function space for spatial
discretization of the weak forms.

Let $\varphi(\mathbf{x})$ denote an arbitrary (test) function.
Multiply $\varphi$ to equation \eqref{equ:CH_psi} and integrate
over the flow domain $\Omega$, and we get
\begin{equation}
\begin{split}
&
\int_{\Omega}\nabla\psi_i^{n+1}\cdot\nabla\varphi
+ (\alpha+S)\int_{\Omega}\psi_i^{n+1}\varphi \\
&= -\int_{\Omega} Q_i \varphi
+ \int_{\Omega}\left[
\nabla\left(\nabla^2c_i^{*,n+1}  - Sc_i^{*,n+1} \right)
+\frac{1}{\mathcal{K}_0}\mathbf{R}_i(\vec{c}^{*,n+1})
\right]\cdot\nabla\varphi \\
& \quad +\int_{\partial\Omega}\left[
 \mathbf{n}\cdot\nabla\psi_i^{n+1} - \mathbf{n}\cdot\nabla\left(\nabla^2c_i^{*,n+1} -Sc_i^{*,n+1}
 \right)
 -\frac{1}{\mathcal{K}_0}\mathbf{n}\cdot\mathbf{R}_i(\vec{c}^{*,n+1})
\right]\varphi, \\
& \quad \forall\varphi, \ \
1\leqslant i\leqslant N-1
\end{split}
\label{equ:psi_trans_1}
\end{equation} 
where we have used integration by part and the divergence theorem.
In light of the equations \eqref{equ:CH_phi} and \eqref{equ:bc_psi_2},
we can transform the above equation into the final weak form
about $\psi_i^{n+1}$,
\begin{equation}
\begin{split}
&
\int_{\Omega}\nabla\psi_i^{n+1}\cdot\nabla\varphi
+ (\alpha+S)\int_{\Omega}\psi_i^{n+1}\varphi \\
&= -\int_{\Omega}Q_i \varphi
+ \int_{\Omega}\left[
\nabla\left(\psi_i^{*,n+1} - (\alpha+S) c_i^{*,n+1}\right) \right. \\
&\qquad\qquad\qquad \left.
+\frac{1}{\mathcal{K}_0}\sum_{j=1}^N m_{ij}(\vec{c}^{*,n+1})\nabla\left(
 -\sum_{k=1}^N\lambda_{jk}\left(\psi_k^{*,n+1}-\alpha c_k^{*,n+1}\right)
 + \mathscr{H}_j(\vec{c}^{*,n+1})
\right)
\right]\cdot\nabla\varphi \\
&\quad + \int_{\partial\Omega}\left[ (\alpha+S) d_{bi}^{n+1}
-\frac{1}{\mathcal{K}_0} d_{ai}^{n+1}\right]\varphi,
 \qquad \forall\varphi, 
\quad 1\leqslant i\leqslant N-1,
\end{split}
\label{equ:psi_weakform}
\end{equation}
where $\psi_N^{n}$ is defined by
$
\psi_N^n = \nabla^2 c_N^n + \alpha c_N^n.
$


Multiplying the test function $\varphi$ to equation
\eqref{equ:CH_phi} and integrating over the domain $\Omega$, we get
the weak form about $c_i^{n+1}$,
\begin{equation}
\int_{\Omega}\nabla c_i^{n+1}\cdot\nabla\varphi
-\alpha\int_{\Omega}c_i^{n+1}\varphi
=-\int_{\Omega}\psi_i^{n+1}\varphi
+\int_{\partial\Omega}d_{bi}^{n+1}\varphi,
\quad \forall\varphi,
\quad 1\leqslant i\leqslant N-1,
\label{equ:phi_weakform}
\end{equation}
where we have used the divergence theorem and
the boundary condition \eqref{equ:phase_3}.


We discretize the domain $\Omega$ using a mesh
of $N_{el}$ non-overlapping conforming spectral elements.
We use the positive integer $K$ to denote 
the element order, which is a measure of the highest
polynomial degree in field expansions within an element.
Let $\Omega_h$ denote the discretized domain, and
$\Omega_h^e$ ($1\leqslant e\leqslant N_{el}$)
denote the element $e$. Define function space
\begin{equation}
H_{\phi} = \left\{\
v\in H^1(\Omega_h) \ : \ v \ \text{is a polynomial of degree characterized
by} \ K \ \text{on} \ \Omega_h^e, \ \text{for} \ 1\leqslant e\leqslant N_{el}
\ \right\}.
\end{equation}
In the following let the subscript in $(\cdot)_h$ denote 
the discretized version of the variable $(\cdot)$.
The fully discretized equations are: \\
\noindent\underline{For $\psi_{hi}^{n+1}$:} \ \
find $\psi_{hi}^{n+1}\in H_{\phi}$ such that
\begin{equation}
\begin{split}
&
\int_{\Omega_h}\nabla\psi_{hi}^{n+1}\cdot\nabla\varphi_h
+ (\alpha+S)\int_{\Omega_h}\psi_{hi}^{n+1}\varphi_h \\
&= -\int_{\Omega_h}Q_{hi} \varphi_h
+ \int_{\Omega_h}\left[
\nabla\left(\psi_{hi}^{*,n+1} - (\alpha+S) c_{hi}^{*,n+1}\right) \right. \\
&\qquad\qquad \left.
+\frac{1}{\mathcal{K}_0}\sum_{j=1}^N m_{hij}(\vec{c}_h^{*,n+1})\nabla\left(
 -\sum_{k=1}^N\lambda_{jk}\left(\psi_{hk}^{*,n+1}-\alpha c_{hk}^{*,n+1}\right)
 + \mathscr{H}_{hj}(\vec{c}_h^{*,n+1})
\right)
\right]\cdot\nabla\varphi_h \\
&\quad + \int_{\partial\Omega_h}\left[ (\alpha+S) d_{bhi}^{n+1}
-\frac{1}{\mathcal{K}_0} d_{ahi}^{n+1}\right]\varphi_h,
 \qquad \forall\varphi_h\in H_{\phi},
\quad 1\leqslant i\leqslant N-1.
\end{split}
\label{equ:psi_weakform_disc}
\end{equation} 
\noindent\underline{For $c_{hi}^{n+1}$:} \ \
find $c_{hi}^{n+1}\in H_{\phi}$ such that
\begin{equation}
\int_{\Omega_h}\nabla c_{hi}^{n+1}\cdot\nabla\varphi_h
-\alpha\int_{\Omega_h}c_{hi}^{n+1}\varphi_h
=-\int_{\Omega_h}\psi_{hi}^{n+1}\varphi_h
+\int_{\partial\Omega_h}d_{bhi}^{n+1}\varphi_h,
\quad \forall\varphi_h\in H_{\phi},
\  1\leqslant i\leqslant N-1.
\label{equ:phi_weakform_disc}
\end{equation}


Therefore, we employ the following steps to
compute $c_i$ ($1\leqslant i\leqslant N$)
within each time step, which will be referred to as the {\bf AdvancePhaseField}
procedure. \\
\noindent\underline{\bf AdvancePhaseField:} 
\begin{itemize}

\item
  Solve equation \eqref{equ:psi_weakform_disc}
  for $\psi_i^{n+1}$ ($1\leqslant i\leqslant N-1$);

\item
  Solve equation \eqref{equ:phi_weakform_disc}
  for $c_i^{n+1}$ ($1\leqslant i\leqslant N-1$);

\item
Compute $c_N^{n+1}$ and $\psi_N^{n+1}$ by
\begin{equation}
c_N^{n+1} = 1-\sum_{i=1}^{N-1}c_i^{n+1}, \quad
\psi_N^{n+1} = \nabla^2c_N^{n+1} + \alpha c_N^{n+1} = \alpha -\sum_{i=1}^{N-1}\psi_i^{n+1}.
\label{equ:def_psi_N}
\end{equation}

\end{itemize}


Combining the above algorithm for
the phase field equations and the algorithm outlined in
Appendix E for the momentum equations, we arrive
at the following overall method for solving the equations \eqref{equ:nse_trans_1},
\eqref{equ:continuity_original} and \eqref{equ:CH_trans_1}
together with the boundary conditions
\eqref{equ:bc_vel}, \eqref{equ:bc_ci_1_trans}
and \eqref{equ:bc_ci_2_trans}.
Given ($\mathbf{u}^n$, $P^n$, $c_i^n$), we compute $c_i^{n+1}$
($1\leqslant i\leqslant N$), $P^{n+1}$ and $\mathbf{u}^{n+1}$
successively in a de-coupled fashion through the following
steps:
\begin{itemize}

\item
  Compute $\psi_i^{n+1}$ and $c_i^{n+1}$ ($1\leqslant i\leqslant N$)
  using the {\bf AdvancePhaseField} procedure;
  
\item
  Solve equation \eqref{equ:p_weakform} for $P^{n+1}$;

\item
  Solve equation \eqref{equ:vel_weakform} for $\mathbf{u}^{n+1}$.

\end{itemize}
Note that this method involves only the solution of linear
algebraic systems with constant and time-independent
coefficient matrices after discretization, even though the governing
equations of the system involve time-dependent field variables such
as $m_{ij}(\vec{c})$, $\rho(\vec{c})$ and $\mu(\vec{c})$.


%% file: Tests.tex
\section{Representative Numerical Examples}
\label{sec:test}


In this section we present 
numerical simulations
of several multiphase flow problems in two dimensions to demonstrate
the accuracy and effectiveness of the formulation
and the algorithm developed in previous sections.
These problems involve multiple fluid components with
large density contrasts and large viscosity contrasts,
and the simulation results will be compared with
theoretical results or exact physical solutions
for certain cases.

\begin{table}
\begin{center}
\begin{tabular*}{1.0\textwidth}{@{\extracolsep{\fill}} l c | l c}
\hline
variables & normalization constant &
variables & normalization constant \\
$\mathbf{x}$, $\eta$ & $L$ 
& $t$, $\Delta t$ & $L/U_0$ \\
$\mathbf{u}$, $\tilde{\mathbf{u}}$, $\mathbf{w}$, $d_{ai}$ & $U_0$
& $p$, $P$, $W(\vec{c},\nabla\vec{c})$, $H(\vec{c})$, $\mathscr{H}_i(\vec{c})$
& $\varrho_dU_0^2$ \\
$\lambda_{ij}$  & $\varrho_dU_0^2L^2$
& $\rho$, $\rho_i$, $\tilde{\rho}_i$, $\rho_0$  & $\varrho_d$ \\
$\mu$, $\tilde{\mu}_i$ & $\varrho_dU_0L$ 
& $m_{ij}(\vec{c})$, $\tilde{m}_{ij}$, $m_0$, $\mathcal{K}_0$ & $\frac{L}{\varrho_d U_0}$ \\
$\tilde{\mathbf{J}}$, $\mathbf{J}_{i}$ & $\varrho_dU_0$
& $d_i$ & $U_0/L$ \\
$\sigma_{ij}$ & $\varrho_dU_0^2L$
& $d_{bi}$ & $1/L$ \\
$\mathbf{f}$ & $\varrho_dU_0^2/L$
& $S$, $\alpha$, $\psi_i$ & $1/L^2$ \\
$c_i$, $\gamma_0$ & $1$
& $\nu_0$ & $U_0L$ \\
 $\mathbf{g}_r$ (gravity) & $U_0^2/L$ \\
\hline
\end{tabular*}
\caption{
Normalization of flow variables and parameters.
$L$: a characteristic length scale; $U_0$: a characteristic velocity scale;
$\varrho_d$: a characteristic density scale.
}
\label{tab:normalization}
\end{center}
\end{table}

A comment on the normalization of physical
variables and parameters is in order.
As discussed in previous works~\cite{Dong2014,Dong2015,Dong2017},
the non-dimensionalized problem (governing equations,
boundary/initial conditions) will retain the same
form as the dimensional problem as long as
the variables are normalized consistently.
We choose a length scale $L$, a velocity scale $U_0$,
and a density scale $\varrho_d$. The values
for these scales will be specified when individual
test problems are investigated.
In Table~\ref{tab:normalization} we list
the normalization constants for the physical variables
encountered in this work.
According to this table, for instance, the non-dimensional
pairwise surface tension is given by
$\frac{\sigma_{ij}}{\varrho_d U_0^2L}$.
Hereafter we will assume that all physical variables have been
appropriately normalized based on Table~\ref{tab:normalization}.
All the variables in subsequent discussions are in non-dimensional
forms unless otherwise specified.

\subsection{Convergence Rates}
\label{sec:conv}

\begin{figure}
  \centerline{
    \includegraphics[width=3in]{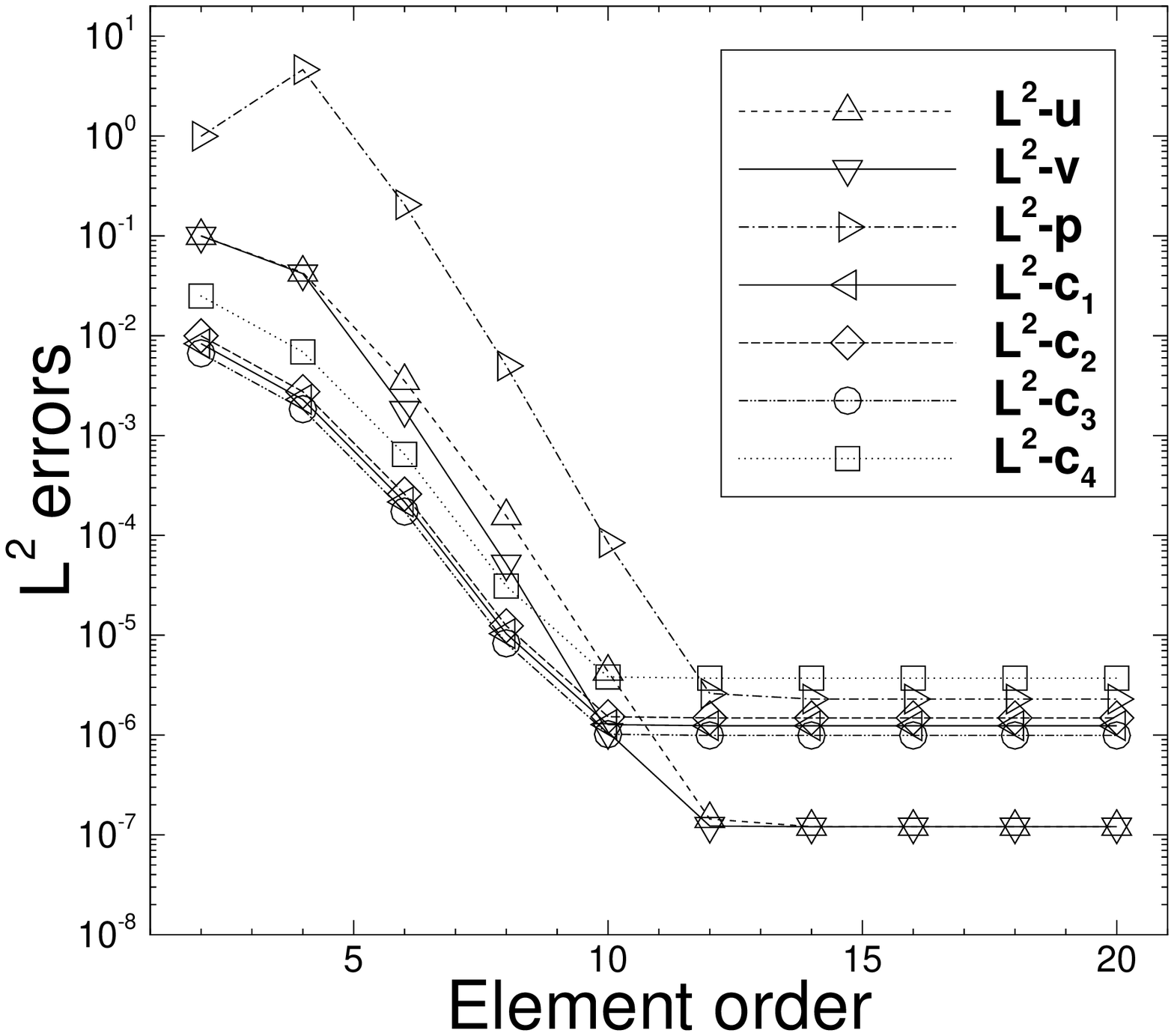}(a)
    \includegraphics[width=3in]{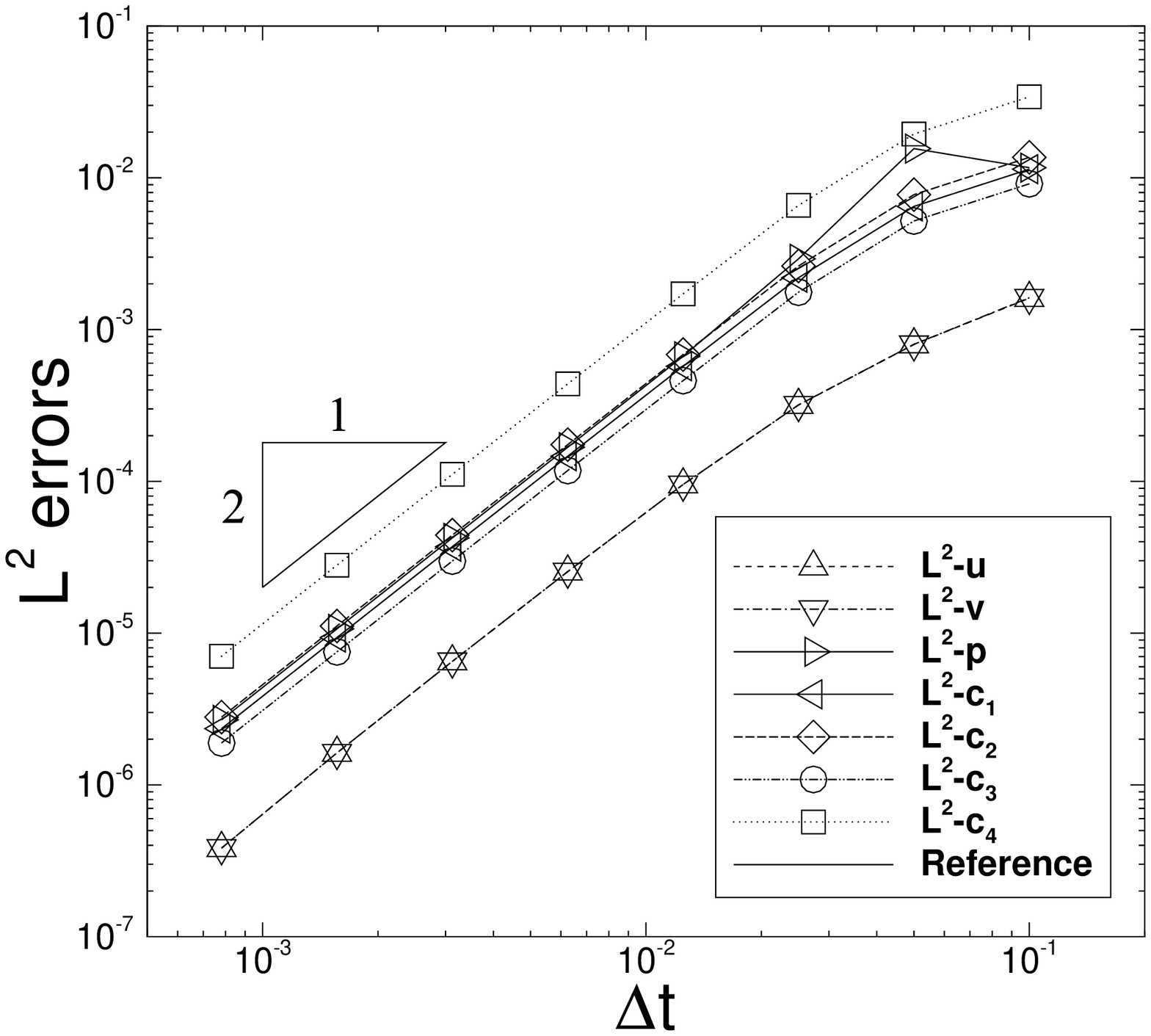}(b)
  }
  \caption{
    Spatial and temporal convergence rates ($4$ fluid components):
    (a) $L^2$ errors of various flow variables versus element order
    (with fixed $\Delta t=0.001$ and $t_f=0.1$) showing spatial exponential
    convergence rate.
    (b) $L^2$ errors versus time step size
    (with fixed Element order $16$ and $t_f=0.2$) showing temporal
    2nd-order convergence rate.
  }
  \label{fig:conv}
\end{figure}

The goal of this subsection is to numerically demonstrate the spatial
and temporal convergence rates of the method developed in
Section \ref{sec:alg} using a contrived analytic solution
to the system of governing equations \eqref{equ:nse_trans_1},
\eqref{equ:continuity_original}
and \eqref{equ:CH_trans_1}.



Consider a rectangular domain, $0\leqslant x\leqslant 2$
and $-1\leqslant y \leqslant 1$, and a four-fluid mixture
contained in this domain.
We assume the following analytic expressions for the flow variables
of this four-phase system,
\begin{equation}
  \left\{
  \begin{split}
    &
    u = A_0 \sin (ax) \cos(\pi y)\sin(\omega_0 t) \\
    &
    v = -(A_0 a/\pi) \cos (ax)\sin(\pi y) \sin(\omega_0 t) \\
    &
    P = A_0\sin(ax)\sin(\pi y) \cos(\omega_0 t) \\
    &
    c_1 = \frac{1}{6}\left[1 + A_1\cos(a_1x)\cos(b_1y)\sin(\omega_1 t)  \right] \\
    &
    c_2 = \frac{1}{6}\left[1 + A_2\cos(a_2x)\cos(b_2y)\sin(\omega_2 t)  \right] \\
    &
    c_3 = \frac{1}{6}\left[1 + A_3\cos(a_3x)\cos(b_3y)\sin(\omega_3 t)  \right], \\
    &
    c_4 = 1-c_1-c_2-c_3
  \end{split}
  \right.
  \label{equ:anal_soln}
\end{equation}
where $(u,v)$ are the two components of
the velocity $\mathbf{u}$.
$A_i$ and $\omega_i$ ($i=0,\dots,3$), $a$,
$a_i$ and $b_i$ ($i=1,2,3$) are 
constant parameters, whose values are to be specified below.
The external force $\mathbf{f}(\mathbf{x},t)$
in \eqref{equ:nse_trans_1} and the source term
$d_i(\mathbf{x},t)$ ($1\leqslant i\leqslant N-1$)
in \eqref{equ:CH_trans_1}
are chosen such that the analytic expressions in
\eqref{equ:anal_soln} exactly satisfy
the equations \eqref{equ:nse_trans_1} and
\eqref{equ:CH_trans_1}.
The above expressions for $(u.v)$
also satisfies the equation \eqref{equ:continuity_original}.

We impose the condition
\eqref{equ:bc_vel} for $\mathbf{u}$ and
the conditions \eqref{equ:bc_ci_1_trans} and
\eqref{equ:bc_ci_2_trans} for $c_i$ ($1\leqslant i\leqslant 3$)
on the domain boundaries,
where the boundary velocity $\mathbf{w}(\mathbf{x},t)$
is chosen based on the analytic expressions given
in \eqref{equ:anal_soln} and
the boundary source terms $d_{ai}(\mathbf{x},t)$ and
$d_{bi}(\mathbf{x},t)$ are chosen such that
the analytic expressions in \eqref{equ:anal_soln}
satisfy the equations \eqref{equ:bc_ci_1_trans}
and \eqref{equ:bc_ci_2_trans} on the boundary.
The initial conditions $\mathbf{u}_{in}$
and $c_i^{in}$ ($1\leqslant i\leqslant 4$)
are chosen based on the analytic expressions
in \eqref{equ:anal_soln} by setting
$t=0$.



To simulate the problem we discretize
the domain using two equal-sized quadrilateral
elements (domain partitioned in the $x$ direction),
and the element order is varied to test
the spatial convergence.
The numerical algorithm from Section \ref{sec:alg}
is employed to integrate in time the governing
equations for this four-phase system
from $t=0$ to $t=t_f$ ($t_f$ to be specified
later). Then the numerical solution and
the exact solution as given by \eqref{equ:anal_soln}
at $t=t_f$ are compared and the errors in the
$L^2$ norms for various flow variables are
computed and recorded. Table \ref{tab:conv_param}
lists the physical and numerical parameters
involved in the simulations of this problem.

\begin{table}
\begin{center}
\begin{tabular}{ l c | l c}
\hline
Parameter & Value & Parameter & Value \\ \hline
$a$, $a_1$, $a_2$, $a_3$ & $\pi$
& $b_1$, $b_2$, $b_3$ & $\pi$ \\
$A_0$ & $2.0$ 
& $A_1$, $A_2$, $A_3$ & $1.0$ \\
$\omega_0$, $\omega_1$ & $1.0$
& $\omega_2$ & $1.2$ \\
$\omega_3$ & $0.8$ 
& $\eta$ & $0.1$ \\
$\tilde{\rho}_1$ & $1.0$
& $\tilde{\rho}_2$ & $3.0$ \\
$\tilde{\rho}_3$ & $2.0$
& $\tilde{\rho}_4$ & $4.0$ \\
$\tilde{\mu}_1$ & $0.01$
& $\tilde{\mu}_2$ & $0.02$ \\
$\tilde{\mu}_3$ & $0.03$
& $\tilde{\mu}_4$ & $0.04$ \\
$\sigma_{12}$ & $6.236E-3$
& $\sigma_{13}$ & $7.265E-3$ \\
$\sigma_{14}$ & $3.727E-3$
& $\sigma_{23}$ & $8.165E-3$ \\
$\sigma_{24}$ & $5.270E-3$
& $\sigma_{34}$ & $6.455E-3$ \\
$m_0$ & $1.0E-3$ & $t_f$ & $0.1$ or $0.2$  \\
$\rho_0$ & $\min(\tilde{\rho}_1,\dots,\tilde{\rho}_4)$ 
& $\nu_0$ & $\max\left(\frac{\tilde{\mu}_1}{\tilde{\rho}_1},\dots,\frac{\tilde{\mu}_4}{\tilde{\rho}_4}\right)$ \\
 $J$ (temporal order) & $2$ 
& Number of elements & $2$ \\
 $\Delta t$ & (varied) 
&  Element order & (varied) \\
\hline
\end{tabular}
\caption{Simulation parameter values for the convergence-rate tests.}
\label{tab:conv_param}
\end{center}
\end{table}


The first group of tests is to examine the spatial
convergence rate of the method.
We fix the final time at $t_f=0.1$ and
the time step size at $\Delta t=0.001$.
The element order is then varied systematically
between $2$ and $20$.
For each element order,
the numerical solution at $t=t_f$ is then obtained
and compared with the exact solution.
Figure \ref{fig:conv}(a) shows
the $L^2$ errors of the velocity, pressure
and the four volume fractions as a function of
the element order from this group of tests.
The error curves approximately exhibit an exponential
rate of decrease with increasing element order,
before the element order reaches a certain value ($10$ or $12$
for this case). This suggests an exponential convergence
rate with respect to the element order.
As the element order increases beyond about $12$,
the error curves essentially level off.
The saturation is due to the fact that the temporal
error becomes dominant as the element order becomes sufficiently large.

Figure \ref{fig:conv}(b) summarizes results for a second
group of tests. In these tests 
we have fixed the element order at $16$ and the final time at $t_f=0.2$,
and then varied the time step size  systematically
between $\Delta t=0.1$ and $\Delta t=0.00078125$.
The figure shows the $L^2$ errors of different
flow variables as a function of $\Delta t$. 
We observe a second-order convergence rate of the errors
when $\Delta t$ becomes sufficiently small.

The above results suggest that the numerical method developed herein
exhibits an exponential convergence rate in space
and a second-order convergence rate in time,
with the reduction-consistent and thermodynamically consistent
formulation for multiple fluid components.


\subsection{Two-Phase Capillary Wave Problem}
\label{sec:capillary_2p}


\begin{figure}
  \centerline{
    \includegraphics[width=3in]{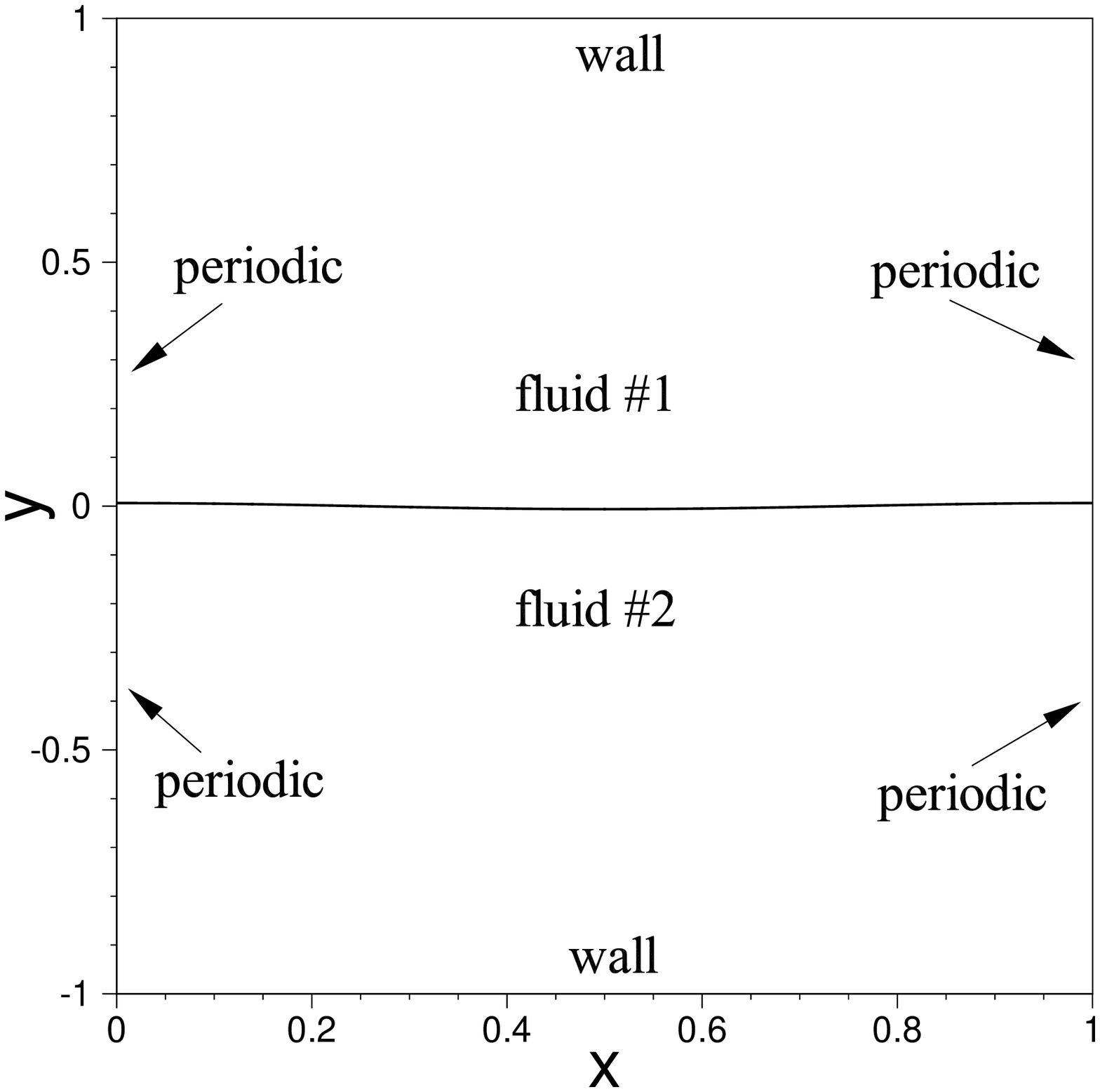}(a)
    \includegraphics[width=3in]{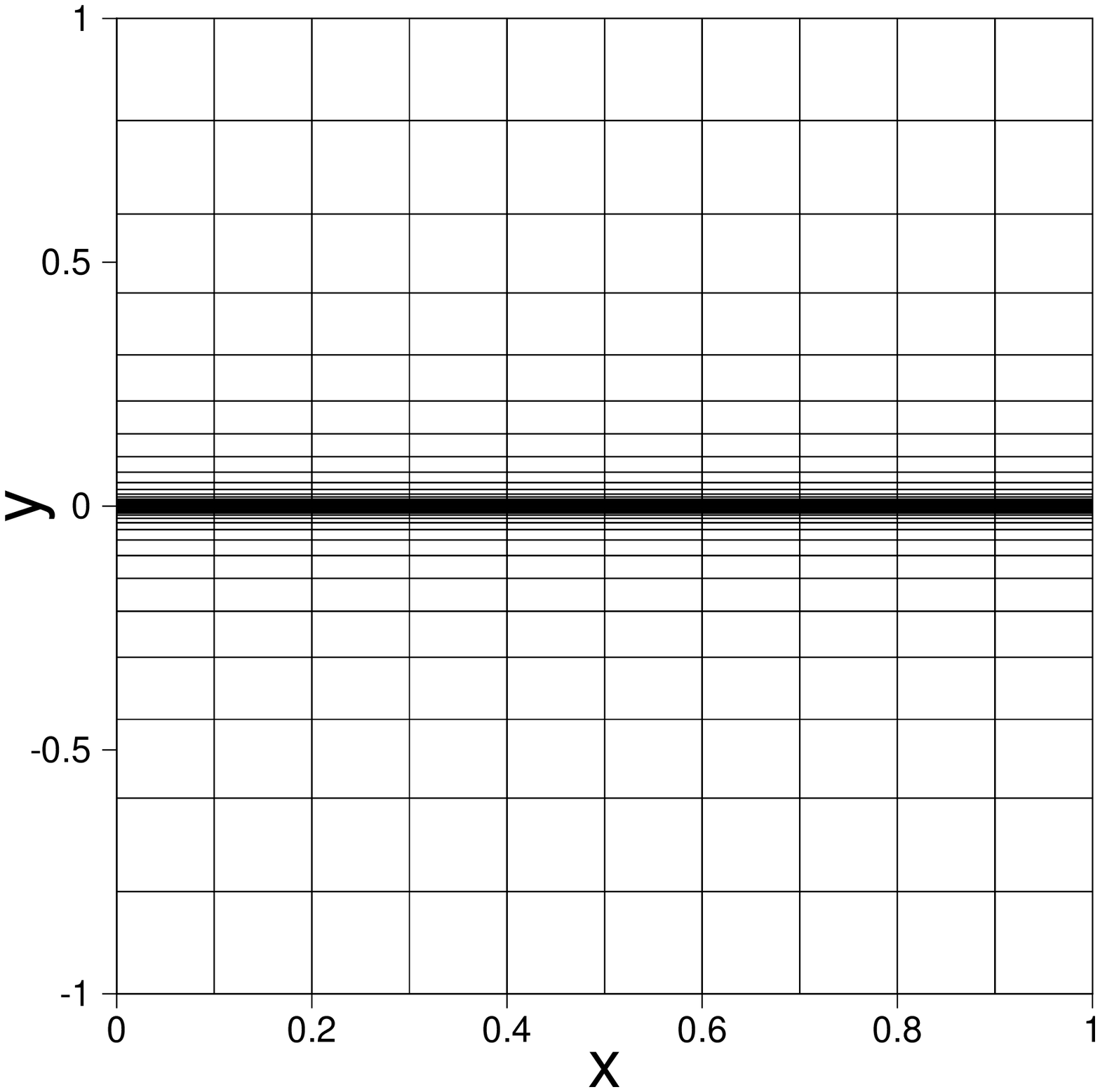}(b)
  }
  \caption{ Capillary wave problem:
    (a) Computational domain and configuration.
    (b) Spectral element mesh of $400$ quadrilateral elements.
  }
  \label{fig:capillary_config}
\end{figure}

The reduction-consistent and thermodynamically consistent
 formulation presented in Section \ref{sec:formulation}
for N-phase systems, with $N=2$, leads to a two-phase
formulation that is different from the usual two-phase
formulations (see e.g.~\cite{YueFLS2004,AbelsGG2012,DongS2012}),
because of the $m_{ij}(\vec{c})$ functions here.
In this subsection we employ the
benchmark two-phase capillary wave
problem (see e.g.~\cite{DongS2012,DongW2016})
to test the physical accuracy of the current method for $N=2$.
Note that both the two-phase formulation and the numerical algorithm
to be tested here are different from those of \cite{DongS2012,DongW2016}.


The problem setting is as follows. Consider two
immiscible incompressible fluids contained in an infinite domain.
The top half of the domain is occupied by the lighter
fluid (fluid \#1), and the bottom half is occupied by
the heavier fluid (fluid \#2).
The gravity is assumed to be in the downward direction.
The interface formed between the two fluids is perturbed from
its horizontal equilibrium position by a small-amplitude sinusoidal
wave form, and starts to oscillate at $t=0$. The objective here
is to study the motion of the interface over time.
In \cite{Prosperetti1981} an exact time-dependent standing-wave
solution to this problem was reported under the condition that
the two fluids must have matched kinematic viscosities (but their
densities and dynamic viscosities can be different).
We will simulate the problem under this condition using
the method developed herein for $N=2$ and compare
simulation results with the exact solution from \cite{Prosperetti1981}.


The simulation setup is illustrated in
Figure \ref{fig:capillary_config}(a).
We consider the computational domain $0\leqslant x\leqslant 1$
and $-1\leqslant y\leqslant 1$.
The top and bottom sides of the domain are solid walls of
neutral wettability. In the horizontal direction the domain and all variables
are assumed to be periodic at $x=0$ and $x=1$.
The equilibrium position of the fluid interface is assumed to
coincide with the $x$-axis. 
The initial perturbed profile of the fluid interface is
given by
$
y = H_0\cos (k_wx),
$
where $H_0=0.01$ is the initial amplitude,
$\lambda_w=1$ is the wave length of the perturbation profile,
and $k_w = \frac{2\pi}{\lambda_w}$ is the wave number.
Note that the initial capillary amplitude $H_0$
is small compared with
the dimension of domain in the vertical direction.
Therefore the effect of the walls at the domain top/bottom
on the motion of the interface will be small.

\begin{figure}
  \centerline{
    \includegraphics[width=3in]{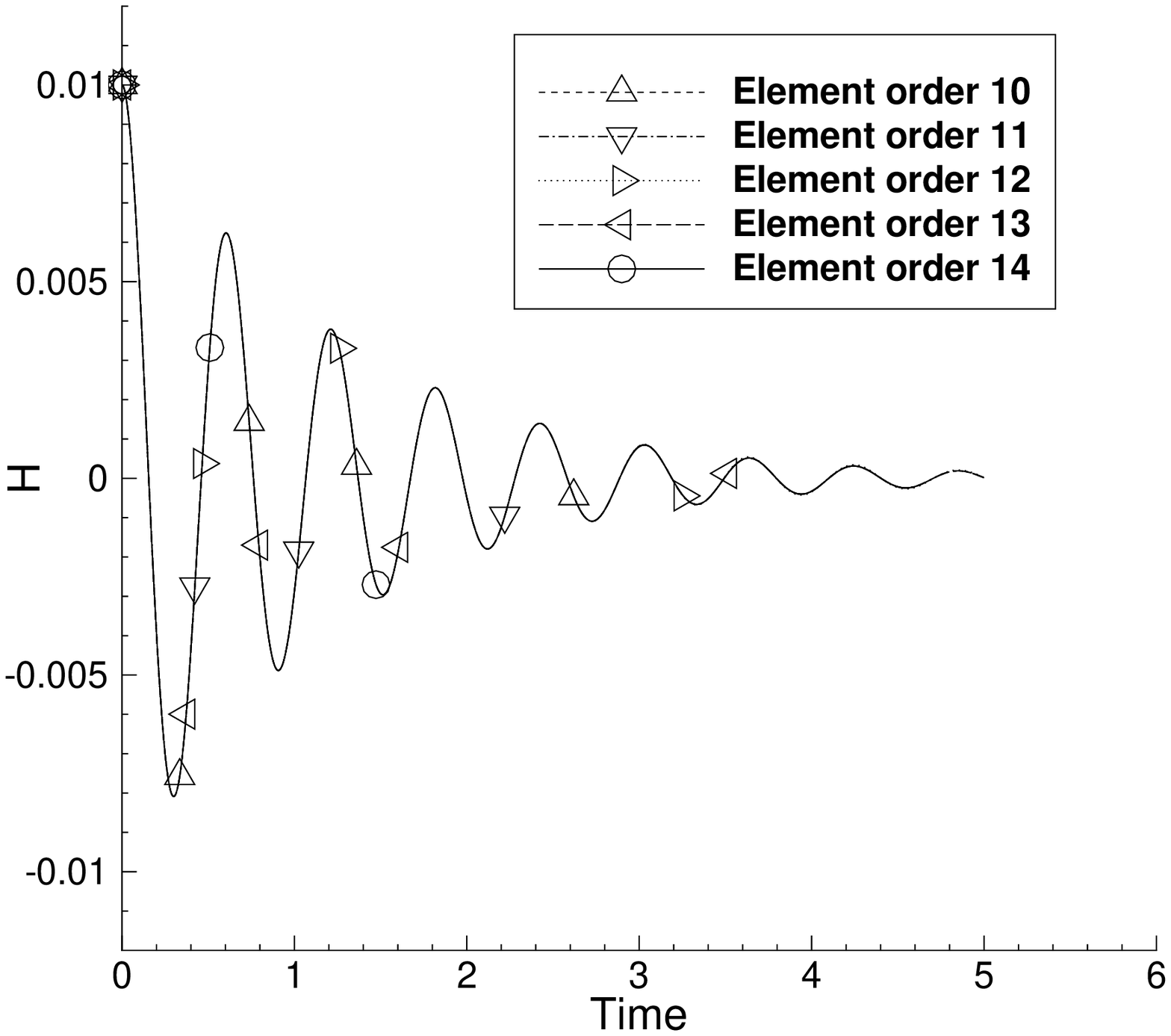}(a)
    \includegraphics[width=3in]{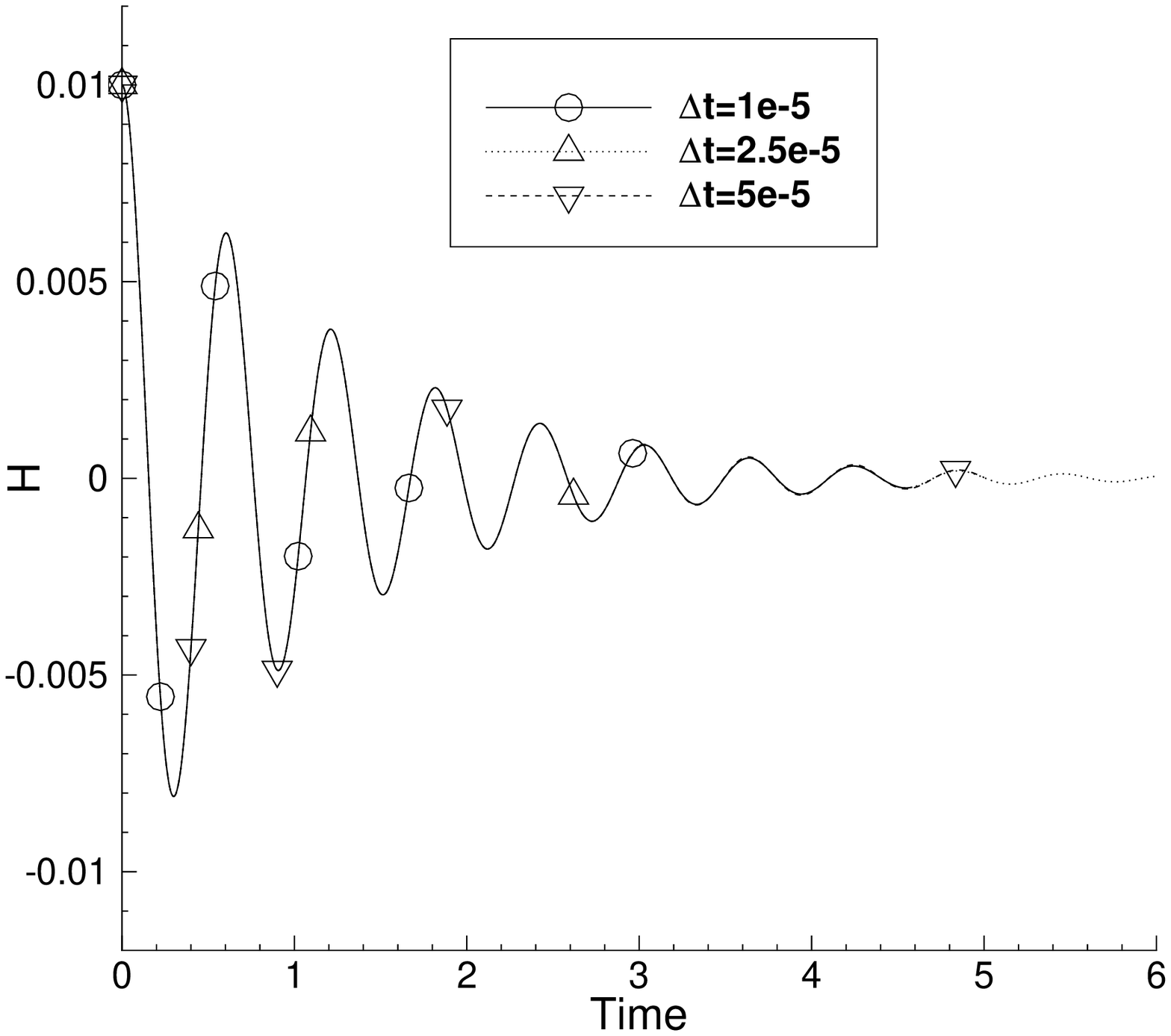}(b)
  }
  \centerline{
    \includegraphics[width=3in]{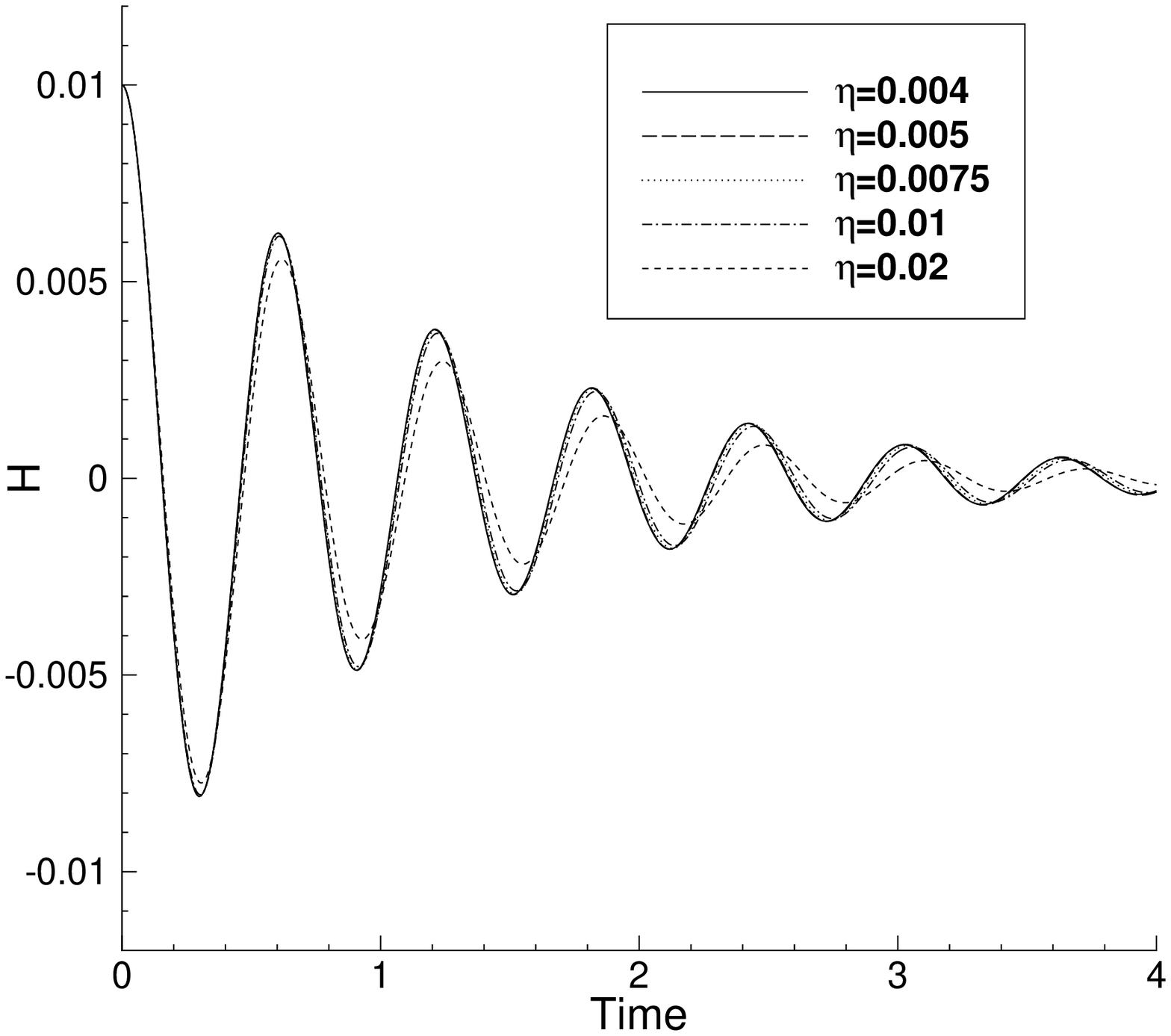}(c)
    \includegraphics[width=3in]{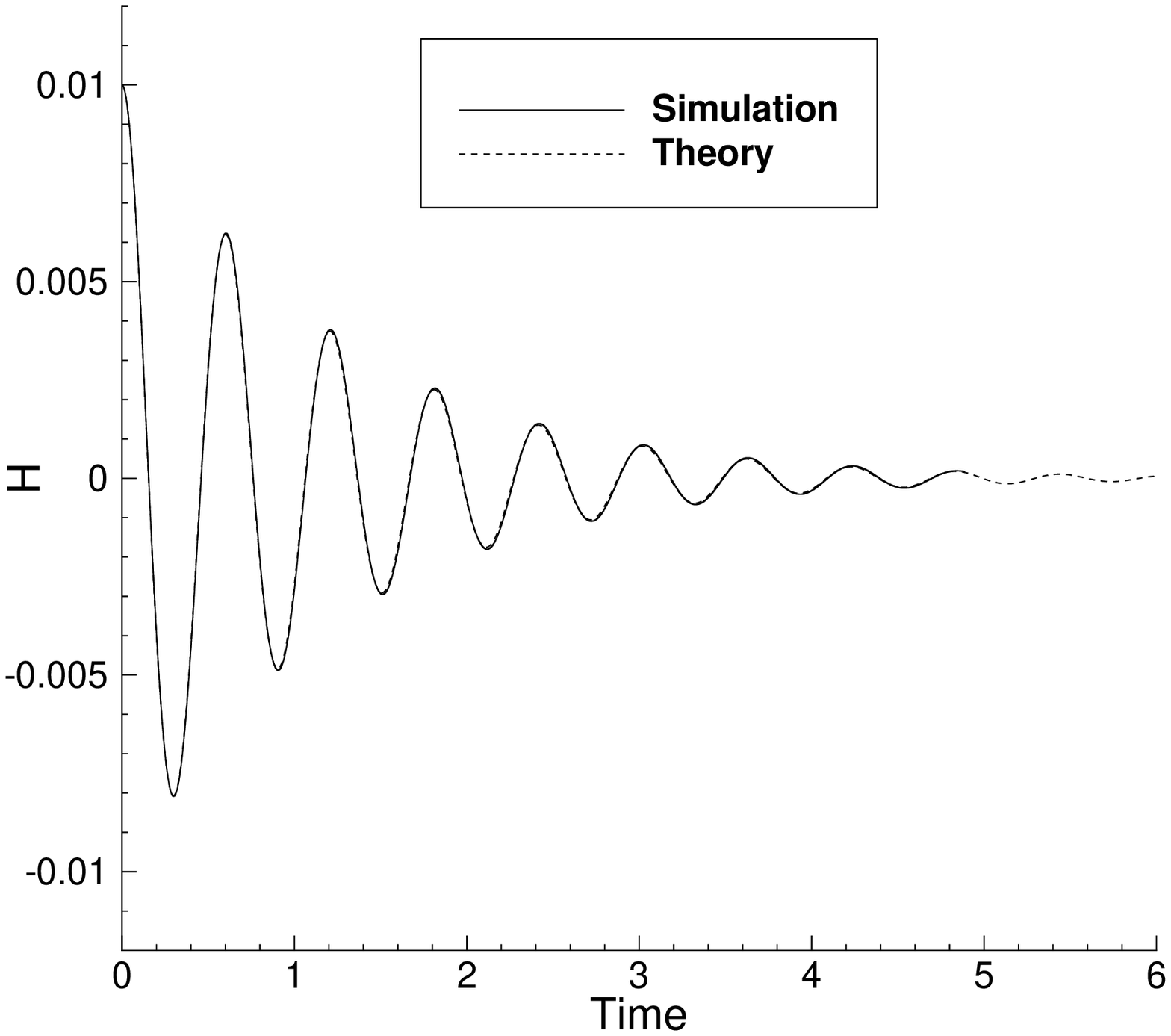}(d)
  }
  \caption{
    Two-phase capillary wave (matched density $\tilde{\rho}_2/\tilde{\rho}_1=1$):
    (a) Effect of spatial resolution (element order) on the capillary amplitude
    history.
    (b) Effect of time step size $\Delta t$ on the capillary amplitude history.
    (c) Effect of interfacial thickness ($\eta$) on the capillary amplitude history.
    (d) comparison between the current simulation (solid curve) and
    Prosperetti's~\cite{Prosperetti1981} exact theoretical solution (dashed curve).
    In (a), results are obtained with $\eta=0.005$, $\Delta t=2.5e-5$ for
    element orders $10$ to $13$
    and $\Delta t=1.0e-5$ for element order $14$.
    In (b), $\eta=0.005$, element order is $12$.
    In (c), element order is $12$, $\Delta t=2.5e-5$.
    In (d), the simulation result
    corresponds to $\eta=0.004$, element order $12$, and $\Delta t=2.5e-5$.
    In all four plots, $m_0=1.0e-5$.
  }
  \label{fig:capillary_dr_1}
\end{figure}


We use the method presented in Section \ref{sec:alg} to simulate
this problem.
The computational domain is discretized using a spectral element mesh
as shown in Figure \ref{fig:capillary_config}(b),
which consists of $400$ quadrilateral elements.
The elements are uniform along the $x$ direction,
but are non-uniform and clustered about the
region $-0.012\lesssim y\lesssim 0.012$ 
in the $y$ direction. The element order is varied to modify
the spatial resolution of the simulations,
and this will be specified below.
The external body force in equation \eqref{equ:nse_trans_1}
is set to
$
  \mathbf{f} = \rho \mathbf{g}_r, 
$
  where $\mathbf{g}_r$ is the gravitational acceleration.
  The source terms in equation \eqref{equ:CH_trans_1} are
  set to $d_i=0$ ($1\leqslant i\leqslant N-1$).
%
On the top/bottom walls, the boundary
condition \eqref{equ:bc_vel} with $\mathbf{w}=0$
is imposed for the velocity,
and the boundary conditions \eqref{equ:bc_ci_1_trans}--\eqref{equ:bc_ci_2_trans}
with $d_{ai}=0$ and $d_{bi}=0$ ($1\leqslant i\leqslant N-1$)
are imposed for the phase field variables.
%
The initial velocity is set to zero, and
the initial volume fractions are set as follows,
\begin{equation}
  \left\{
  \begin{split}
    &
    c_1 = \frac{1}{2}\left[1 + \tanh\frac{y-H_0\cos (k_wx)}{\sqrt{2}\eta}  \right], \\
    &
    c_2 = 1-c_1 = \frac{1}{2}\left[1 - \tanh\frac{y-H_0\cos (k_wx)}{\sqrt{2}\eta}  \right].
  \end{split}
  \right.
  \label{equ:capillary_ic}
\end{equation}
We list in Table \ref{tab:capillary_param}
the values for the physical and simulation
parameters involved in this problem.

\begin{table}
\begin{center}
\begin{tabular}{ l c | l c}
\hline
Parameter & Value & Parameter & Value \\ \hline
$H_0$ & $0.01$
& $\lambda_w$ & $1.0$ \\
$\sigma_{12}$ & $1.0$
& $|\mathbf{g}_r|$ (gravity) & $1.0$ \\
$\tilde{\rho}_1$ & $1.0$
& $\tilde{\mu}_1$ & $0.01$ \\
$\frac{\tilde{\mu}_2}{\tilde{\rho}_2}$ & $\frac{\tilde{\mu}_1}{\tilde{\rho}_1}$
& $\tilde{\rho}_2$ & (varied) \\
$\tilde{\mu}_2$ & $\frac{\tilde{\mu}_1}{\tilde{\rho}_1}\tilde{\rho}_2$ &
$\eta$ & (varied) \\
$m_0$ & $1.0E-5$ &
$\rho_0$ & $\min(\tilde{\rho}_1,\tilde{\rho}_2)$ \\
$\nu_0$ & $\max\left(\frac{\tilde{\mu}_1}{\tilde{\rho}_1},\frac{\tilde{\mu}_2}{\tilde{\rho}_2} \right)$ &
$J$ (integration order) & $2$ \\
$\Delta t$ & (varied) &
Number of elements in mesh & $400$ \\
 Element order & (varied) \\
\hline
\end{tabular}
\caption{Simulation parameter values for the two-phase
  capillary wave problem.}
\label{tab:capillary_param}
\end{center}
\end{table}


We have varied the element order, the time step size ($\Delta t$)
and the interfacial thickness scale ($\eta$) systematically
in the simulations to ensure the convergence of the
simulation results.
Figure \ref{fig:capillary_dr_1} summarizes some of the test results
with matched densities ($\tilde{\rho}_2/\tilde{\rho}_1=1$) for
the two fluids.
Figure \ref{fig:capillary_dr_1}(a) compares
the time histories of the capillary wave amplitude
obtained with element orders ranging from $10$ to $14$ in the simulation.
The history curves corresponding to different
element orders overlap with one another, suggesting
independence of the results with respect to the
grid resolutions.
Figure \ref{fig:capillary_dr_1}(b) is a comparison of
the capillary amplitude histories computed using
several time step sizes. The results indicate the convergence
with respect to $\Delta t$.
Figure \ref{fig:capillary_dr_1}(c) shows the time histories
of the capillary amplitude obtained with the interfacial
thickness scale parameter ranging from $\eta=0.02$ to $\eta=0.004$.
Note that the simulations become much more
challenging when $\eta$ becomes small, taxing the
grid resolution and the time step size.
We initially observe some  influence on the amplitude
and the phase of the history curves as $\eta$  decreases from
$0.02$ to $0.01$. As $\eta$ decreases further to
$\eta=0.0075$ and below, on the other hand, the history curves essentially overlap
with one another and little difference can be
observed among them, suggesting a convergence of the results with respect to
$\eta$.
%
In Figure \ref{fig:capillary_dr_1}(d) we compare
the capillary amplitude history from the current simulation
(corresponding to $\eta=0.004$) and the exact theoretical solution
given by \cite{Prosperetti1981}.
The history curve from the simulation essentially overlaps
with the theoretical curve, attesting to the physical accuracy
of the simulation results.
%
The above results are obtained with a mobility parameter $m_0=1e-5$.
Some other values for $m_0$ have also been considered.
The tests suggest that the computation would be unstable if $m_0$ is
too large (larger
than a certain value). With decreasing $m_0$ values, the simulation
tends to require a smaller interfacial thickness  $\eta$ value
for stability or accuracy, which in turn increases the computational
challenge and demand.

\begin{figure}
  \centerline{
    \includegraphics[width=3in]{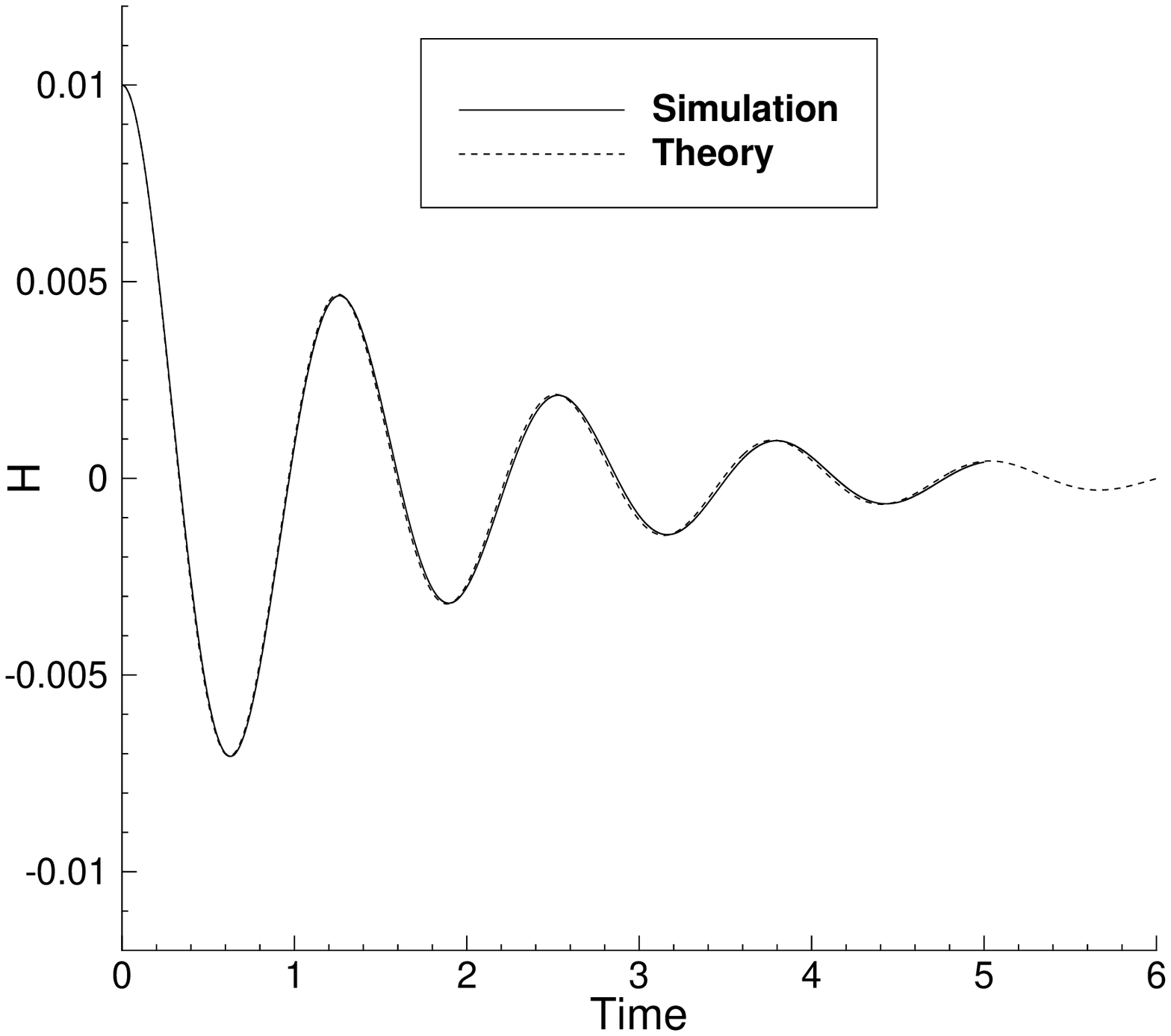}(a)
    \includegraphics[width=3in]{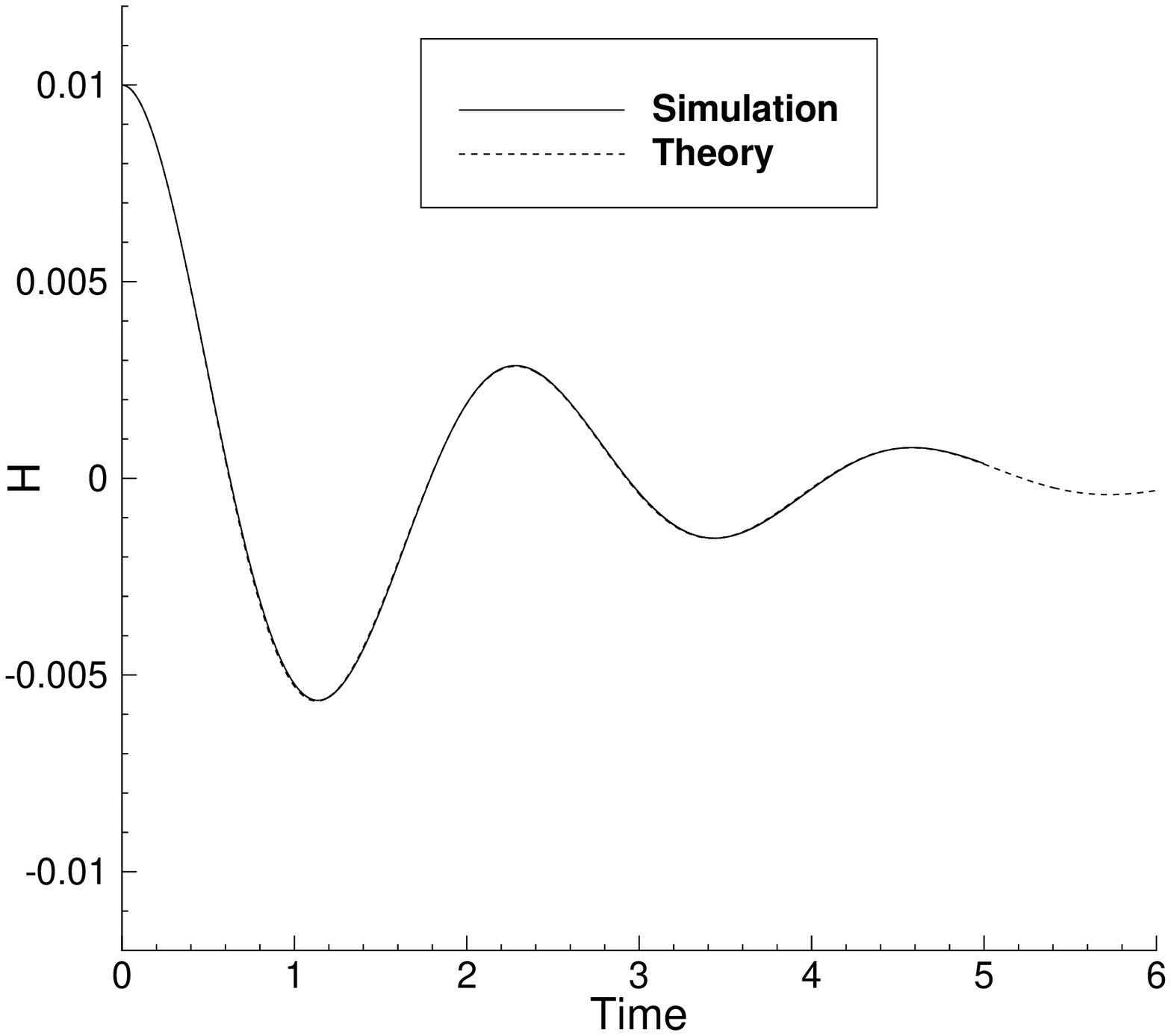}(b)
  }
  \centerline{
    \includegraphics[width=3in]{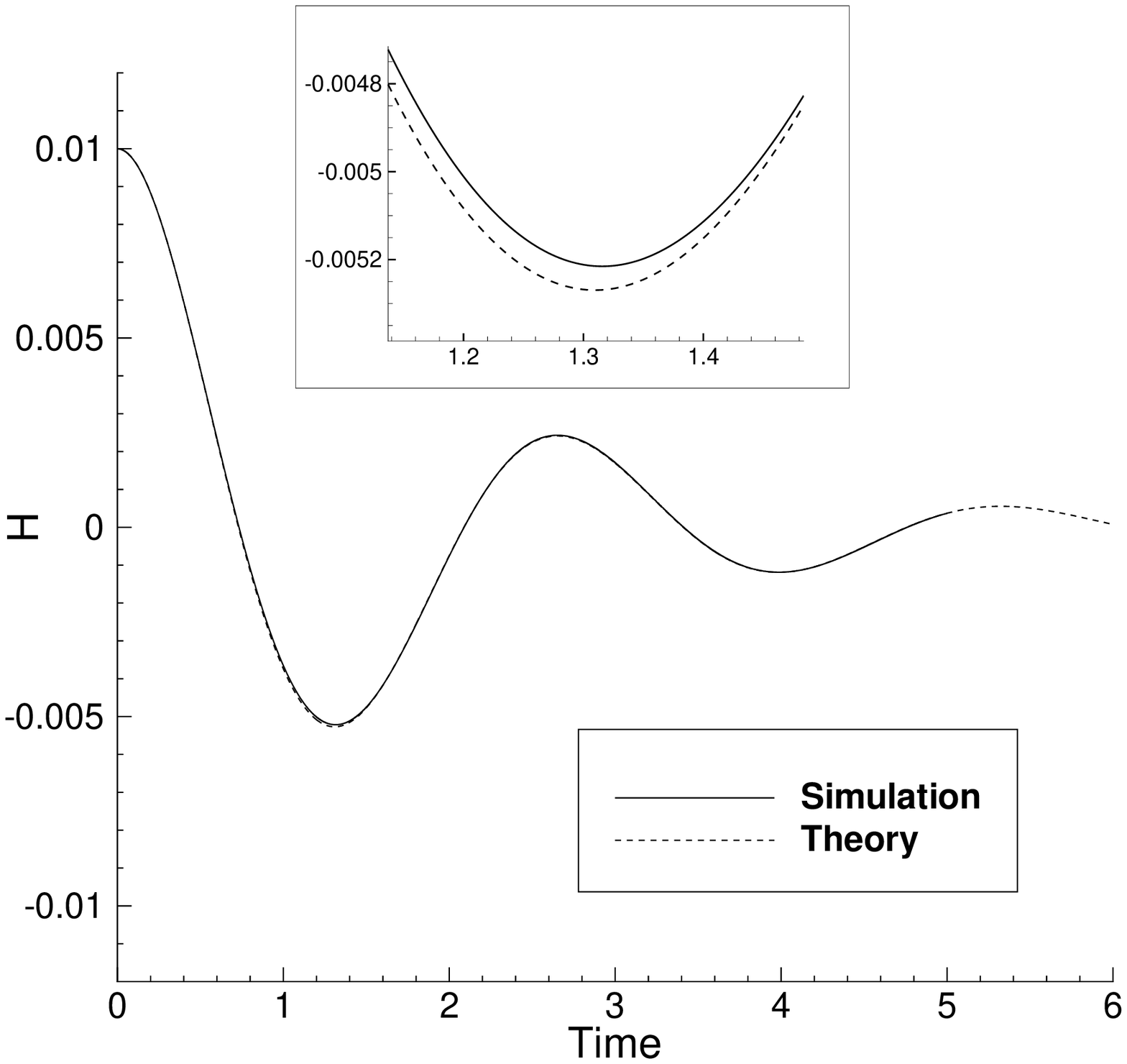}(c)
  }
  \caption{
    Two-phase capillary wave (larger density ratios):
    Comparison of capillary amplitude as a function of time between
    current simulations (solid curves) and the exact solutions~\cite{Prosperetti1981}
    (dashed curves) for density ratios $\tilde{\rho}_2/\tilde{\rho}_1=10$ (a),
    $100$ (b), and $1000$ (c).
    The inset of plot (c) shows a magnified view of a section of
    the curves.
  }
  \label{fig:capillary_dr_effect}
\end{figure}


To investigate
the density ratio effect on the motion of the fluid interface,
the density and dynamic viscosity of the second fluid ($\tilde{\rho}_2$
and $\tilde{\mu}_2$) have been varied systematically 
while the relation $\tilde{\mu}_2/\tilde{\rho}_2=\tilde{\mu}_1/\tilde{\rho}_1$
is maintained as required by the exact solution of \cite{Prosperetti1981}.
In Figure \ref{fig:capillary_dr_effect} we show the time histories of
the capillary amplitude corresponding to three larger density ratios
$\tilde{\rho}_2/\tilde{\rho}_1=10$, $100$ and $1000$ from
our simulations, and compare them with the exact solutions
from \cite{Prosperetti1981}.
The simulation results correspond to an element order $12$,
time step size $\Delta t=5.0e-5$, interfacial thickness $\eta=0.004$,
and $m_0=1.0e-5$ in the simulations.
The history curves from the simulations essentially overlap
with those of the exact solutions.
The inset of Figure \ref{fig:capillary_dr_effect}(c) is a zoomed-in
view of the curves for the density ratio $\tilde{\rho}_2/\tilde{\rho}_1=1000$,
showing some but small difference between the simulation
and the theoretical solution.
These comparisons suggest that our simulation results are in good agreement
with the physical solution for the whole range of density ratios
considered here. 


The two-phase capillary wave problem and in particular
the comparisons with Prosperetti's exact solution for
this problem demonstrate that the reduction-consistent
formulation and the numerical method developed herein
(with $N=2$) have produced physically accurate results
for a wide range of density ratios (up to density ratio $1000$
tested here) and at large density ratios.
This provides a reference, for two fluid phases,
when the method is subject to
subsequent tests involving multiple fluid components.


\subsection{Three-/Four-Phase Capillary Wave With Absent Fluid Components}
\label{sec:capillary_34p}

In this subsection we consider a three-phase system and a four-phase
system, but with some fluid components absent, so that they
are physically equivalent to a system containing a smaller number of fluids.
We employ a setting similar to that of Section \ref{sec:capillary_2p},
so these three-phase and four-phase problems are physically
equivalent to the two-phase capillary wave problem.
This allows us to compare the three-phase and four-phase
simulation results with Prosperetti's exact physical
solution~\cite{Prosperetti1981} for two-phase problems.

\begin{figure}
  \centerline{
    \includegraphics[width=3in]{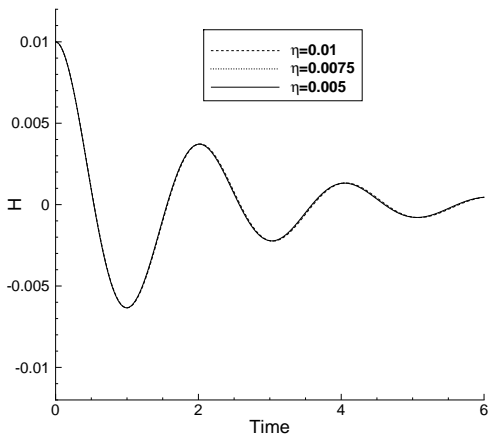}
  }
  \caption{
    Effect of interfacial thickness $\eta$ on the capillary amplitude
    as a function of time for the four-phase capillary wave
    problem with two absent fluids, where $\tilde{\rho}_i=1$ ($i=1,\dots,4$).
  }
  \label{fig:capillary_4p_eta}
\end{figure}


More specifically, we consider the same computational domain
and the same mesh as in Section \ref{sec:capillary_2p},
as shown in Figure \ref{fig:capillary_config}.
Consider a system of three immiscible incompressible fluids
contained in this domain, but {\em with fluid \#2 absent}. So this
three-phase system is physically equivalent to a two-phase
system containing fluid \#1 and fluid \#3.
Similar to in Section \ref{sec:capillary_2p}, the interface
between fluid \#1 and fluid \#3 is perturbed by a sinusoidal wave
form of a small amplitude ($H_0=0.01$) from its equilibrium
position, and our goal is to study the motion of
the interface  over time.
In the three-phase simulations, we employ periodic conditions
for all flow variables in the horizontal direction.
On the top/bottom walls we impose the no-slip condition,
i.e.~equation \eqref{equ:bc_vel} with $\mathbf{w}=0$,
for the velocity and the boundary conditions
\eqref{equ:bc_ci_1_trans}--\eqref{equ:bc_ci_2_trans}
with $d_{ai}=0$ and $d_{bi}=0$ ($i=1,2,3$) for
the volume fractions.
The source terms in equation \eqref{equ:CH_trans_1}
are set to $d_i=0$ ($1\leqslant i\leqslant N-1$).
The initial velocity is zero, and the initial volume
fractions are set to
\begin{equation}
  \left\{
  \begin{split}
    &
    c_1 = \frac{1}{2}\left[1 + \tanh\frac{y-H_0\cos (k_wx)}{\sqrt{2}\eta}  \right], \\
    &
    c_2 = 0, \\
    &
    c_3 = 1-c_1-c_2 = \frac{1}{2}\left[1 - \tanh\frac{y-H_0\cos (k_wx)}{\sqrt{2}\eta}  \right].
  \end{split}
  \right.
  \label{equ:capillary_3p_ic}
\end{equation}
Note that the initial volume fraction of fluid \#2 is $c_2=0$ (absent fluid).
Therefore, the solution to this three-phase problem
physically consists of the exact solution given
by \cite{Prosperetti1981} for fluid \#1 and fluid \#3
and $c_2(\mathbf{x},t)=0$ for fluid \#2.

\begin{table}
\begin{center}
\begin{tabular}{ l c | l c}
\hline
Parameter & Value & Parameter & Value \\ \hline
$H_0$ & $0.01$
& $\lambda_w$ & $1.0$ \\
$\sigma_{ij}$ ($1\leqslant i\neq j\leqslant 4$) & $0.1$
& $|\mathbf{g}_r|$ (gravity) & $0.1$ \\
$\tilde{\rho}_1$ & $1.0$
& $\tilde{\mu}_1$ & $0.01$ \\
$\frac{\tilde{\mu}_2}{\tilde{\rho}_2}$ & $\frac{\tilde{\mu}_1}{\tilde{\rho}_1}$
& $\tilde{\rho}_2$ & (varied) \\
$\frac{\tilde{\mu}_3}{\tilde{\rho}_3}$ (three-/four-phase) & $\frac{\tilde{\mu}_1}{\tilde{\rho}_1}$
& $\frac{\tilde{\mu}_4}{\tilde{\rho}_4}$ (four-phase) & $\frac{\tilde{\mu}_1}{\tilde{\rho}_1}$ \\
$\tilde{\rho}_3$ (three-/four-phase) & $\tilde{\rho}_2$
& $\tilde{\rho}_4$ (four-phase) & $\tilde{\rho}_2$ \\
$\tilde{\mu}_2$ & $\frac{\tilde{\mu}_1}{\tilde{\rho}_1}\tilde{\rho}_2$ 
& $\tilde{\mu}_3$ (three-/four-phase) & $\frac{\tilde{\mu}_1}{\tilde{\rho}_1}\tilde{\rho}_3$ \\
$\tilde{\mu}_4$ (four-phase) & $\frac{\tilde{\mu}_1}{\tilde{\rho}_1}\tilde{\rho}_4$ 
& $\eta$ & $0.005$ (or varied) \\
$m_0$ & $1.0E-4$ &
$\rho_0$ & $\min(\tilde{\rho}_1,\tilde{\rho}_2,\tilde{\rho}_3,\tilde{\rho}_4)$ \\
$\nu_0$ & $\max\left(\frac{\tilde{\mu}_1}{\tilde{\rho}_1},\frac{\tilde{\mu}_2}{\tilde{\rho}_2},\frac{\tilde{\mu}_3}{\tilde{\rho}_3},\frac{\tilde{\mu}_4}{\tilde{\rho}_4} \right)$ &
$J$ (integration order) & $2$ \\
$\Delta t$ & $1E-4$ &
Number of elements in mesh & $400$ \\
 Element order & $10$ \\
\hline
\end{tabular}
\caption{Simulation parameter values for the three-phase and four-phase
  capillary wave problems.}
\label{tab:capillary_3p_param}
\end{center}
\end{table}

In addition to the above three-phase problem, we also consider
a four-phase system contained in this domain, in which {\em fluid \#2  
and fluid \#3 are absent}. Therefore this four-phase system is
physically equivalent to a two-phase system that consists of fluid \#1
and fluid \#4 only. We consider the motion of the interface
between fluid \#1 and fluid \#4 after a perturbation from its
equilibrium horizontal position, similar to in Section \ref{sec:capillary_2p}.
The boundary conditions are set in an analogous way to
the three-phase problem. We employ a zero initial velocity
and the following initial volume fraction distributions:
\begin{equation}
  \left\{
  \begin{split}
    &
    c_1 = \frac{1}{2}\left[1 + \tanh\frac{y-H_0\cos (k_wx)}{\sqrt{2}\eta}  \right], \\
    &
    c_2 = 0, \\
    & 
    c_3 = 0, \\
    &
    c_4 = 1-c_1-c_2-c_3 = \frac{1}{2}\left[1 - \tanh\frac{y-H_0\cos (k_wx)}{\sqrt{2}\eta}  \right].
  \end{split}
  \right.
  \label{equ:capillary_4p_ic}
\end{equation}
Note that $c_2$ and $c_3$ are both identically zero initially,
so physically they should be zero over time.
%
In Table \ref{tab:capillary_3p_param} we have listed the values of
the physical and simulation parameters for the above
three-phase and four-phase capillary-wave problems considered here.


We have varied the interfacial thickness $\eta$ to look into
its effect on the simulation result.
Figure \ref{fig:capillary_4p_eta} compares the time histories
of the capillary amplitude of the interface formed between
fluid \#1 and fluid \#4 for the four-phase capillary
wave problem with $\tilde{\rho}_i=1$ ($i=1,2,3,4$),
obtained using $\eta=0.01$, $0.0075$ and $0.005$ in the simulations.
Note that the pairwise surface tension values $\sigma_{ij}$ 
and the gravity value employed here are different from those
of Section \ref{sec:capillary_2p}, and
one can observe that this has a notable effect on
the period and attenuation of the capillary wave history
(e.g.~compare Figures \ref{fig:capillary_4p_eta} and  \ref{fig:capillary_dr_1}(a)).
We observe from Figure \ref{fig:capillary_4p_eta} that 
there is little difference in the simulation results corresponding to
these different $\eta$ values, suggesting the independence of
the results with respect to $\eta$.
The simulation results reported below for
the three-/four-phase capillary wave problem
are obtained with $\eta=0.005$.

\begin{figure}
  \centerline{
    \includegraphics[width=3in]{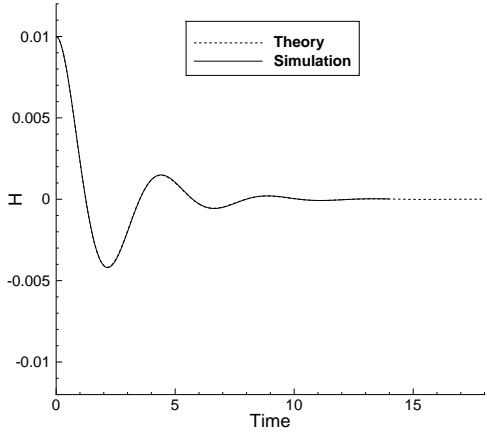}(a)
    \includegraphics[width=3in]{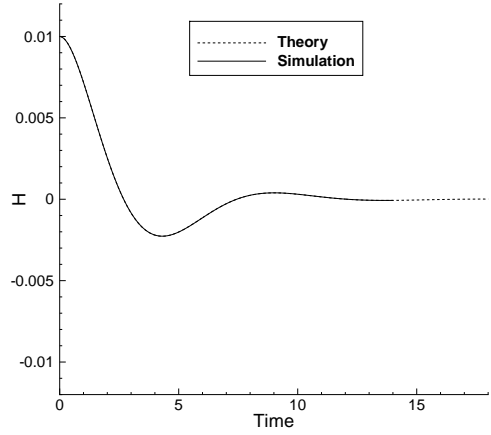}(b)
  }
  \caption{
    Three-phase capillary wave problem with one absent fluid component:
    Comparison of capillary amplitude versus time between simulations
    and the exact solutions~\cite{Prosperetti1981}, corresponding to fluid densities:
    (a) $\tilde{\rho}_1=1$, $\tilde{\rho}_2=\tilde{\rho}_3=10$;
    (b) $\tilde{\rho}_1=1$, $\tilde{\rho}_2=\tilde{\rho}_3=100$.
    In these simulations the second fluid is absent.
  }
  \label{fig:capillary_3p_dr}
\end{figure}

\begin{figure}
  \centerline{
    \includegraphics[width=3in]{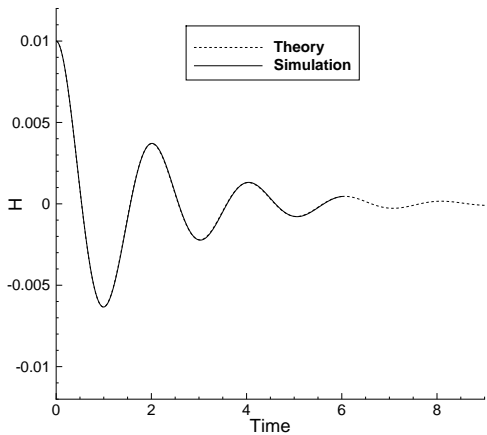}(a)
    \includegraphics[width=3in]{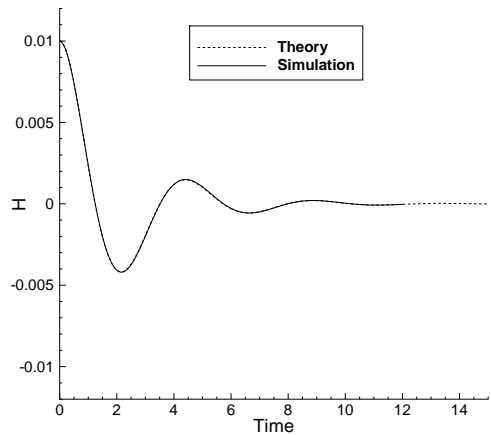}(b)
  }
  \caption{
    Four-phase capillary wave problem with two absent fluid components:
    comparison of capillary amplitude versus time between simulations
    and the exact solutions~\cite{Prosperetti1981} with fluid densities:
    (a) $\tilde{\rho}_1=\tilde{\rho}_2=\tilde{\rho}_3=\tilde{\rho}_4=1$;
    (b) $\tilde{\rho}_1=1$, $\tilde{\rho}_2=\tilde{\rho}_3=\tilde{\rho}_4=10$.
    Fluid two and fluid three are absent in the simulations.
  }
  \label{fig:capillary_4p_dr}
\end{figure}

Let us next compare the simulations for the three-/four-phase
capillary wave problems and Prosperetti's exact solutions~\cite{Prosperetti1981}
to study the accuracy of the simulation results.
We have varied the fluid density values to look into
their effect on the simulation results.
Figure \ref{fig:capillary_3p_dr} shows a comparison of
the capillary amplitude history (of the interface formed
between fluids \#1 and \#3) between the simulation and
the exact solution from \cite{Prosperetti1981}
for the three-phase capillary wave problem,
corresponding to two density ratios
$\frac{\tilde{\rho}_2}{\tilde{\rho}_1} = \frac{\tilde{\rho}_3}{\tilde{\rho}_1} =10$
and $\frac{\tilde{\rho}_2}{\tilde{\rho}_1} = \frac{\tilde{\rho}_3}{\tilde{\rho}_1} =100$.
Figure \ref{fig:capillary_4p_dr} is
a comparison of the capillary amplitude
histories between the simulation of the four-phase
capillary wave problem and the exact solution~\cite{Prosperetti1981},
corresponding to density ratios
$
\frac{\tilde{\rho}_2}{\tilde{\rho}_1}
= \frac{\tilde{\rho}_3}{\tilde{\rho}_1}
= \frac{\tilde{\rho}_4}{\tilde{\rho}_1}
= 1
$
and 
$
\frac{\tilde{\rho}_2}{\tilde{\rho}_1}
= \frac{\tilde{\rho}_3}{\tilde{\rho}_1}
= \frac{\tilde{\rho}_4}{\tilde{\rho}_1}
=10
$.
It can be observed that the history curves from the simulations
almost exactly overlap with those of the physical solutions.
This indicates that our simulations of the three- and four-phase
capillary wave problems with absent fluid components have
captured the motion of the fluid interface accurately.

\begin{figure}
  \centerline{
    \includegraphics[width=3in]{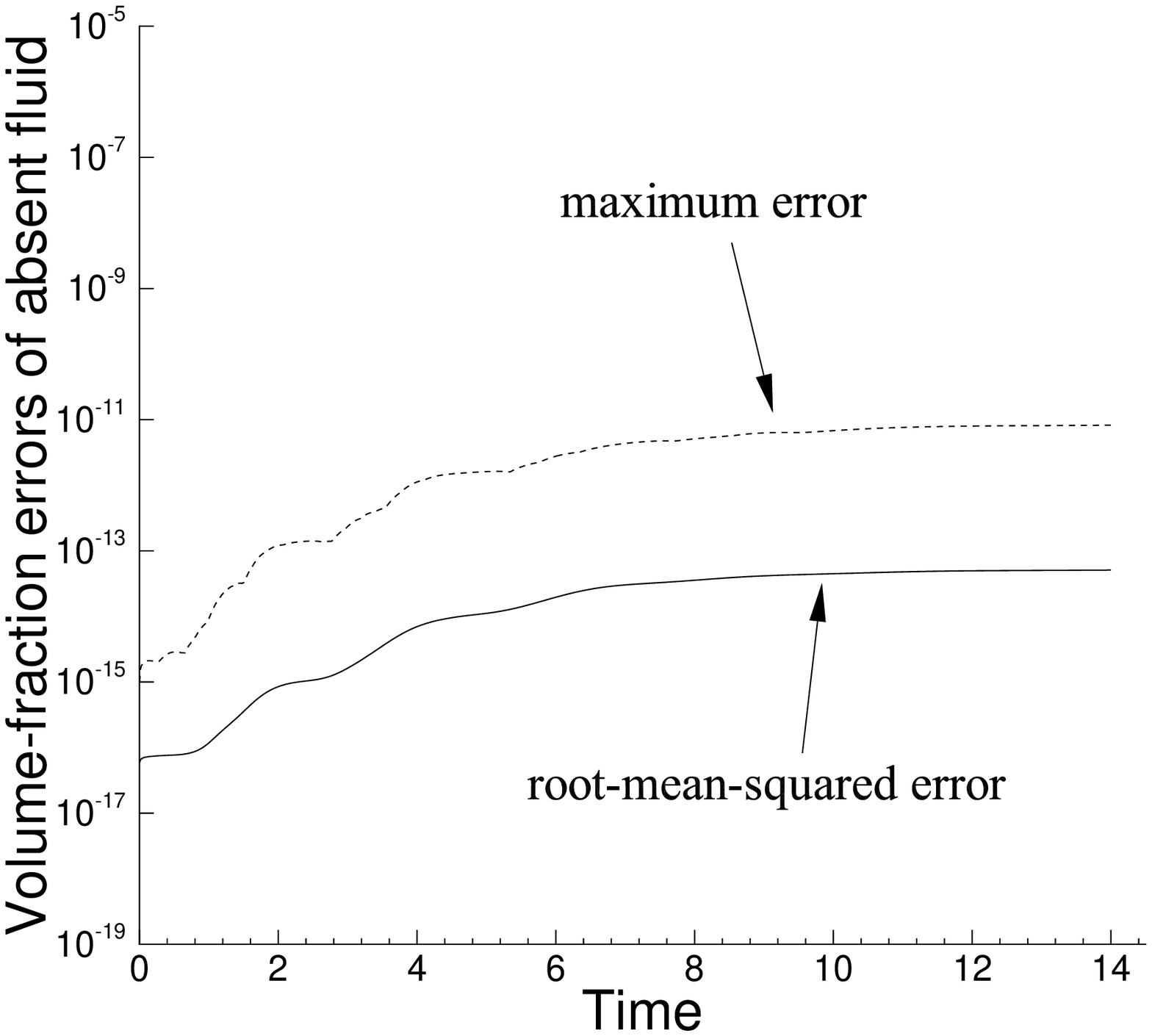}(a)
    \includegraphics[width=3in]{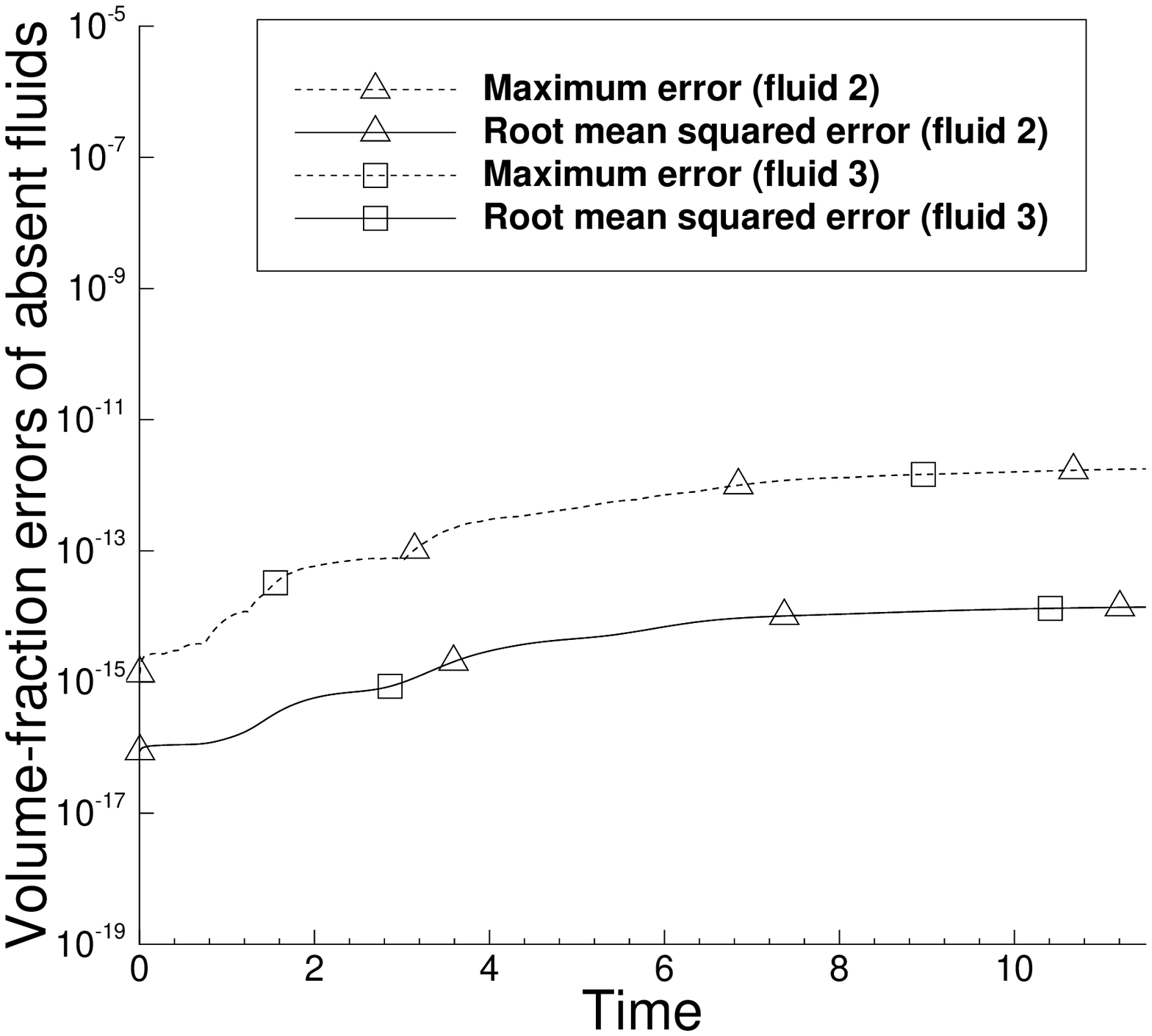}(b)
  }
  \caption{
    Time histories of volume-fraction errors of the absent fluids for
    the three-/four-phase capillary wave problems:
    (a) three-phase problem with one absent fluid,
    corresponding to $\tilde{\rho}_1=1$, $\tilde{\rho}_2=\tilde{\rho}_3=10$;
    (b) four-phase problem with two absent fluids, corresponding to
    $\tilde{\rho}_1=1$, $\tilde{\rho}_2=\tilde{\rho}_3=\tilde{\rho}_4=10$.
  }
  \label{fig:capillary_34p_error}
\end{figure}


In the three-phase and four-phase capillary wave
problems considered here, the physical solution for
the absent fluids corresponds to a zero volume-fraction
field. In the simulations, however, owing to
the numerical errors the computed volume-fraction fields
corresponding to the absent fluids will not be exactly zero,
but contain very small yet non-zero values.
The results in Figure \ref{fig:capillary_34p_error}
demonstrate this point.
Figure \ref{fig:capillary_34p_error}(a) 
shows time histories of the maximum error,
$
\max_{\mathbf{x}\in\Omega}|c_2(\mathbf{x},t)|,
$
and the root-mean-squared (RMS) error,
$
\sqrt{\frac{1}{V_{\Omega}}\int_{\Omega}|c_2(\mathbf{x},t)|^2d\mathbf{x}}
$
($V_{\Omega}=\int_{\Omega}d\mathbf{x}$ denoting
the volume of domain $\Omega$),
of the volume fraction of fluid \#2 (absent fluid)
for the three-phase capillary wave problem
with density ratios
$\frac{\tilde{\rho}_2}{\tilde{\rho}_1} = \frac{\tilde{\rho}_3}{\tilde{\rho}_1} =10$.
It is observed that the errors increase initially
and gradually level off over time.
The maximum error approximately levels off
on the order of magnitude $10^{-11}$, and
the RMS error levels off at a level $10^{-14}$.
%
Figure \ref{fig:capillary_34p_error}(b) shows
time histories of the maximum errors and the RMS errors
of the volume fractions of fluid \#2 and fluid \#3
(the absent fluids) for the four-phase capillary
wave problem with density ratios
$
\frac{\tilde{\rho}_2}{\tilde{\rho}_1}
= \frac{\tilde{\rho}_3}{\tilde{\rho}_1}
= \frac{\tilde{\rho}_4}{\tilde{\rho}_1}
=10
$.
We observe a general behavior in the errors similar to that of
the thee-phase
case. The curves for the maximum and RMS errors of fluid \#2 basically
overlap with those for fluid \#3.
The maximum error curves appear to level off
on the order of magnitude $10^{-11}$ and the RMS error
curves appear to level off on the order of magnitude $10^{-14}$.


\subsection{Floating Liquid Lens}

In this subsection we employ the so-called floating liquid lens problem
to test the method developed herein.
The basic goal is to simulate and
study the equilibrium configuration of
an oil drop floating on the air-water
interface. There exist theoretical results about such
three-phase problems in the literature, in particular,
quantitative relations 
about the oil-drop thickness expressed in terms of
the other physical parameters have been developed
for the case when the gravity is dominant
by e.g.~Langmuir and de Gennes~\cite{Langmuir1933,deGennesBQ2003}.
We will compare our simulation results with
the Langmuir-de Gennes theory to evaluate the accuracy
of our method. 
The floating liquid lens problem has also been considered
in some of our previous works (see e.g.~\cite{Dong2014,Dong2015}).
It should be noted that the method to be tested here,
in terms of both the formulation and the algorithm,
is very different from those of \cite{Dong2014,Dong2015}.

\subsubsection{Floating Liquid Lens as a Three-Phase Problem}
\label{sec:floatlens_3p}


\begin{figure}
  \centerline{
    \includegraphics[width=3.5in]{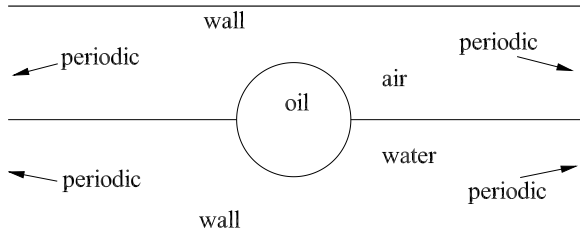}
  }
  \caption{
    Floating liquid lens: problem setup and initial configuration.
  }
  \label{fig:float_lens_config}
\end{figure}

We first simulate the floating lens problem in the natural way,
by treating it as a three-phase system consisting of
air, water and oil.
Specifically, we consider the domain sketched in
Figure \ref{fig:float_lens_config},
$-L\leqslant x\leqslant L$ and $0\leqslant y\leqslant \frac{4}{5}L$,
where $L=4cm$. The top and bottom sides
of the domain are solid walls, and in
the horizontal direction the domain is
periodic at $x=\pm L$.
The walls are of neutral wettability, i.e.~if any
fluid interface intersects the top or bottom walls
the contact angle at the wall will be $90^0$.
The top half of the domain is filled with air,
and the bottom half is filled with water.
An oil drop, initially circular with a radius $R_0=\frac{1}{5}L$,
is held at rest on the water surface,
and its center is located at
$\mathbf{x}_c=(x_c,y_c)=(0,\frac{2}{5}L)$.
The gravitational acceleration $\mathbf{g}_r$
is assumed to be in the $-y$ direction.
At $t=0$ the system is released and evolves
due to the interactions among the three
surface tensions (air/water, air/oil, water/oil)
and the gravity, reaching an equilibrium state
eventually. The objective of this
problem is to study the equilibrium configuration
of this three-phase system.

\begin{table}
  \begin{center}
    \begin{tabular}{llll}
      \hline
      Density [$kg/m^3$]: & air -- $1.2041$ & water -- $998.207$ & oil -- $577$ \\
      Dynamic viscosity [$kg/(m\cdot s)$]: & air -- $1.78\times 10^{-5}$
      & water -- $1.002\times 10^{-3}$ & oil -- $9.15\times 10^{-2}$ \\
      Surface tension [$kg/s^2$]: & air/water -- $0.0728$
      & air/oil -- $0.055$ & oil/water -- $0.04$ \\
      Gravity [$m/s^2$]: & varied from $0$ to $9.8$ \\
      \hline
    \end{tabular}
  \end{center}
  \caption{
    Physical property values  of air, water and oil.
  }
  \label{tab:float_lens_property}
\end{table}


The physical properties (including the densities, viscosities,
and surface tensions) of air, water and oil employed in this problem
are listed in Table \ref{tab:float_lens_property}.
We choose $L$ as the length scale,
the velocity scale as
$
U_0 = \sqrt{g_{r0}L}
$
where $g_{r0}=1 m/s^2$,
and the air density as the density scale $\varrho_d$.
The physical variables and parameters are
then normalized according to Table \ref{tab:normalization}.
In the following simulations water, oil and air are assigned
as the first, the second, and the third fluid,
respectively.

We employ the method described in Section  \ref{sec:alg}
to simulate this problem.
The flow domain is discretized using a mesh of $360$ equal-sized
quadrilateral elements, with $30$ elements along the horizontal
direction and $12$ elements along the vertical direction.
The element orders are varied between $9$ and $13$,
but the majority of results reported below are computed using
an element order $13$ within each element.
The source terms in the phase field equations \eqref{equ:CH_trans_1}
are set to $d_i=0$ ($i=1,\dots,N-1$).
%
On the top/bottom walls the no-slip condition,
equation \eqref{equ:bc_vel} with $\mathbf{w}=0$,
is imposed for the velocity, and the boundary
conditions \eqref{equ:bc_ci_1_trans}--\eqref{equ:bc_ci_2_trans}
with $d_{ai}=0$ and $d_{bi}=0$ are imposed for
the volume fractions.
In the horizontal direction all flow variables are
set to be periodic at $x=\pm L$.
%
The initial velocity is set to zero, and the initial
volume fraction distributions are set as
\begin{equation}
  \left\{
  \begin{split}
    &
    c_1 = \frac{1}{2}\left[1 - \tanh\left(\frac{y-y_c}{\sqrt{2}\eta}\right)  \right]
    \frac{1}{2}\left[1 + \tanh\left(\frac{|\mathbf{x}-\mathbf{x}_c|-R_0}{\sqrt{2}\eta}\right)  \right] \\
    &
    c_2 = \frac{1}{2}\left[1 - \tanh\left(\frac{|\mathbf{x}-\mathbf{x}_c|-R_0}{\sqrt{2}\eta}\right)  \right] \\
    &
    c_3 = 1-c_1-c_2.
  \end{split}
  \right.
  \label{equ:float_lens_ic}
\end{equation}
The simulation parameter values are summarized in
Table \ref{tab:float_lens_param}.

\begin{table}
  \begin{center}
    \begin{tabular}{ll}
      \hline
      Parameters & Values \\
      $\lambda_{ij}$ & given by equation \eqref{equ:free_energy_param} \\
      $\eta/L$ & $0.01$ and $0.0075$ \\
      $m_0\varrho_d U_0/L$ & $10^{-8}$ \\
      $U_0\Delta t/L$ & $1.0 \times 10^{-5}$ \\
      $\rho_0$ & $\min(\tilde{\rho}_1,\tilde{\rho}_2, \tilde{\rho}_3)$ \\
      $\nu_0$ & $5\max\left(\frac{\tilde{\mu}_1}{\tilde{\rho}_1},
      \frac{\tilde{\mu}_2}{\tilde{\rho}_2}, \frac{\tilde{\mu}_3}{\tilde{\rho}_3} \right)$ \\
      $\mathcal{K}_0$ & given by equation \eqref{equ:def_K0} \\
      $S$ & $\sqrt{\frac{4\gamma_0}{\mathcal{K}_0\Delta t}}$ \\
      $\alpha$ & computed by equation \eqref{equ:def_alpha} \\
      $J$ (temporal order) & $2$ \\
      Number of elements & $360$ \\
      Element order & $9 \sim 13$ (mostly $13$)
      \\
      \hline
    \end{tabular}
  \end{center}
  \caption{Simulation parameter values for air/water/oil three-phase
    floating lens problem.}
  \label{tab:float_lens_param}
\end{table}


We observe that a smooth field for the initial volume fractions (such as
those given above)
is important for the current method.
Note that in the current formulation $m_{ij}(\vec{c})$ are functions
dependent on the volume fraction distributions.
This places a more stringent requirement on the smoothness
of the  initial volume fractions.
Since the initial volume fraction distributions are unknown physically
and must be prescribed, any discontinuity in the prescribed
initial volume-fraction distributions will affect the dynamics
and may influence the time to reach the equilibrium state.
For example, in \cite{Dong2014} some Heaviside step functions
are involved in the prescribed initial volume fractions for
the floating liquid lens problem, inducing discontinuities
in the distributions. Those initial volume-fraction
distributions do not work well with the current method.

\begin{figure}
  \centerline{
    \includegraphics[width=4.5in]{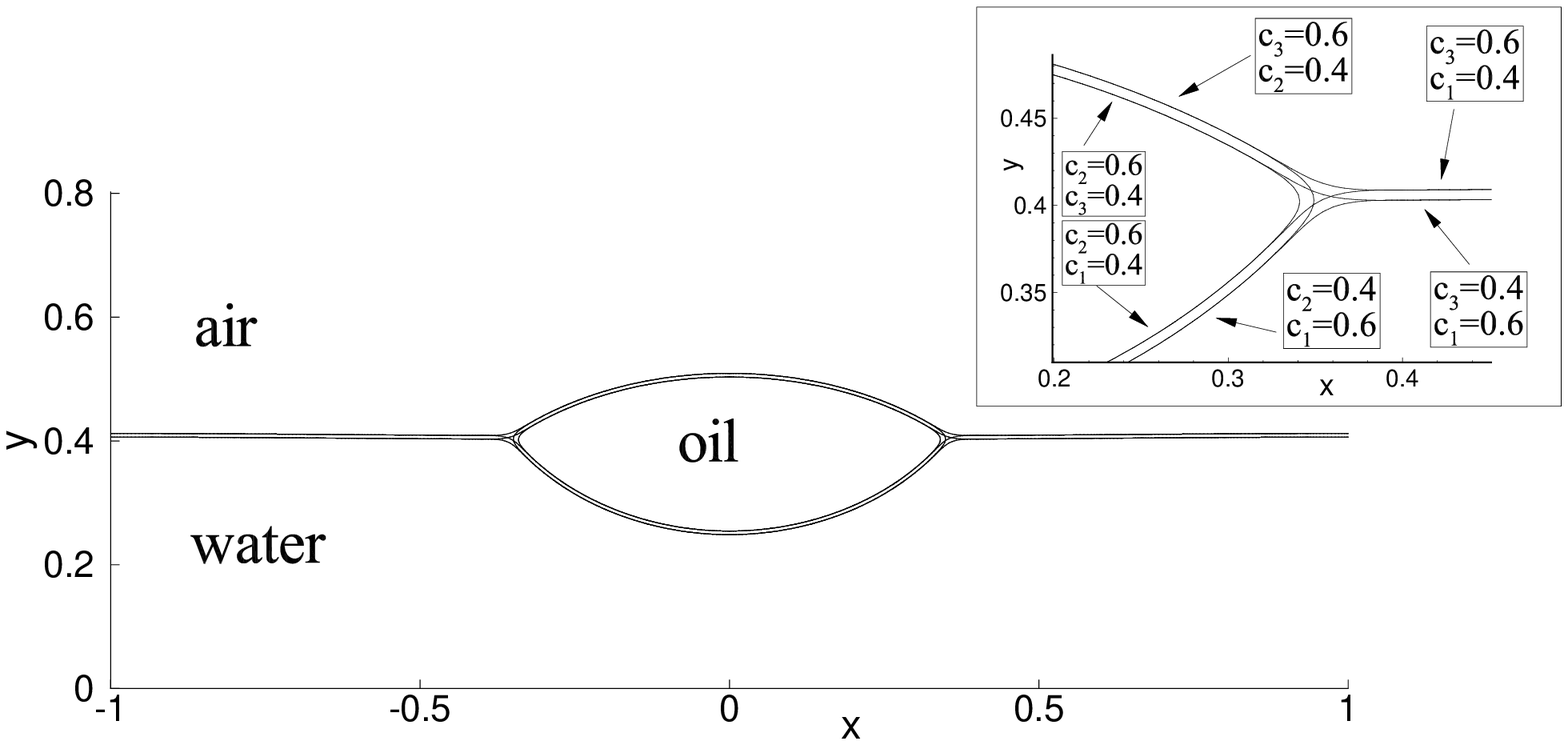}(a)
  }
  \centerline{
    \includegraphics[width=3in]{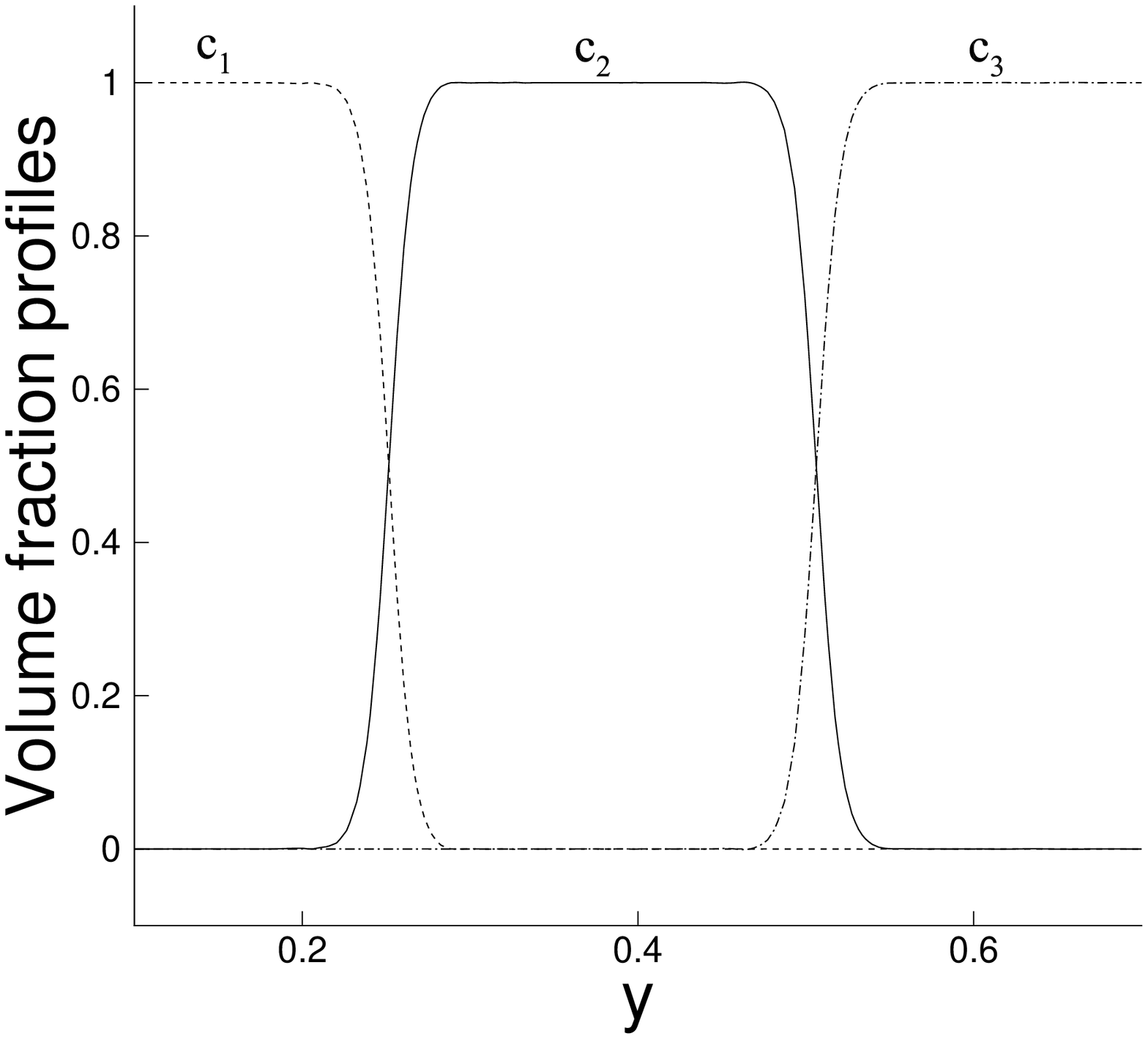}(b)
    \includegraphics[width=3in]{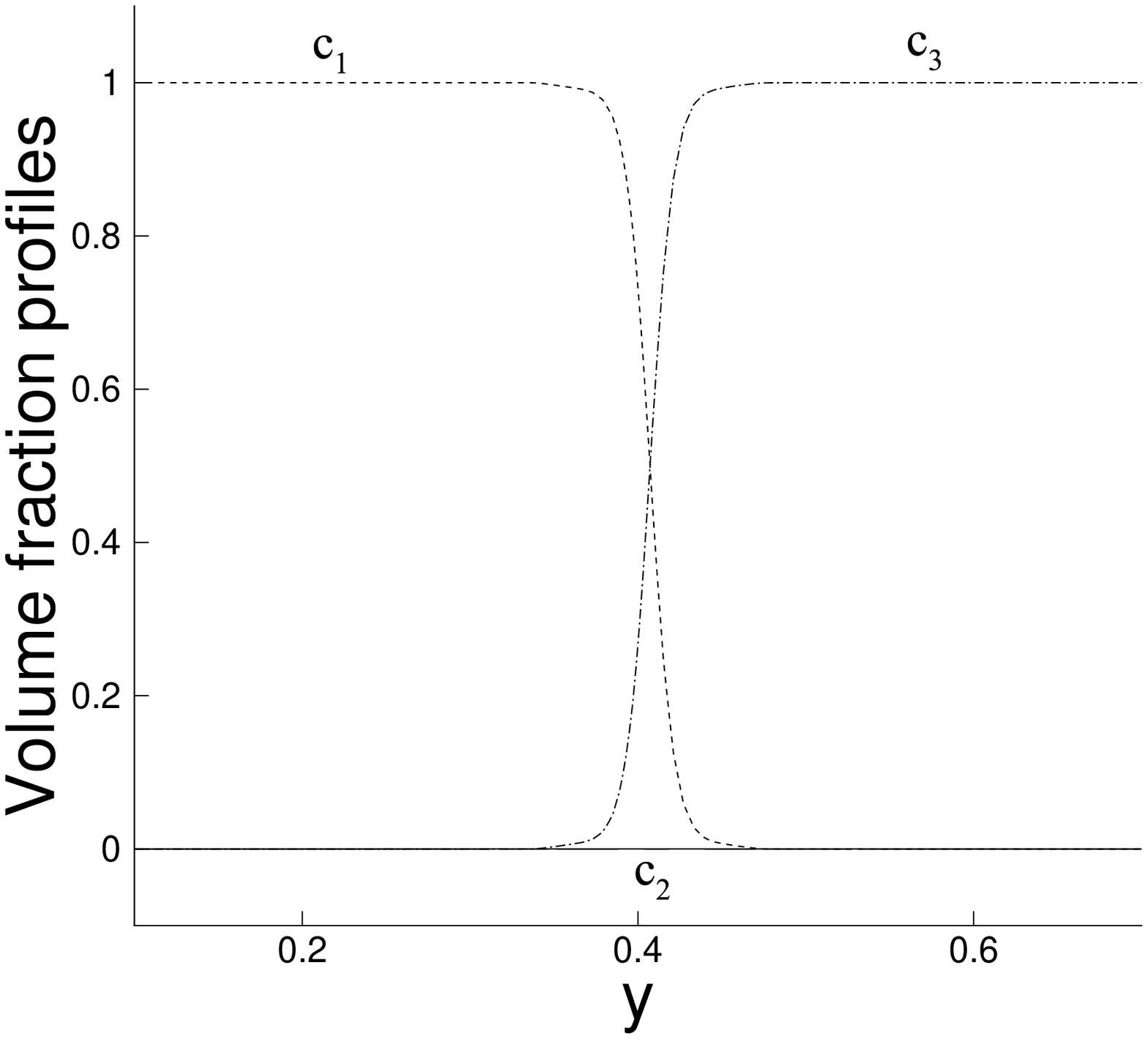}(c)
  }
  \caption{
    Floating liquid lens (gravity $|\mathbf{g}_r|=0.5m/s^2$):
    (a) Contour levels $c_i=0.4$ and $0.6$ ($i=1,2,3$), and
    the inset shows a magnified view of the 3-phase line region.
    Profiles of the volume fractions $c_i$ ($i=1,2,3$)
    along the centerline $x=0$ (b) and along the vertical line $x=0.6$ (c).
  }
  \label{fig:float_lens_profiles}
\end{figure}


We first look into the distribution characteristics of
different fluids within the domain.
Figure \ref{fig:float_lens_profiles}(a) shows two contour levels
$c_i=0.4$ and $c_i=0.6$ ($i=1,2,3$) of
the three volume fractions for the equilibrium state of
this three-phase system with
a gravitational acceleration $0.5 m/s^2$.
Note that $c_1$, $c_2$ and $c_3$ correspond to water, oil and
air, respectively.
The result is computed using an interfacial thickness scale
$\eta/L=0.01$ and an element order $11$.
The inset of Figure \ref{fig:float_lens_profiles}(a) 
is a zoomed-in view around the three-phase line
region. It can be observed that,
along the air/oil interface and away from the three-phase line region,
the contours $c_3=0.6$ and $c_2=0.4$ coincide with each other
and the contours $c_3=0.4$ and $c_2=0.6$ coincide with
each other. This is consistent with the intuition as water is not present
(i.e.~$c_1=0$) on the air-oil interface
away from the three-phase region.
Similar distribution characteristics can be observed on
the water/oil and air/water interfaces away from the three-phase
line region.
Figure \ref{fig:float_lens_profiles}(b) shows profiles
of the three volume fractions along the centerline of the domain ($x=0$).
One can observe that in the bulk of the water region (oil region,
air region) $c_1=1$ (resp.~$c_2=1$, $c_3=1$) while the other
two volume fractions are zeros.
At the water/oil interface
$c_1$ decreases from the unit value to zero
and $c_2$ increases from zero to the unit value,
while $c_3=0$ in this region.
At the air/oil interface $c_2$ decreases from the unit value to zero
and $c_3$ increases from zero to the unit value,
while $c_1=0$ in this region.
Figure \ref{fig:float_lens_profiles}(c) shows profiles of
the three volume fractions along another vertical line $x/L=0.6$.
Since only the air and water exist in this region,
one observes that $c_2=0$ and
that $c_1$ transitions to $c_3$ as the air/water
interface  is crossed.

\begin{figure}
  \centering
  \includegraphics[width=4in]{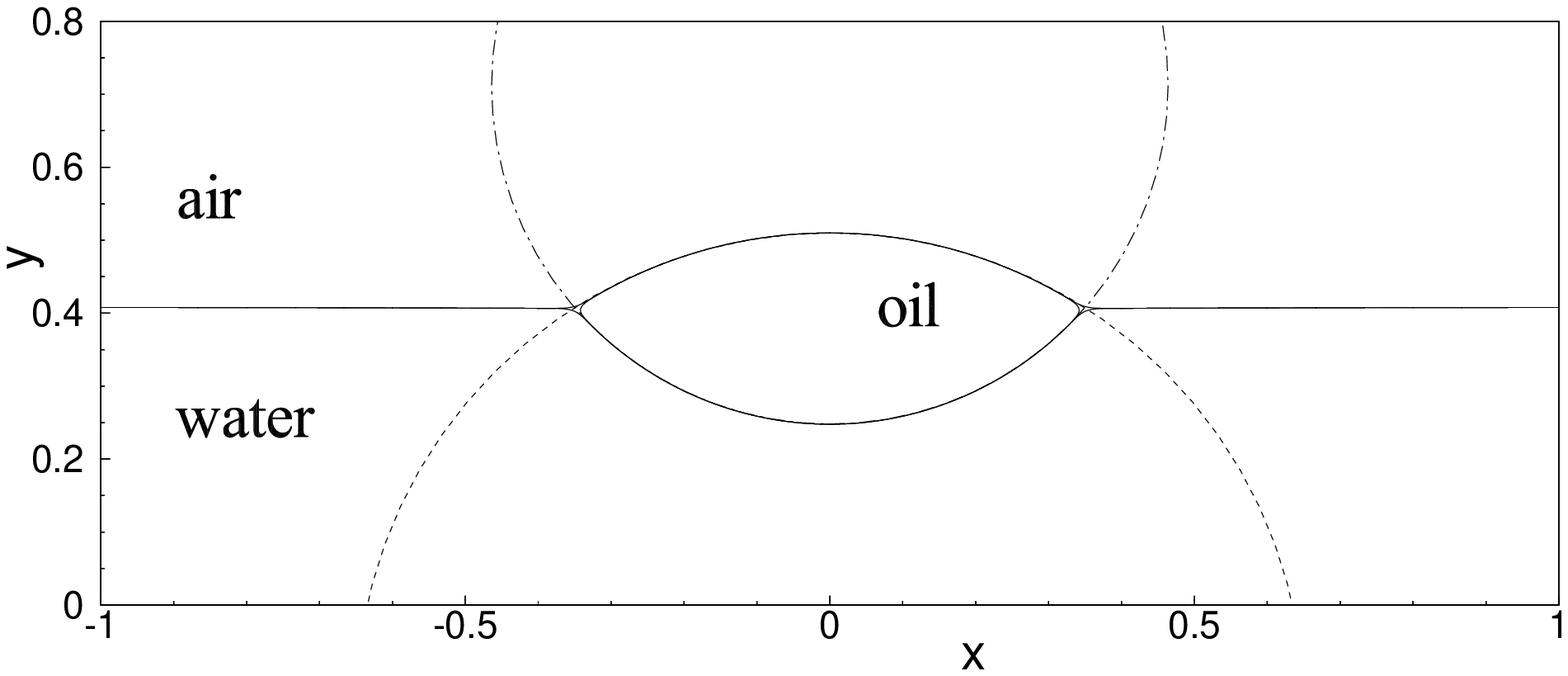}(a)
  \includegraphics[width=4in]{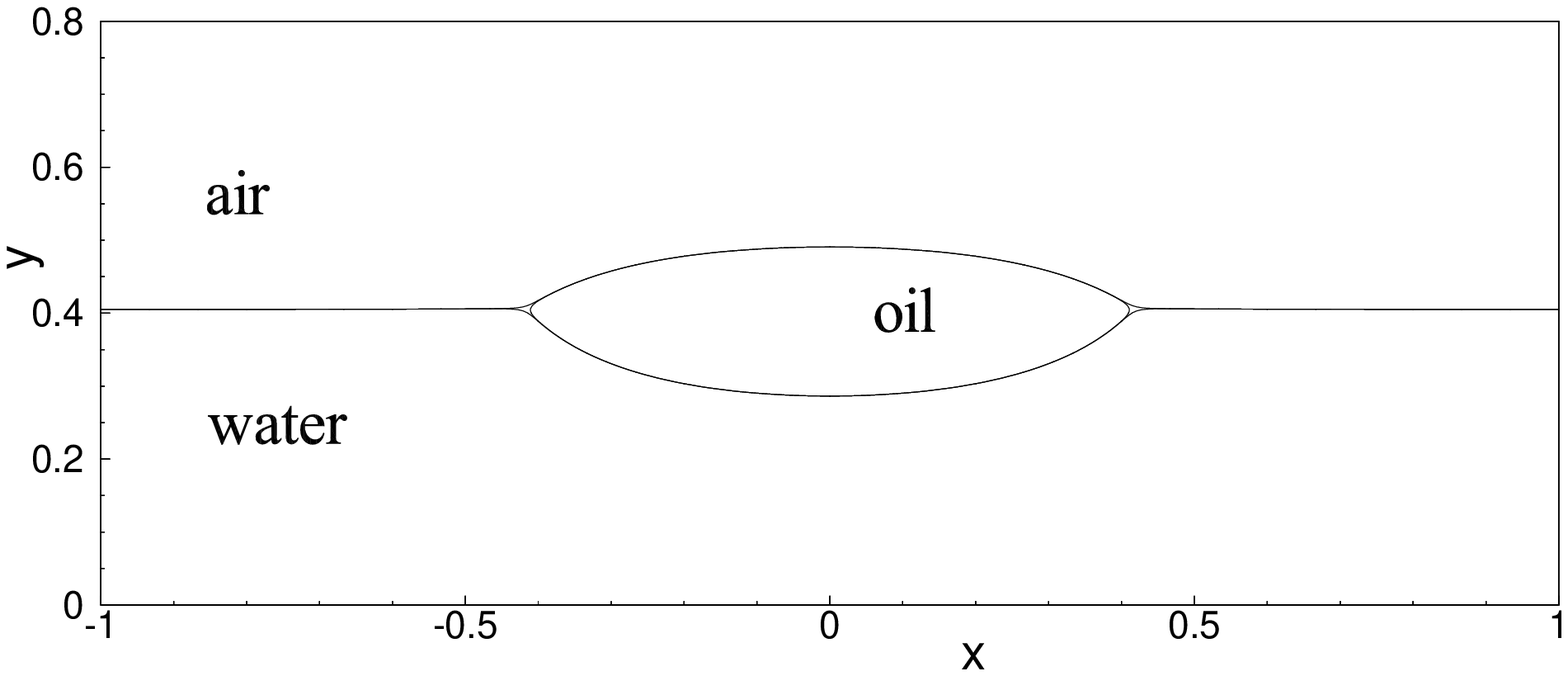}(b)
  \includegraphics[width=4in]{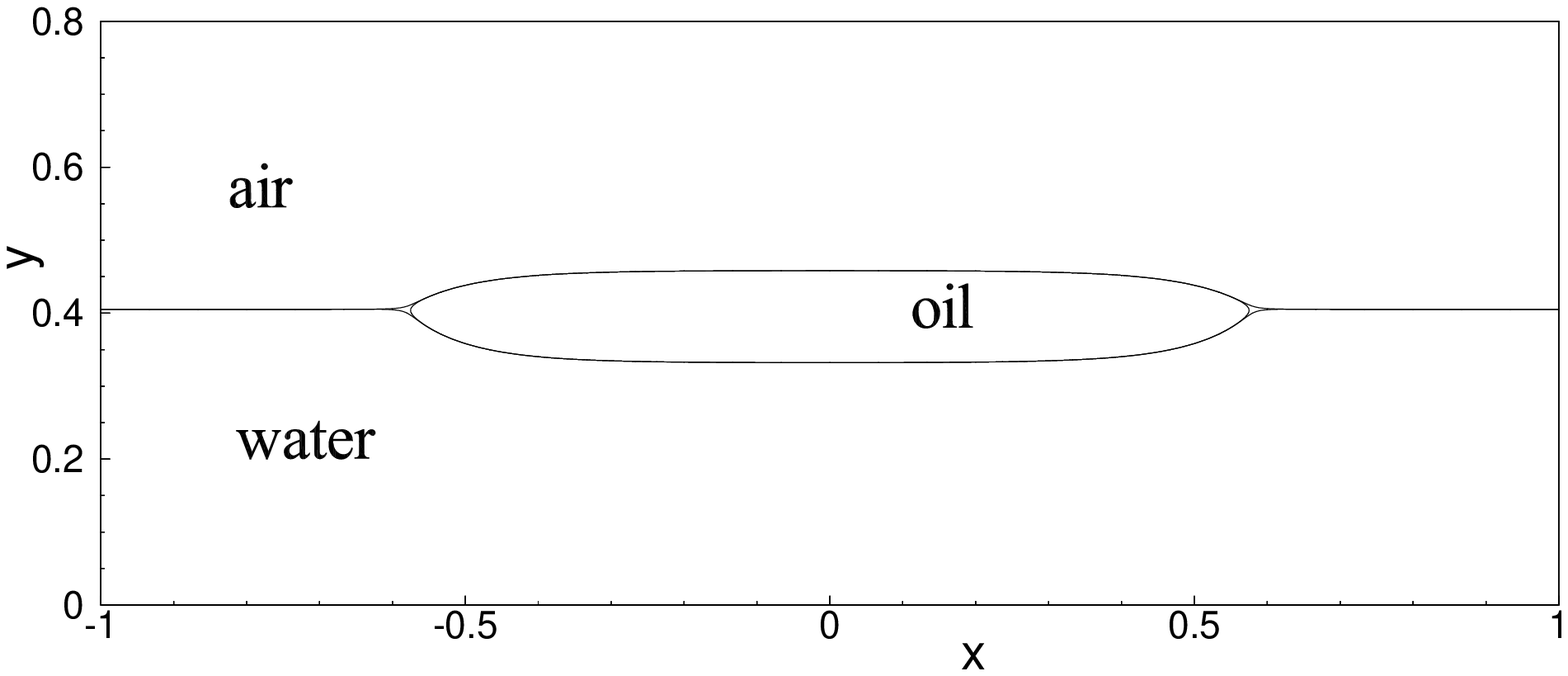}(c)
  \caption{
    Floating liquid lens: equilibrium oil-drop profiles  on air/water
    interface with
    (a) no gravity,
    (b) $|\mathbf{g}_r|=2m/s^2$, and
    (c) $|\mathbf{g}_r|=7.5m/s^2$.
    Fluid interfaces are visualized by contour levels $c_i=1/2$ ($i=1,2,3$).
    In (a) dashed and dashed-dot curves are part of two circles.
  }
  \label{fig:float_lens_equil_profile}
\end{figure}

Let us now look into the equilibrium configuration of this three-phase
system. The physics of floating liquid lenses was explained
in~\cite{Langmuir1933,deGennesBQ2003}. The equilibrium shape
of the oil drop is determined by the interplay of the three
pairwise surface tensions and the gravity, and it is also
affected by the three densities.
If the surface tension effects dominate (e.g.~when the oil drop is small),
the equilibrium drop shape comprises two circular caps in two
dimensions (or two spherical caps in three dimensions).
On the other hand, if the gravity effect dominates (e.g.~when 
the oil drop is large) the oil forms a puddle at equilibrium.
To determine which effect dominates, one can compare
the drop size with the three capillary lengths associated
with the three fluid interfaces; see \cite{deGennesBQ2003}
for details.

We have varied the magnitude of the gravitational
acceleration systematically, and simulated the equilibrium
configurations of this system corresponding to these gravity values.
In Figure \ref{fig:float_lens_equil_profile} we
show the equilibrium profiles of the oil drop corresponding
to gravity: $|\mathbf{g}_r|=0$, $2m/s^2$ and $7.5m/s^2$.
The fluid interfaces are visualized by the contour
levels $c_i=1/2$ ($i=1,2,3$).
These results correspond to an interfacial thickness
scale $\eta/L=0.0075$ and an element order $13$ within
each element in the simulations.
In these plots one can observe a small star-shaped region around
the three-phase line (where the three fluid regions intersect),
which is due to the fact that within this region no fluid has
a volume fraction larger than $1/2$.
In Figure \ref{fig:float_lens_equil_profile}(a) (zero gravity),
we have also shown two reference circles (dashed and dashed-dot curves),
which overlap with the upper and lower pieces of the
oil-drop profile. This indicates that with zero gravity (surface tensions
dominant)
the computed oil-drop profile indeed consists of two
circular caps, consistent with the theory~\cite{deGennesBQ2003}.
With increasing gravity the oil lens tends to spread out on
the water surface (Figure \ref{fig:float_lens_equil_profile}(b)).
With a gravity $|\mathbf{g}_r|=7.5m/s^2$ the oil forms
a puddle on the water surface under the conditions
considered here, with flat upper and lower
surfaces (Figure \ref{fig:float_lens_equil_profile}(c)).
The simulation results are qualitatively consistent with 
the Langmuir-de Gennes theory~\cite{Langmuir1933,deGennesBQ2003}.

\begin{figure}
  \centerline{
    \includegraphics[width=3in]{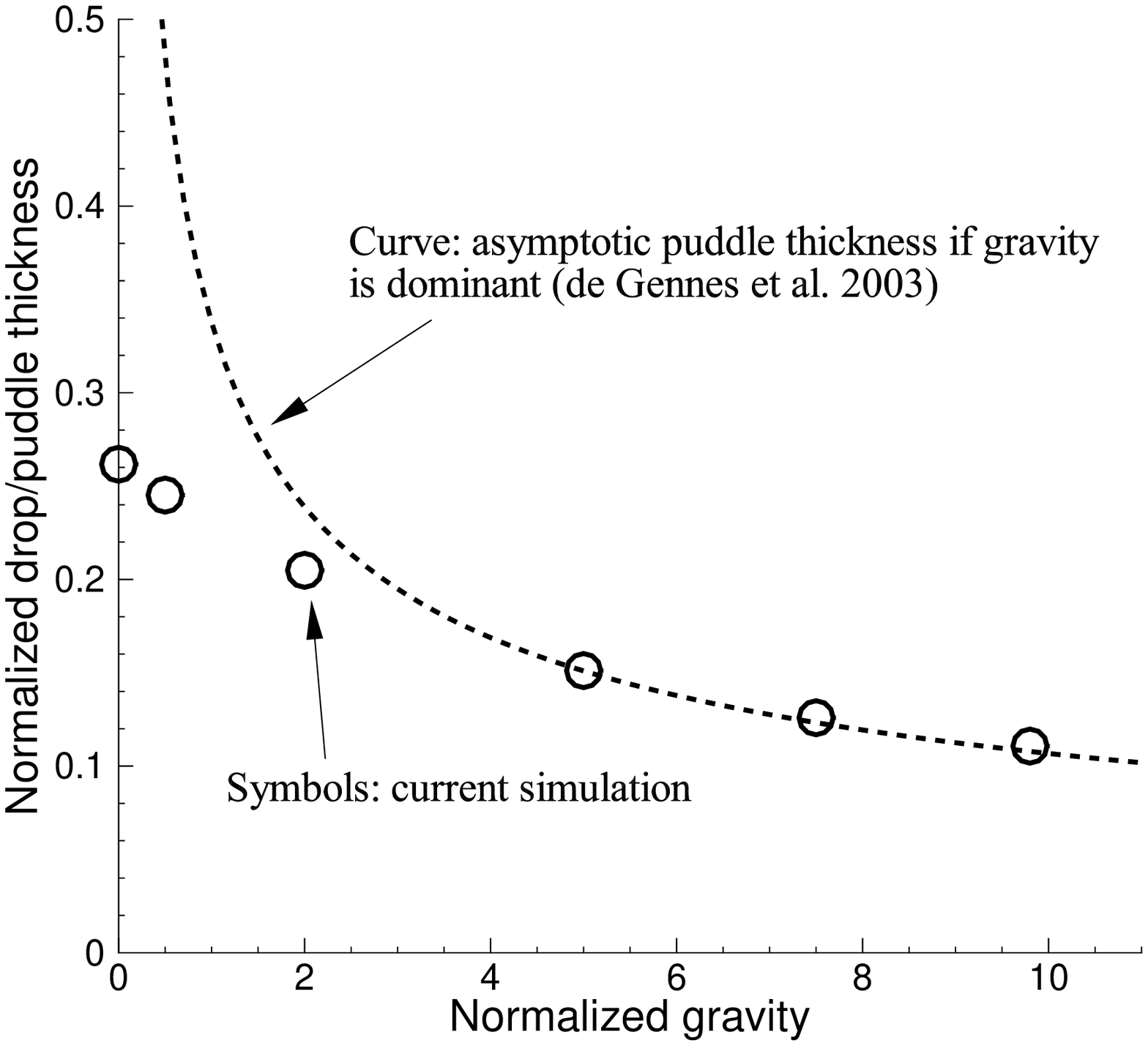}
  }
  \caption{
    Comparison of oil-drop/-puddle thickness as a function
    of the gravity between current simulations and the de Gennes
    theory~\cite{deGennesBQ2003}.
  }
  \label{fig:float_lens_gravity}
\end{figure}

We next show some quantitative comparisons with the Langmuir-de Gennes
theory. Following \cite{Dong2014}, we define the oil-drop/-puddle
thickness as the largest distance between the upper and lower
boundaries of the equilibrium drop/puddle profile along
the vertical direction. 
When the gravity is dominant, the asymptotic thickness of
the oil puddle (denoted by $e_c$)
is given by the following expression~\cite{deGennesBQ2003}
\begin{equation}
  e_c = \sqrt{\frac{2(\sigma_{ao}+\sigma_{ow}-\sigma_{aw})\rho_w}
    {\rho_o(\rho_w-\rho_o)|\mathbf{g}_r|} }
  \label{equ:puddle_thickness}
\end{equation}
where $\rho_w$ and $\rho_o$ are the water and oil densities respectively,
$\sigma_{aw}$, $\sigma_{ao}$ and $\sigma_{ow}$ are
the air/water, air/oil and oil/water surface tensions respectively,
and $|\mathbf{g}_r|$ is the magnitude of the gravitational
acceleration.
We have computed the oil-drop/-puddle thickness corresponding
to different gravity magnitudes.
In Figure \ref{fig:float_lens_gravity} we plot
the oil-drop/-puddle thickness as a function of
the normalized gravity
$
\frac{|\mathbf{g}_r|}{U_0^2/L} = \frac{|\mathbf{g}_r|}{g_{r0}}
$
where $g_{r0}=1m/s^2$.
The symbols denote results from current simulations,
and the dashed curve denotes the relation
given by equation \eqref{equ:puddle_thickness}.
The simulation results correspond to $\eta/L=0.0075$
and element order $13$ in the simulations.
It is observed that when the gravity becomes large ($|\mathbf{g}_r|=5m/s^2$
or larger) the puddle thickness values from the simulations
are in good agreement with
the asymptotic puddle thickness values from the Langmuir-de Gennes
theory~\cite{deGennesBQ2003}.

\subsubsection{Floating Liquid Lens as a Four-Phase Problem with One
Absent Fluid}



The floating liquid lens problem can also be physically
considered as a multiphase system consisting of more than
three fluid components, in which however only the three fluids
air, water and oil are present. 
We will next treat and simulate
the floating liquid lens problem as a four-phase system,
comprising air, water, oil, and another liquid referred to
as $F_A$, in which the liquid $F_A$ is absent however.
We assume that these four fluids are
mutually immiscible. 
Thanks to the reduction consistency of our formulation,
we expect that the simulation of this four-phase problem
using the method developed herein will produce the same results
as the three-phase simulation.

In addition to the physical parameters given in
Table \ref{tab:float_lens_property} for the properties of air/water/oil,
we assume the following physical parameters involving
$F_A$:
\begin{equation*}
  \left\{
  \begin{array}{llll}
    F_A \ \text{density}: &  100kg/m^3 \\
    F_A \ \text{dynamic viscosity}: & 9.0\times 10^{-3} kg/(m\cdot s) \\
    \text{Surface tension} \ [kg/s^2]: & \text{air/}F_A - 0.045,
    & \text{water/}F_A - 0.05, & \text{oil/}F_A - 0.052.
    \\
  \end{array}
  \right.
\end{equation*}
We assign water, oil, $F_A$, and air as the first, the second, the third,
and the fourth fluid in the simulations.
We employ the same length scale, velocity scale and the density scale
as in the three-phase simulations for the normalization of
the problem.

\begin{figure}
  \centerline{
    \includegraphics[width=3.5in]{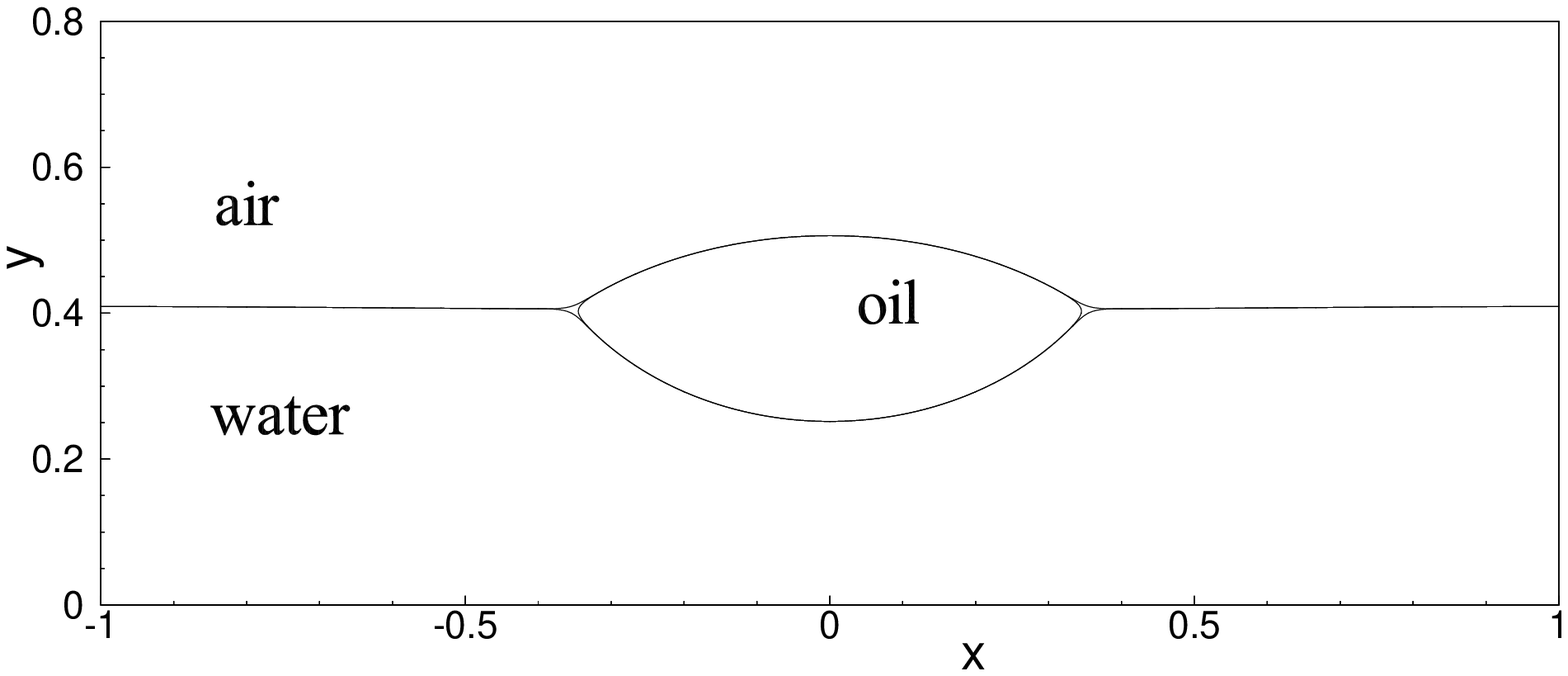}(a)
    \includegraphics[width=2.6in]{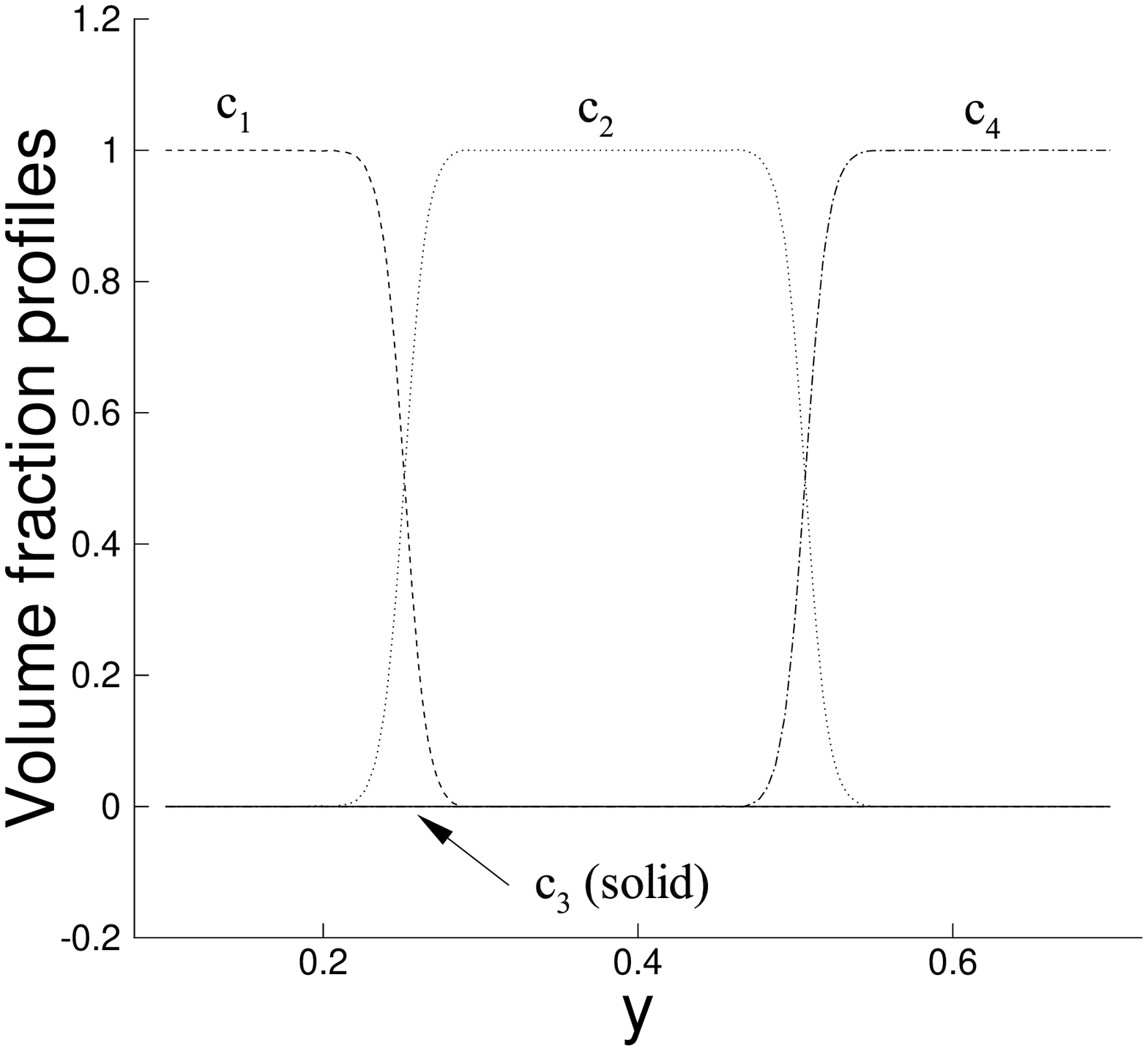}(b)
  }
  \centerline{
    \includegraphics[width=2.6in]{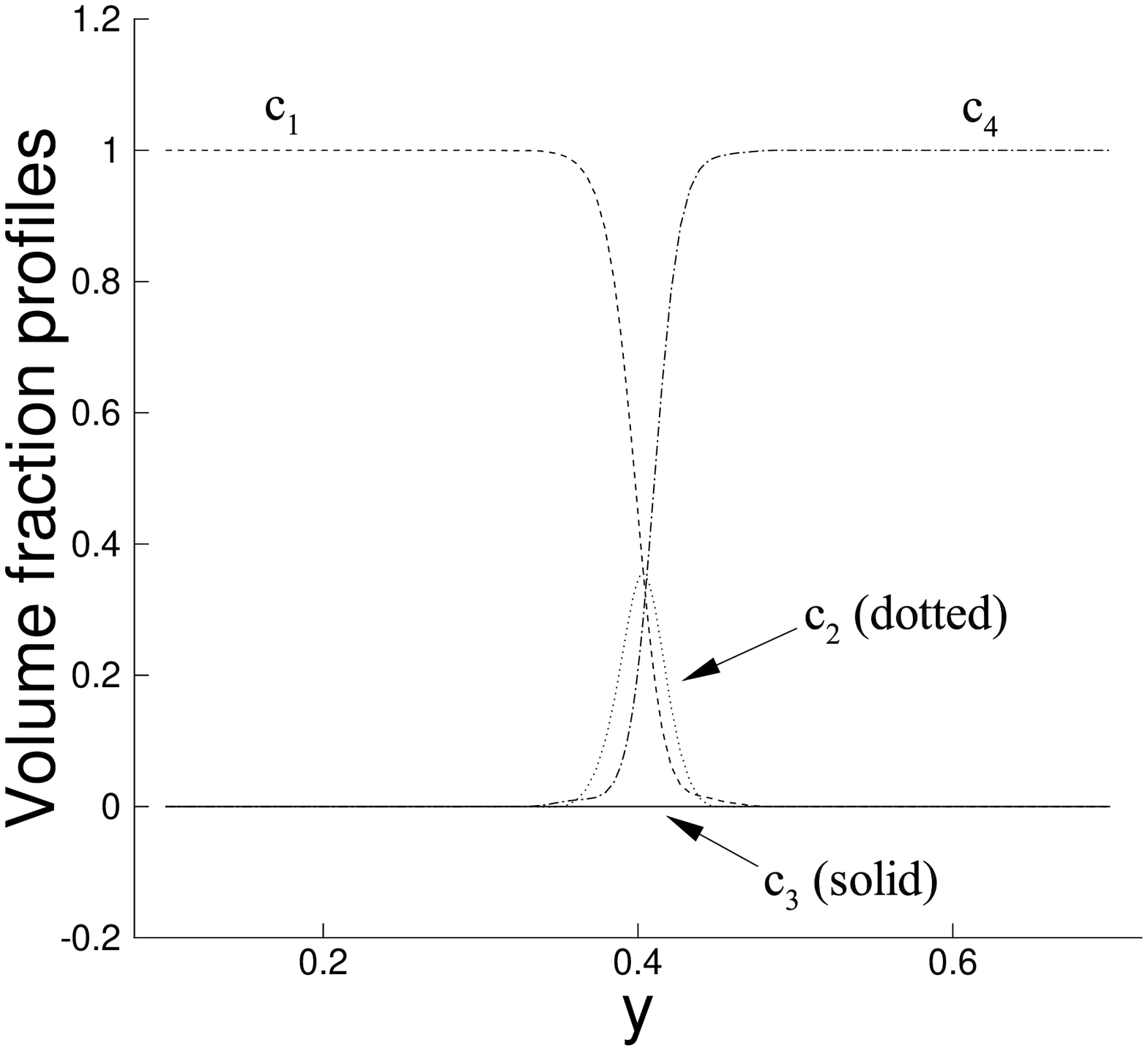}(c)
    \includegraphics[width=2.6in]{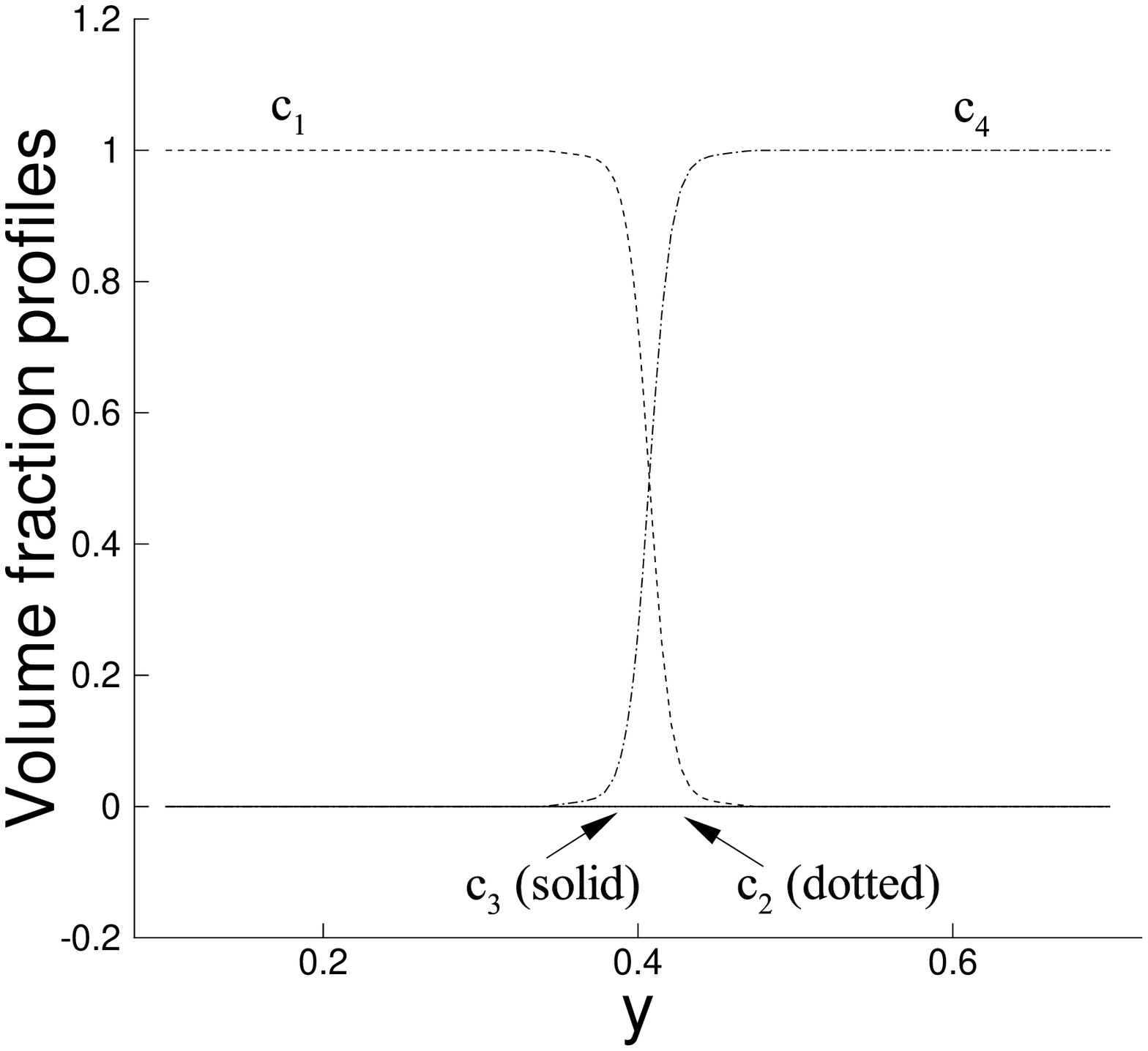}(d)
  }
  \caption{
    Floating liquid lens simulated as a four-phase problem with
    one absent fluid (gravity $0.5m/s^2$):
    (a) Equilibrium configuration visualized by contour levels
    $c_i=1/2$ ($i=1,2,4$).
    (b)--(d): Profiles of volume fractions $c_i$ ($i=1,2,3,4$)
    along the vertical lines $x=0$ (b), $x=0.351$ (c), and
    $x=0.6$ (d). $c_3$ corresponds to the absent fluid.
  }
  \label{fig:floatlens_4p1pabs_profiles}
\end{figure}

The flow domain and the problem setting will be the same as
those of the three-phase simulations.
We use the same spectral-element mesh and the same boundary
conditions for the four-phase simulations.
The initial velocity is zero, and 
the initial volume fractions for the four-phase simulation
are as follows:
\begin{equation}
  \left\{
  \begin{split}
    &
    c_1 = \frac{1}{2}\left[1 - \tanh\left(\frac{y-y_c}{\sqrt{2}\eta}\right)  \right]
    \frac{1}{2}\left[1 + \tanh\left(\frac{|\mathbf{x}-\mathbf{x}_c|-R_0}{\sqrt{2}\eta}\right)  \right] \\
    &
    c_2 = \frac{1}{2}\left[1 - \tanh\left(\frac{|\mathbf{x}-\mathbf{x}_c|-R_0}{\sqrt{2}\eta}\right)  \right] \\
    &
    c_3 = 0 \\
    &
    c_4 = 1-c_1 - c_2 - c_3 = 1-c_1 - c_2
  \end{split}
  \right.
  \label{equ:float_lens_4p1pabs_ic}
\end{equation}
where $y_c$, $R_0$, $\mathbf{x}_c$ are the same as those in
the three-phase simulations.
Note that $c_3$ corresponds to the liquid $F_A$. It is set to zero (absent)
initially, and so physically $F_A$ should remain absent over time.

We consider only one case for the four-phase simulations,
with a gravity $0.5m/s^2$,
and employ the following simulation parameters:
$\eta/L=0.01$, and an element order $11$ for all elements.
The rest of the simulation parameters are the same as
given by Table \ref{tab:float_lens_param}.
We will compare the four-phase simulation results with
the three-phase simulations using the same simulation parameter values.


Figure \ref{fig:floatlens_4p1pabs_profiles}(a)
shows the equilibrium configuration of the system
(corresponding to a gravity $0.5m/s^2$)
from the four-phase simulations. The fluid interfaces
are visualized by the contour lines
$c_i=\frac{1}{2}$ ($i=1,2,4$).
Figures \ref{fig:floatlens_4p1pabs_profiles}(b)--(d)
show the profiles of the four volume
fractions $c_i$ ($i=1,2,3,4$) along three vertical
lines located at $x=0$, $x=0.351$ and $x=0.6$.
Note that the vertical line $x=0.351$ passes through
the right star-shaped region around the three-phase line
in Figure \ref{fig:floatlens_4p1pabs_profiles}(a).
We observe that the distributions for
$c_1$ (water), $c_2$ (oil) and $c_4$ (air)
are very similar to those from the three-phase
simulations (see e.g.~Figure \ref{fig:float_lens_profiles}(b)-(c)).
The volume fraction $c_3$ (liquid $F_A$) is
practically zero,
with a maximum value on the order of magnitude 
$10^{-13}$ in the entire domain.

\begin{figure}
  \centerline{
    \includegraphics[width=3in]{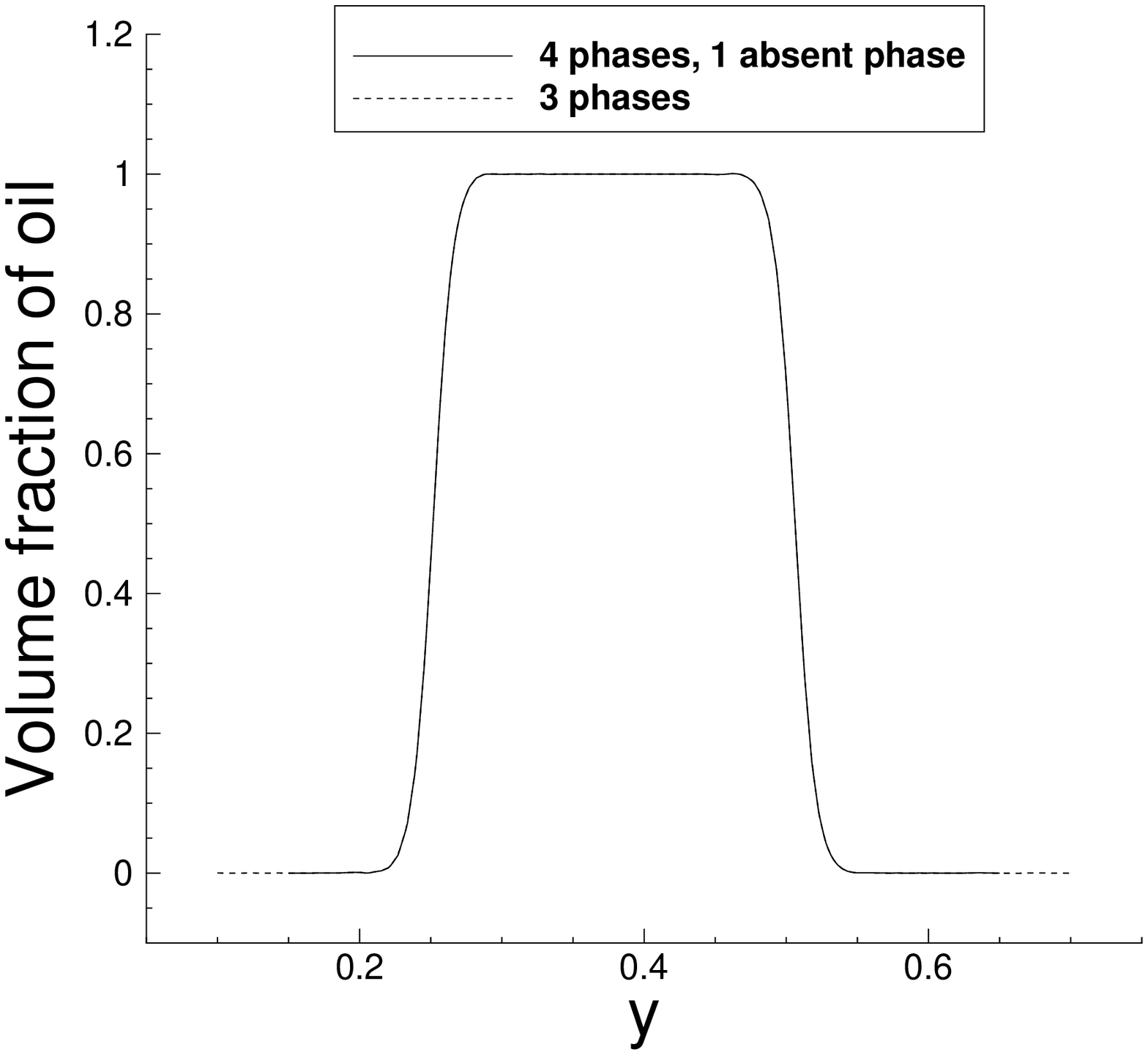}(a)
    \includegraphics[width=3in]{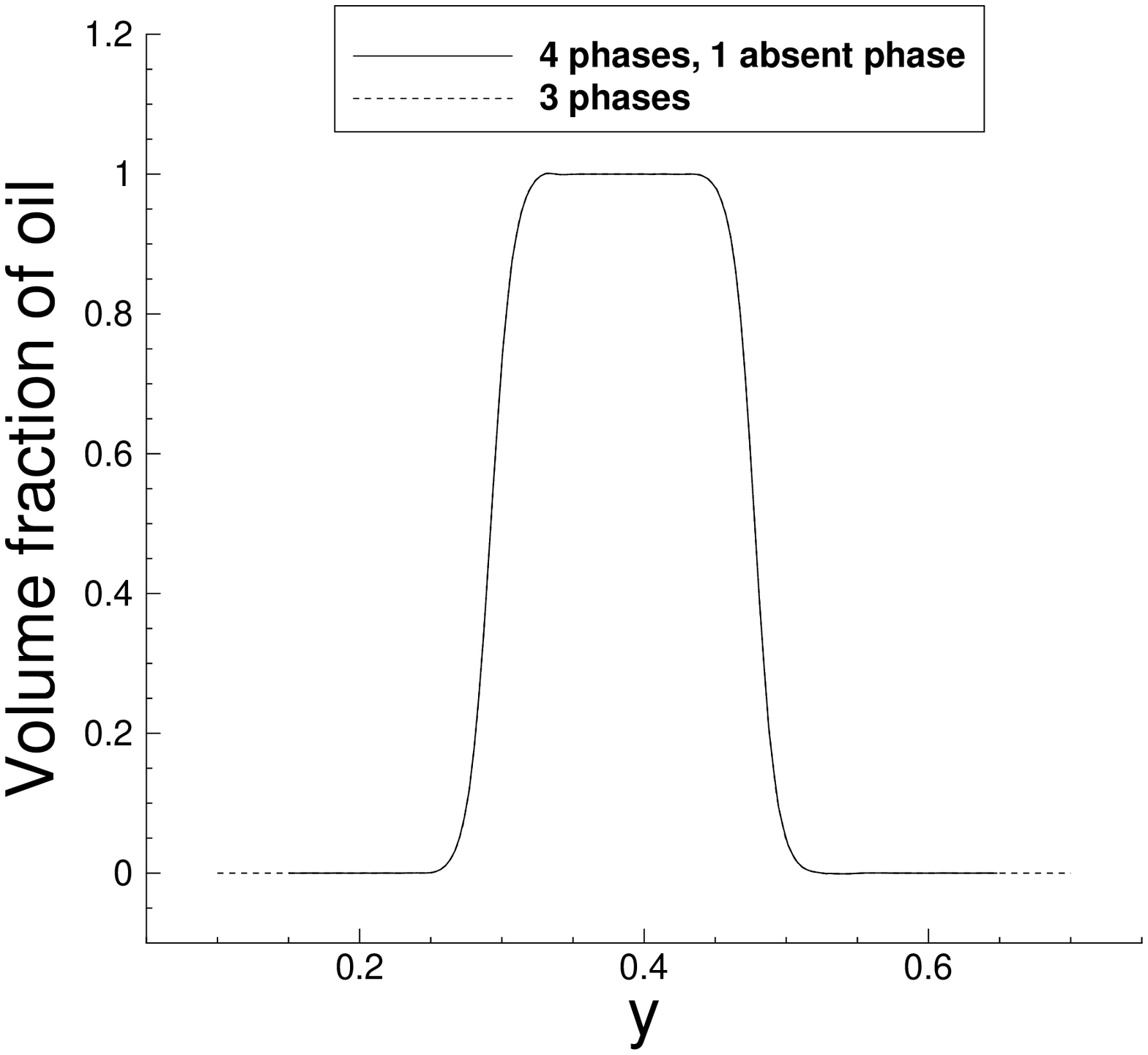}(b)
  }
  \centerline{
    \includegraphics[width=3in]{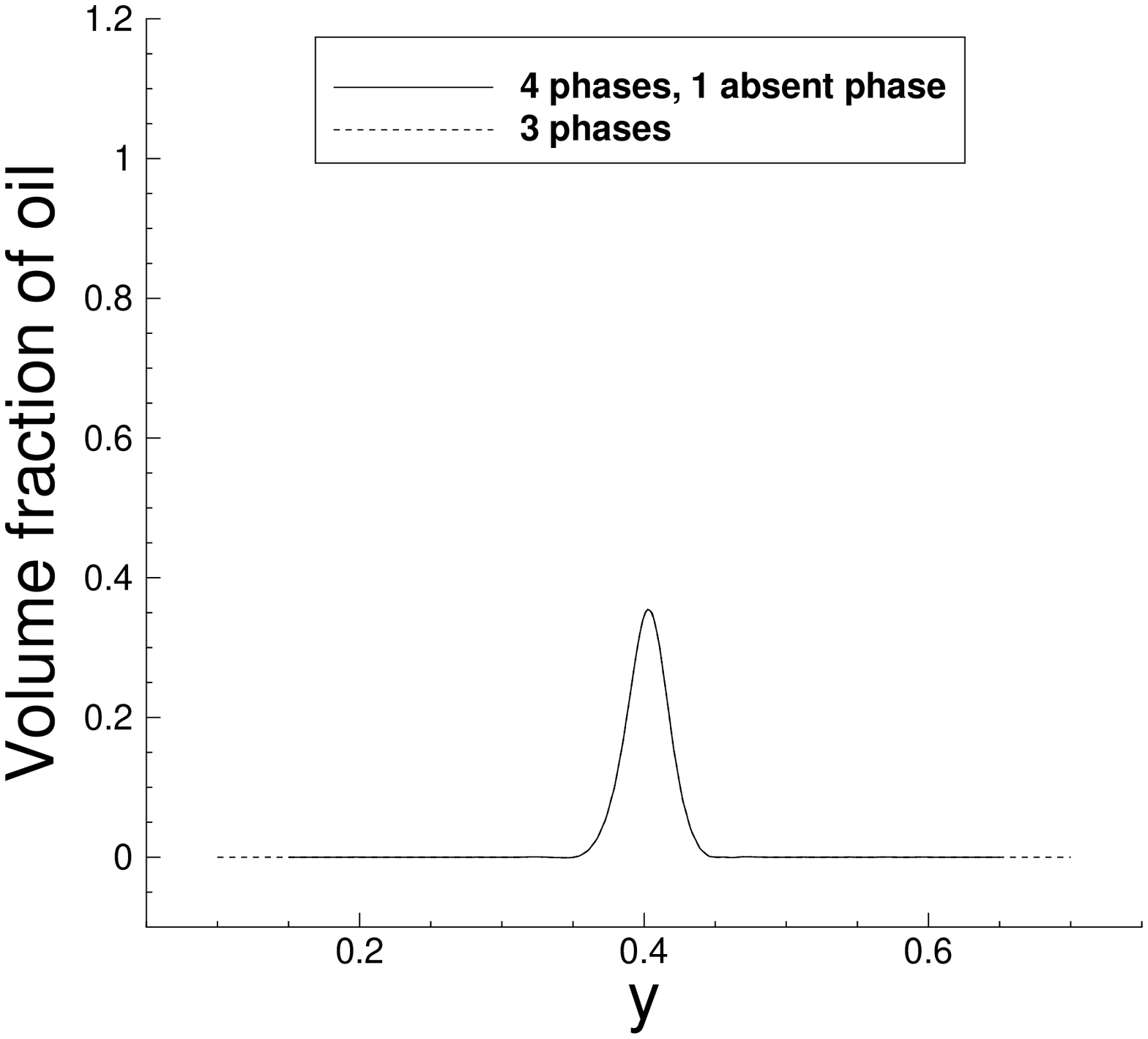}(c)
    \includegraphics[width=3in]{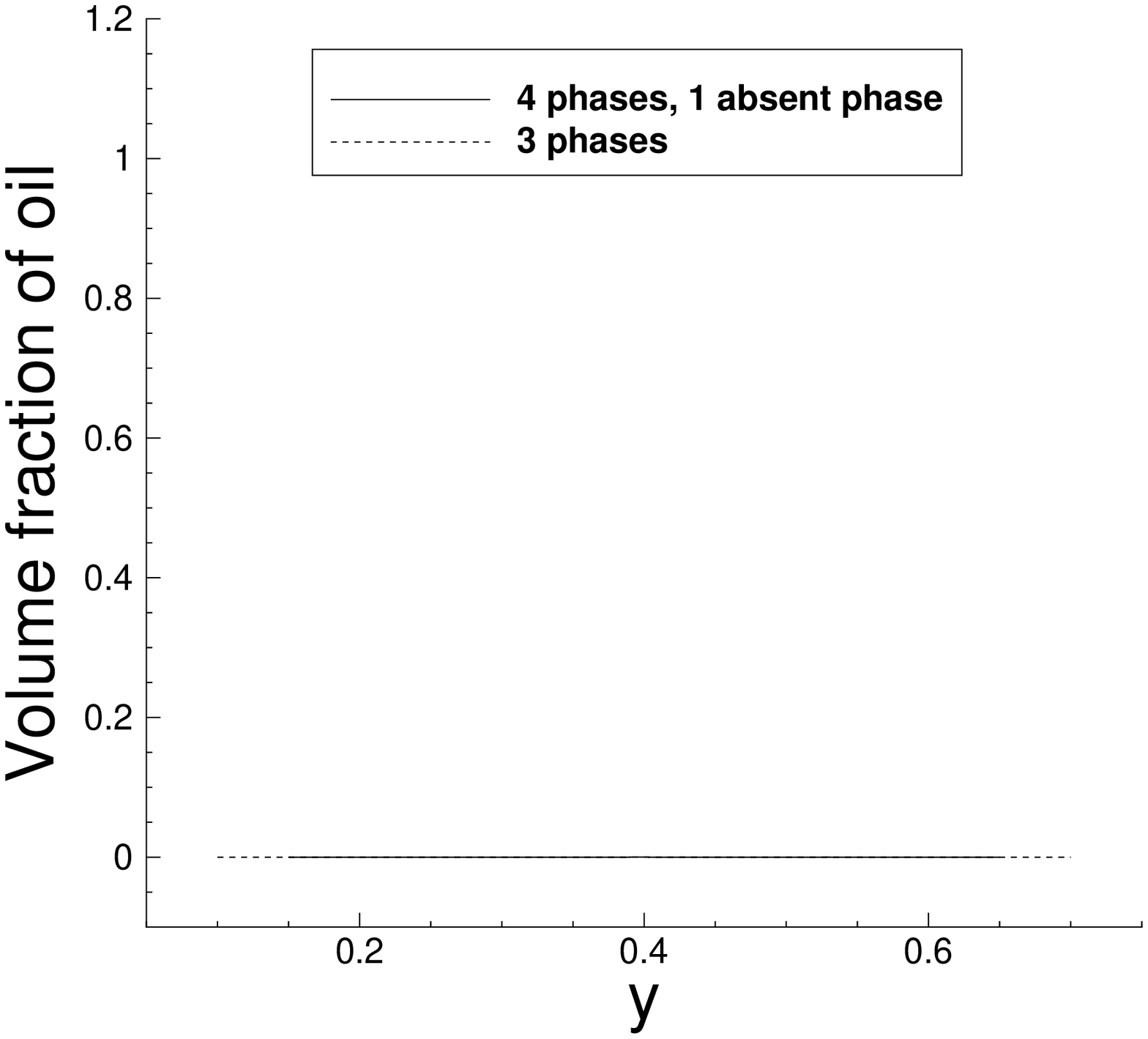}(d)
  }
  \caption{
    Comparison of volume-fraction profiles of the oil
    (gravity $0.5m/s^2$) simulated
    as a three-phase problem and  as a four-phase problem
    with one absent fluid, along  vertical lines located at
    (a) $x=0$,
    (b) $x=0.2$,
    (c) $x=0.351$,
    (d) $x=0.6$.
  }
  \label{fig:floatlens_4p1pabs_compare_3p}
\end{figure}

In Figure \ref{fig:floatlens_4p1pabs_compare_3p}
we compare the volume-fraction profiles of the
oil ($c_2$) obtained from the four-phase simulation
(with one absent fluid) and from the three-phase
simulation, computed under the same simulation
parameter values.
The four plots correspond to the profiles along
several vertical lines
located at $x=0$, $0.2$, $0.351$ and $0.6$.
The profiles from the four-phase simulation (solid
curves) almost exactly overlap with
those from the three-phase simulation (dashed curves),
suggesting that the four-phase simulation (with one
absent fluid) has produced the same results
as the three-phase simulation for
the floating liquid lens problem.



\subsection{Dynamics of a Four-Phase Problem}

\begin{figure}
  \centerline{
    \includegraphics[width=1.5in]{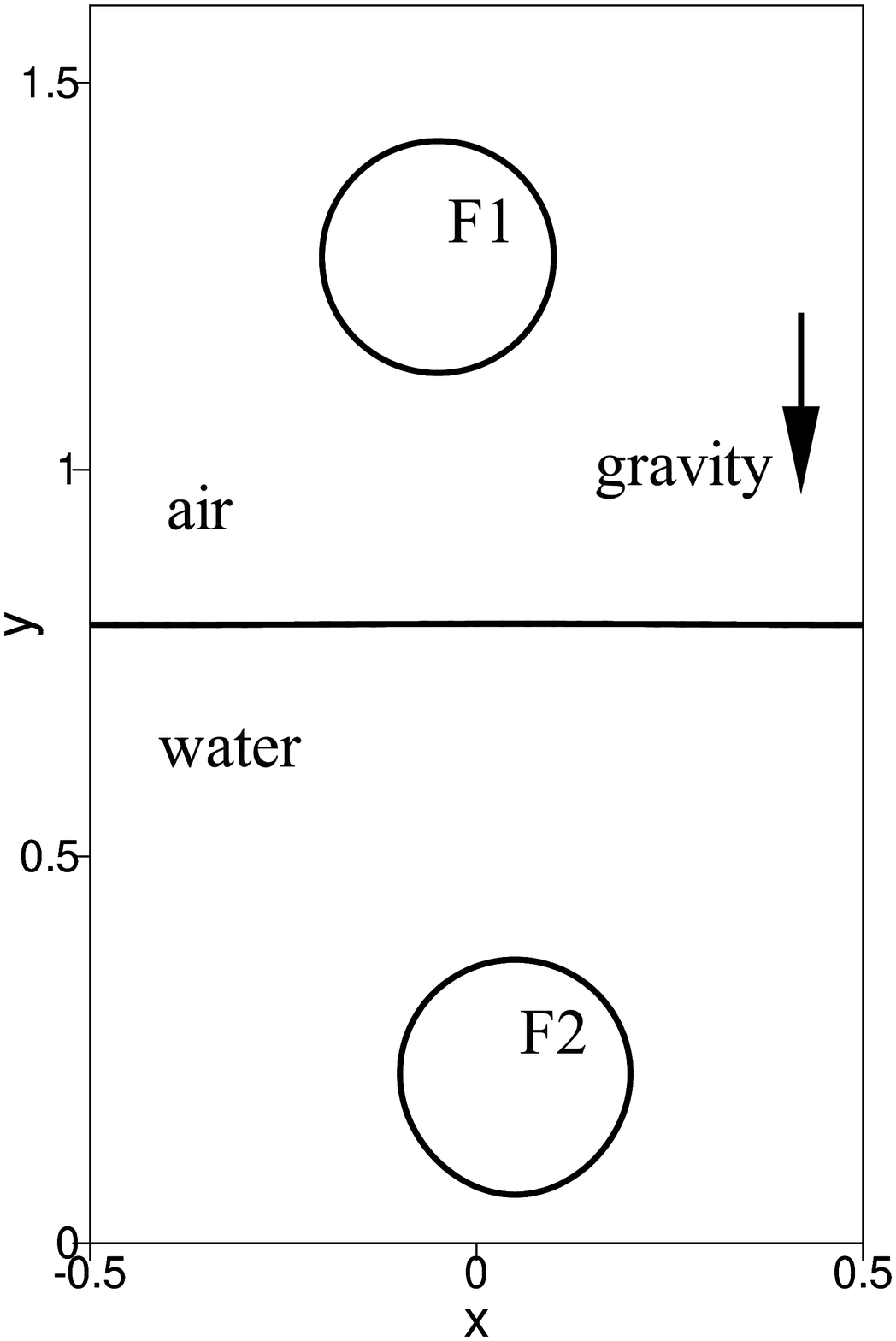}(a)
    \includegraphics[width=1.5in]{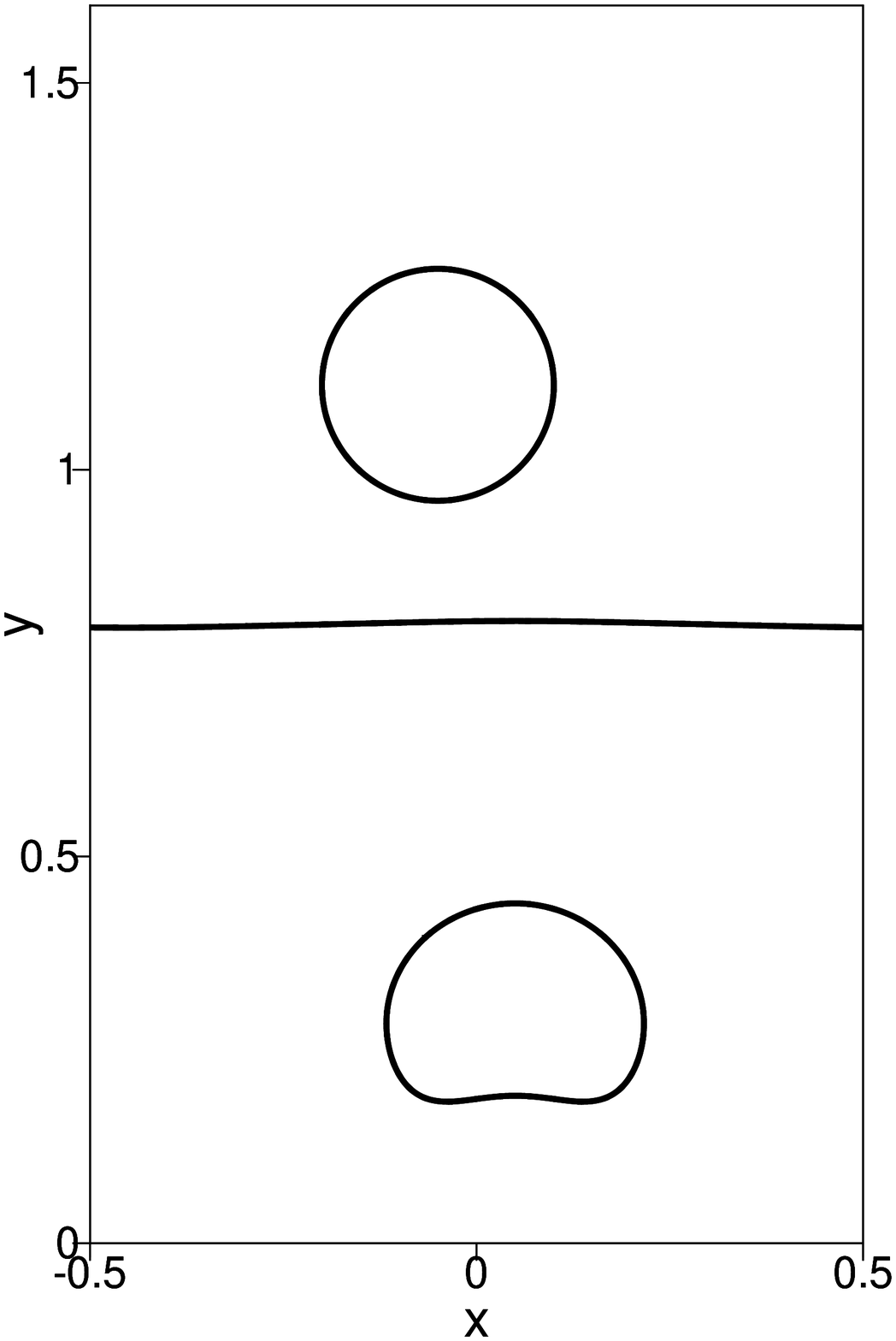}(b)
    \includegraphics[width=1.5in]{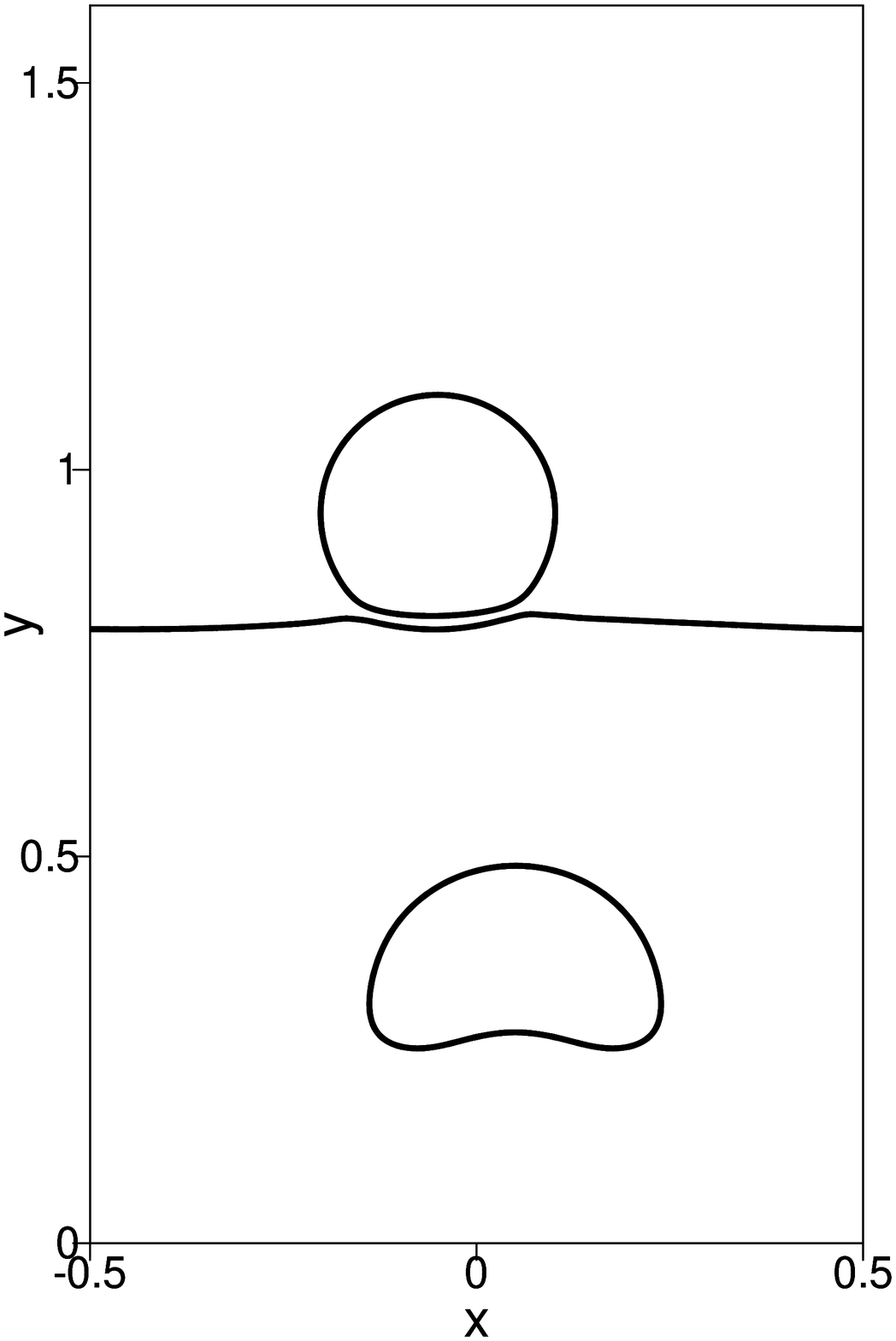}(c)
    \includegraphics[width=1.5in]{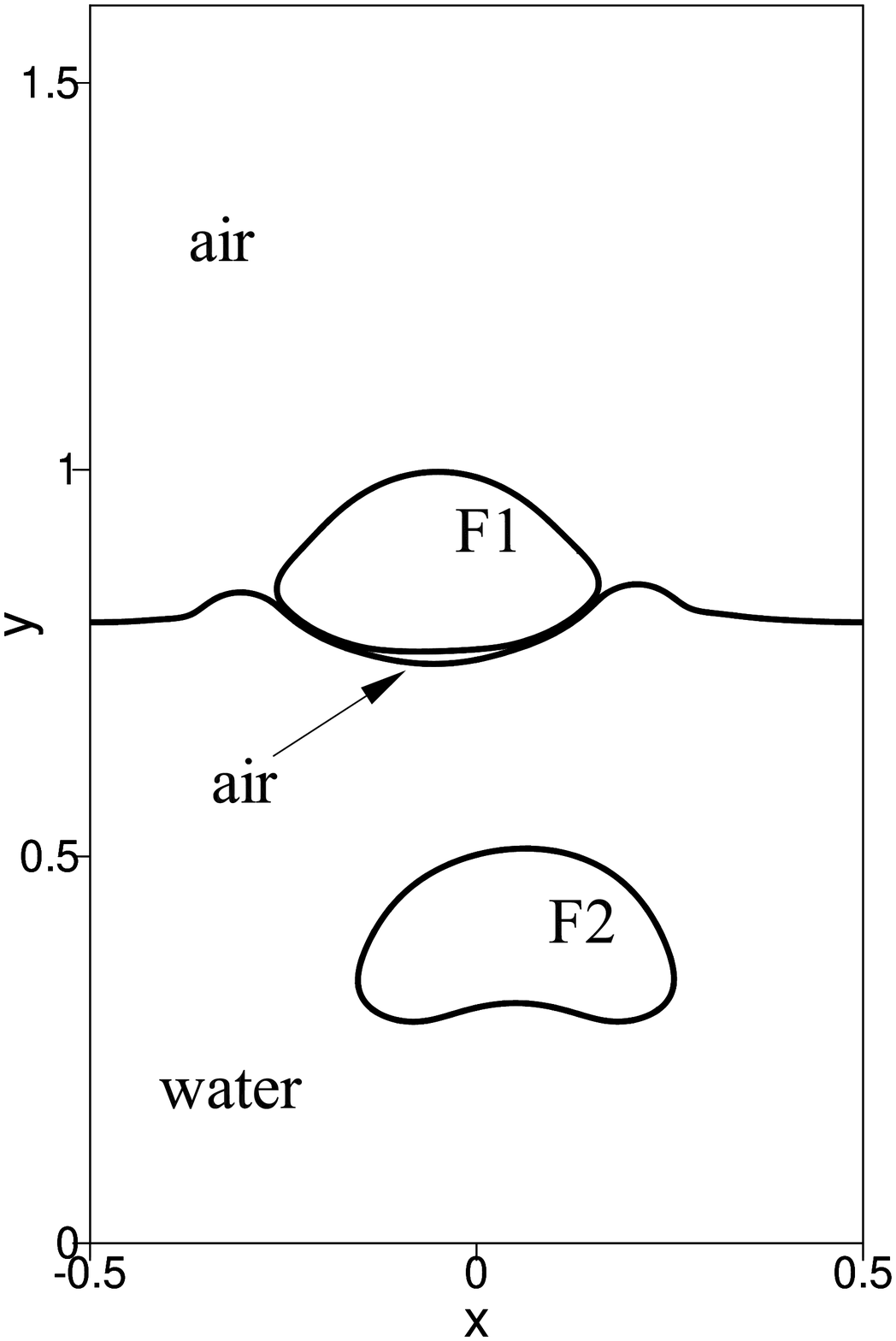}(d)
  }
  \centerline{
    \includegraphics[width=1.5in]{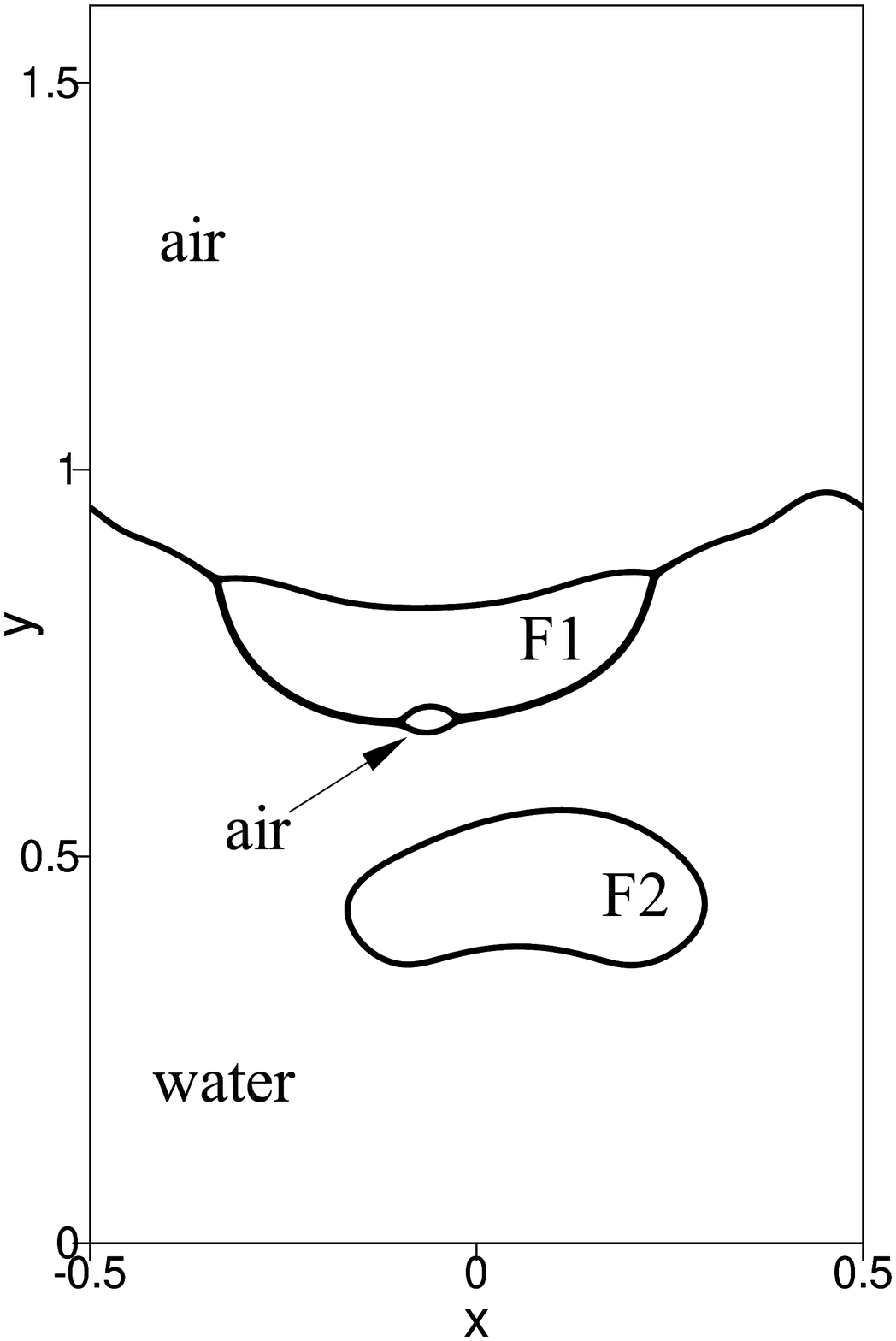}(e)
    \includegraphics[width=1.5in]{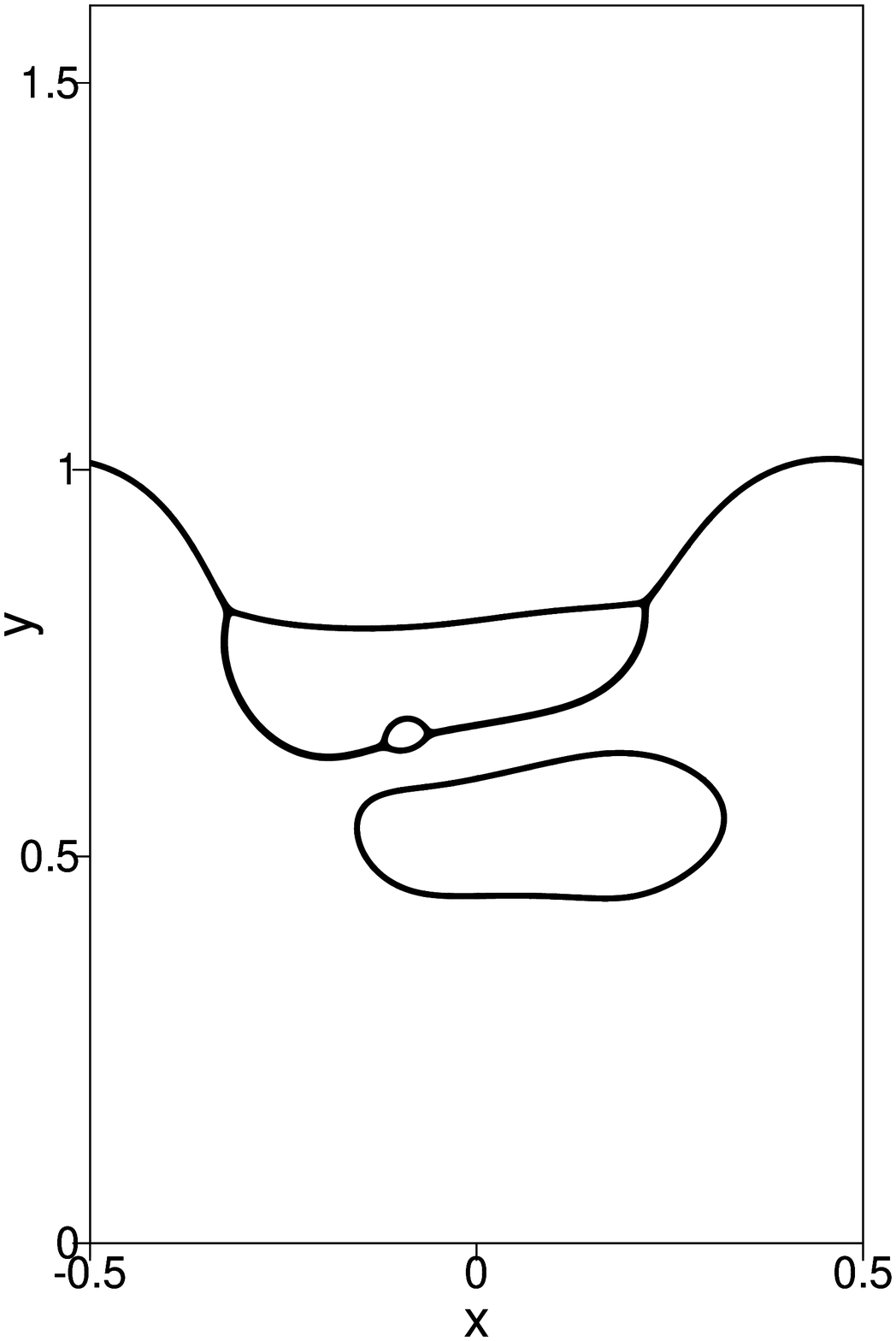}(f)
    \includegraphics[width=1.5in]{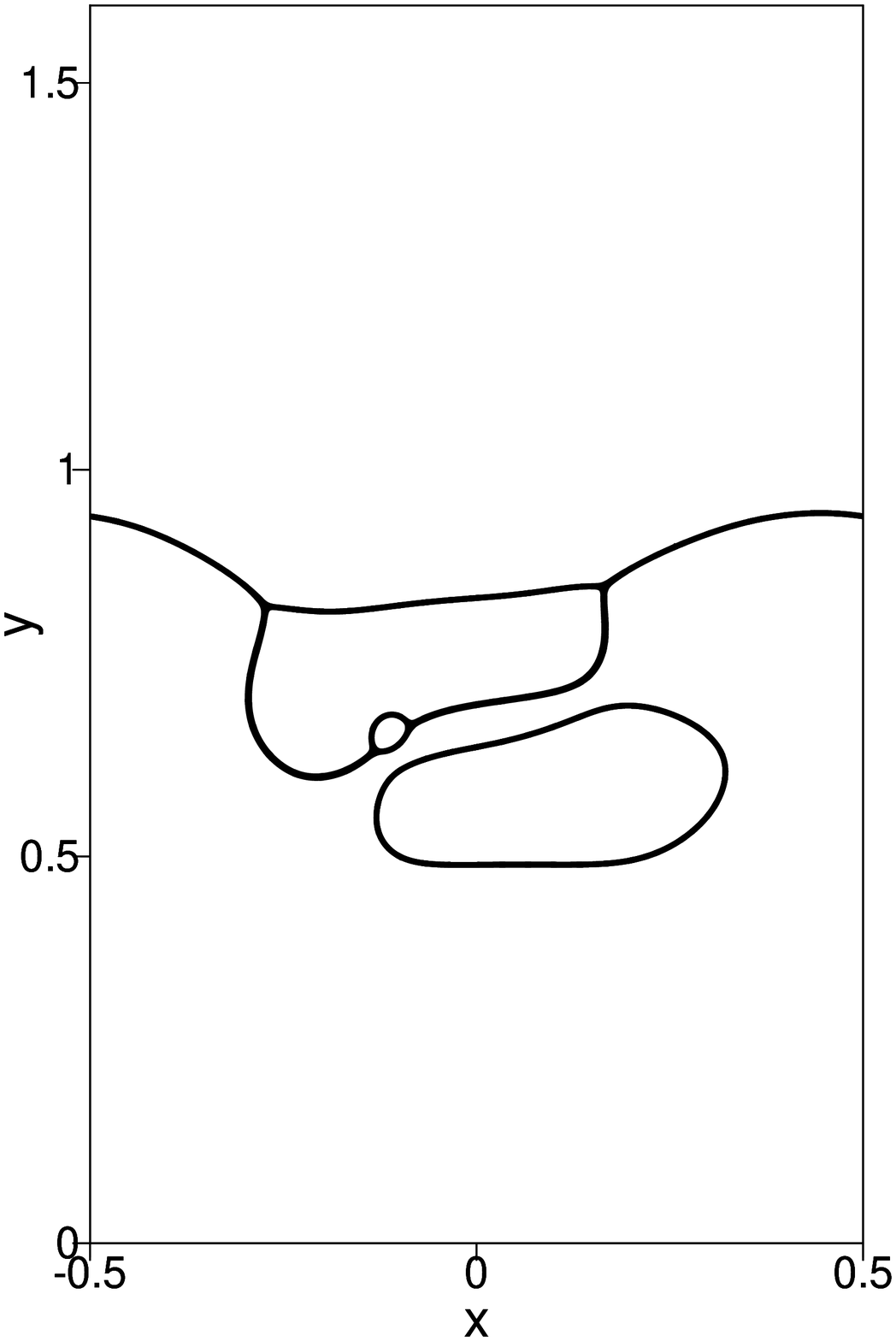}(g)
    \includegraphics[width=1.5in]{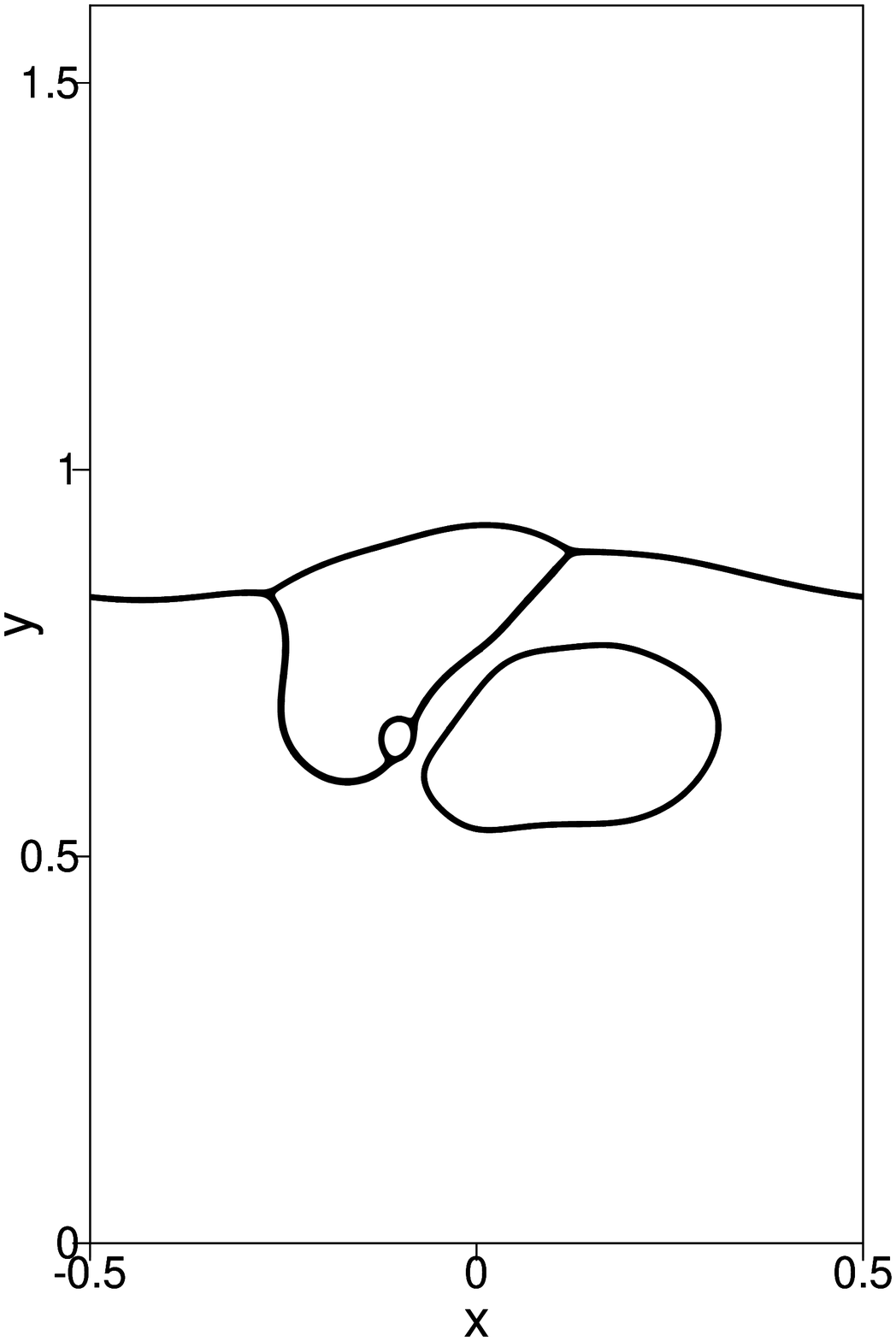}(h)
  }
  \centerline{
    \includegraphics[width=1.5in]{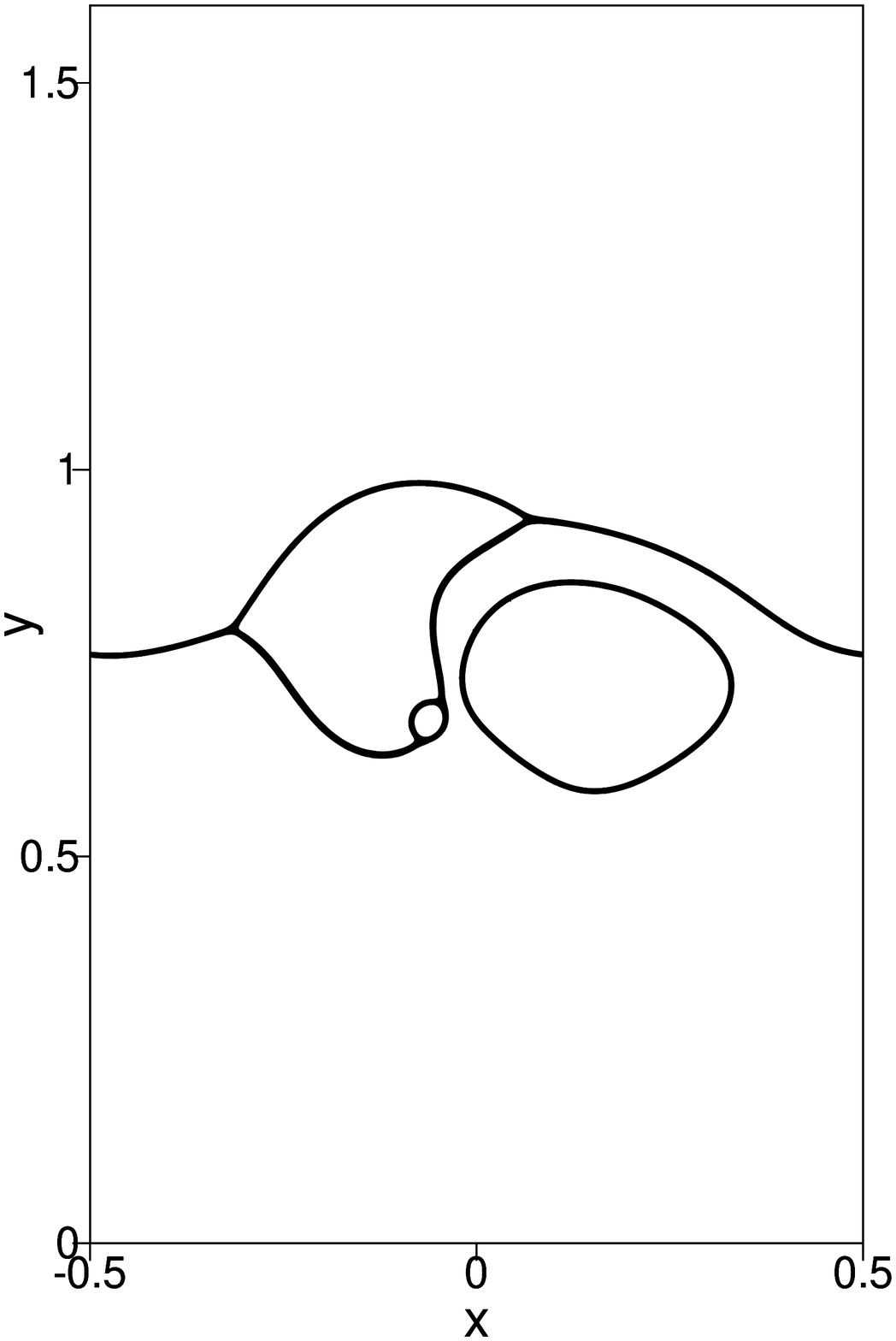}(i)
    \includegraphics[width=1.5in]{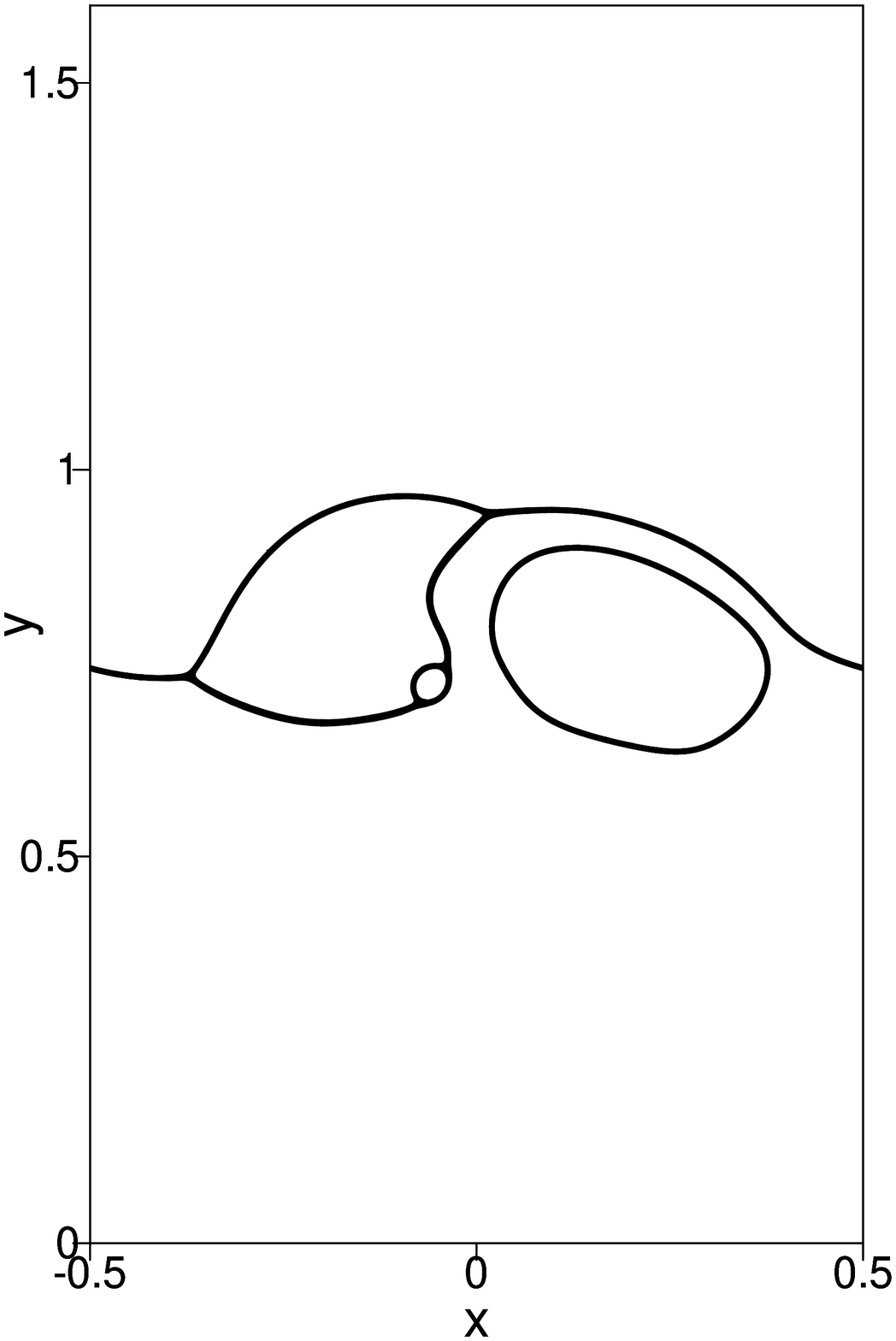}(j)
    \includegraphics[width=1.5in]{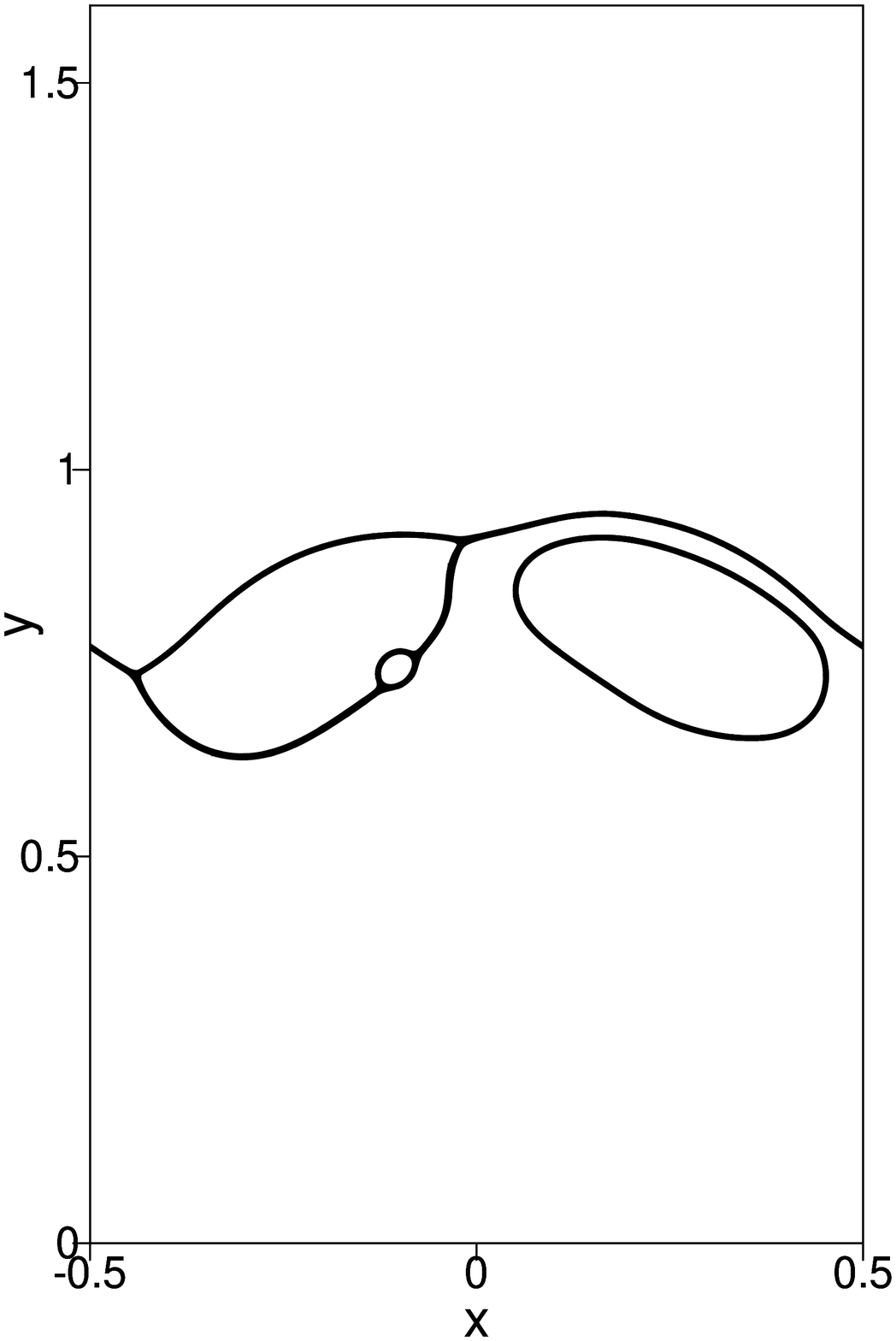}(k)
    \includegraphics[width=1.5in]{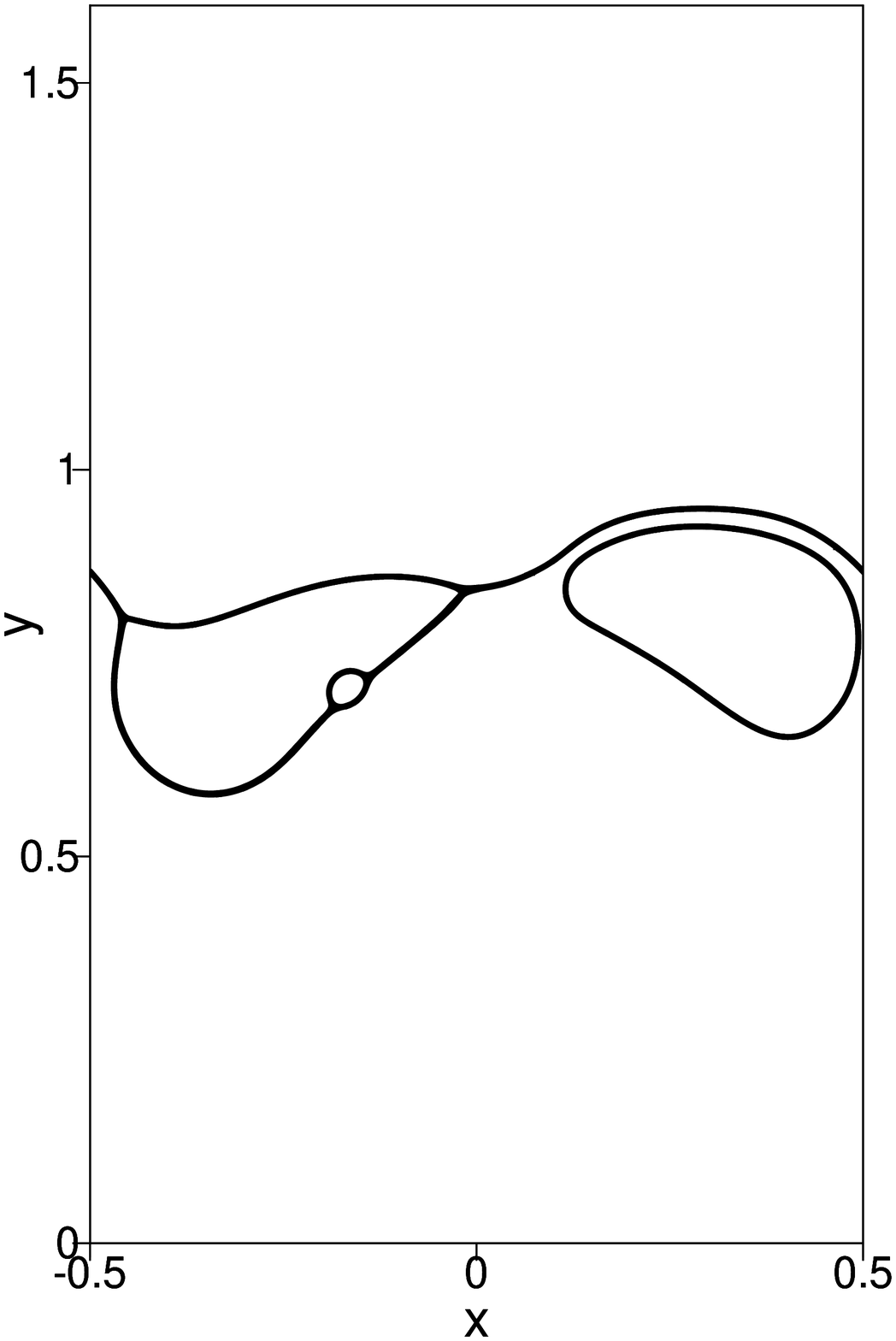}(l)
  }
  \caption{
    Temporal sequence of snapshots of fluid interfaces, visualized by
    the volume-fraction contours $c_i=1/2$ ($i=1,\dots,4$),
    showing two fluid drops impacting water surface
    ($4$ fluid components):
    (a) $t=0.072$,
    (b) $t=0.198$,
    (c) $t=0.27$,
    (d) $t=0.31$,
    (e) $t=0.41$,
    (f) $t=0.538$,
    (g) $t=0.614$,
    (h) $t=0.702$,
    (i) $t=0.79$,
    (j) $t=0.862$,
    (k) $t=0.934$,
    (l) $t=1.01$.
  }
  \label{fig:dyna_4p_phase}
\end{figure}

In this subsection we study a dynamic problem involving
four fluid components as another test for the method
developed in this work.
The problem setting is illustrated by Figure \ref{fig:dyna_4p_phase}(a).
We consider a rectangular domain
$-L/2\leqslant x\leqslant L/2$ and $0\leqslant y\leqslant 1.6L$,
where $L=2cm$,
and four immiscible incompressible fluids contained
in this domain: air, water, liquid ``F1'', and liquid ``F2''.
F1 and F2 are both heavier than air and lighter than water. 
The domain is bounded by two solid walls of neutral wettability on the top
and bottom sides, and is periodic in the horizontal
direction. The gravitational acceleration $\mathbf{g}_r$
is in the $-y$ direction.
The top half of the domain is initially filled with air,
and the bottom half is filled with water.
A drop of the liquid F1, initially circular with a diameter
$0.3L$, is suspended in the air and held at rest.
A drop of the liquid F2, initially circular with a
diameter of $0.3L$ also, is trapped in the water and held
at rest. The centers of the two drops are located at
\begin{equation*}
  \left\{
  \begin{split}
    &
    \mathbf{x}_{F1} = (x_{F1}, y_{F1}) = (-0.05L, 1.3L) \quad \text{(F1 drop)} \\
    &
    \mathbf{x}_{F2} = (x_{F2}, y_{F2}) = (0.05L, 0.2L) \quad \text{(F2 drop)}
  \end{split}
  \right.
\end{equation*}
At $t=0$, the two liquid drops are released,
and they fall through the air and rise through the water,
and impact the water surface.
Our goal is to study this dynamic process.

\begin{table}
  \begin{center}
    \begin{tabular}{lllll}
      \hline
      Density [$kg/m^3$]: & air -- $1.2041$ & water -- $998.207$
      & F1 -- $870$ & F2 -- $50$ \\
      Dynamic viscosity [$kg/(m\cdot s)$]: & air -- $1.78E-5$
      & water -- $1.002E-3$ & F1 -- $0.0915$
      & F2 -- $0.01$ \\
      Surface tension [$kg/s^2$]: & air/water -- $0.0728$ &
      air/F1 -- $0.055$ & air/F2 -- $0.06$ \\
      & water/F1 -- $0.044$ & water/F2 -- $0.045$
      & F1/F2 -- $0.048$ \\
      Gravity [$m/s^2$]: & 9.8
      \\
      \hline
    \end{tabular}
  \end{center}
  \caption{Physical parameter values for the air/water/F1/F2 four-phase problem.}
  \label{tab:4p_param}
\end{table}


The values for the physical properties of the four fluid components
employed in this problem
are listed in Table \ref{tab:4p_param}, including the densities,
dynamic viscosities, pair-wise surface tensions and
the gravity.
We assign the air, water, F1 and F2 as the first, the second, the third
and the fourth fluid in the simulations, respectively.
We choose the air density as the density scale $\varrho_d$,
$L$ as the length scale, and $U_0=\sqrt{g_{r0}L}$ as
the velocity scale, where $g_{r0}=1m/s^2$.
All the variables are then normalized according to
Table \ref{tab:normalization}.
The source terms in the phase field equations
\eqref{equ:CH_trans_1} are set to $d_i = 0$ ($1\leqslant i\leqslant 3$).


We discretize the domain using a spectral element mesh of
$1440$ quadrilateral elements of equal sizes, with $30$ elements
along the $x$ direction and $48$ elements along the $y$ direction.
An element order $9$ is used in the simulations for all elements.
On the top/bottom walls the no slip condition, equation
\eqref{equ:bc_vel} with $\mathbf{w}=0$, is imposed on the velocity,
and the boundary conditions \eqref{equ:bc_ci_1_trans}
and \eqref{equ:bc_ci_2_trans} with $d_{ai}=0$ and $d_{bi}=0$
are imposed on the volume fractions $c_i$ ($1\leqslant i\leqslant 3$).
Periodic conditions are employed for all flow variables
at $x=\pm L/2$.
%
We set the initial velocity to zero, and the initial
volume fractions to the following functions:
\begin{equation}
  \left\{
  \begin{split}
    &
    c_1 = \frac{1}{2}\left(1 + \tanh\frac{y-y_w}{\sqrt{2}\eta}  \right)
    \left[1 - \frac{1}{2}\left(1 - \tanh\frac{|\mathbf{x}-\mathbf{x}_{F1}|-R_0}{\sqrt{2}\eta}  \right)  \right] \\
    &
    c_2 = \frac{1}{2}\left(1 - \tanh\frac{y-y_w}{\sqrt{2}\eta}  \right)
    \left[1 - \frac{1}{2}\left(1 - \tanh\frac{|\mathbf{x}-\mathbf{x}_{F2}|-R_0}{\sqrt{2}\eta}  \right)  \right] \\
    &
    c_3 = \frac{1}{2}\left(1 - \tanh\frac{|\mathbf{x}-\mathbf{x}_{F1}|-R_0}{\sqrt{2}\eta}  \right) \\
    &
    c_4 = 1-c_1 - c_2 - c_3
  \end{split}
  \right.
\end{equation}
where $y_w = 0.8L$ is the y coordinate of the initial water surface,
and $R_0=0.15L$ is the initial radius of the F1 and F2 drops.
Table \ref{tab:4p_simul_param} lists the values
of the simulation parameters for this problem.

\begin{table}
  \begin{center}
    \begin{tabular}{ll}
      \hline
      Parameters & Values \\
      $\lambda_{ij}$ & computed by equation \eqref{equ:free_energy_param} \\
      $\eta/L$ & $0.005$ \\
      $m_0\varrho_d U_0/L$ & $10^{-8}$ \\
      $U_0\Delta t/L$ & $1.0 \times 10^{-6}$ \\
      $\rho_0$ & $\min(\tilde{\rho}_1,\tilde{\rho}_2, \tilde{\rho}_3, \tilde{\rho}_4)$ \\
      $\nu_0$ & $10\max\left(\frac{\tilde{\mu}_1}{\tilde{\rho}_1},
      \frac{\tilde{\mu}_2}{\tilde{\rho}_2}, \frac{\tilde{\mu}_3}{\tilde{\rho}_3},
      \frac{\tilde{\mu}_4}{\tilde{\rho}_4} \right)$ \\
      $\mathcal{K}_0$ & computed by equation \eqref{equ:def_K0} \\
      $S$ & $\sqrt{\frac{4\gamma_0}{\mathcal{K}_0\Delta t}}$ \\
      $\alpha$ & computed by equation \eqref{equ:def_alpha} \\
      $J$ (temporal order) & $2$ \\
      Number of elements & $1440$ \\
      Element order & $9$      
      \\
      \hline
    \end{tabular}
  \end{center}
  \caption{Simulation parameter values for the
    air/water/F1/F2 four-phase problem.}
  \label{tab:4p_simul_param}
\end{table}


Let us look into the dynamics of this four-phase system.
Figure \ref{fig:dyna_4p_phase} shows a temporal sequence
of snapshots of the fluid interfaces of the system.
The interfaces are visualized by the contour lines
of the volume fractions $c_i=\frac{1}{2}$ ($i=1,\dots,4$).
From Figures \ref{fig:dyna_4p_phase}(a)--(c), we observe that
as the system is released the F1 drop falls rapidly through
the air, with little deformation in this process.
But as the F1 drop approaches the water surface just before the impact
(Figure \ref{fig:dyna_4p_phase}(c)), 
a depression on the water surface and
a deformation of the lower side of the F1 drop
can be clearly noticed.
At the same time, the F2 drop rises through
the water at a much slower speed. The deformation
of the F2 drop is substantial, and its shape
resembles a circular ``cap'' (Figures \ref{fig:dyna_4p_phase}(b)--(c)).
Subsequently, the falling F1 drop impacts the water,
causing a ripple on the water surface (Figure \ref{fig:dyna_4p_phase}(d)).
It can be observed that the F1 drop has trapped a thin 
cushion of air between its underside and the water
surface (Figure \ref{fig:dyna_4p_phase}(d)).
The impact causes the F1 drop to deform severely,
and it forms a pool of the F1 liquid floating
on the surface of water (Figure \ref{fig:dyna_4p_phase}(e)--(h)).
The air trapped between the F1 drop and the water
surface forms a small air bubble at the underside of
the pool of F1 fluid (Figure \ref{fig:dyna_4p_phase}(e)--(l)).
As the F2 drop rises further and approaches
the pool of F1 liquid that now covers a portion of
the water surface, it experiences significant
deformation and its shape has become
highly irregular (Figure \ref{fig:dyna_4p_phase}(d)--(e)).
Subsequently, it can be observed
from Figures \ref{fig:dyna_4p_phase}(f)--(j) that
the interaction between the F1 fluid and the F2 drop
appears to cause both fluids to move sideways away from each
other while the F2 drop rises further.
The F2 drop appears to glide past the pool of F1 fluid
(Figures \ref{fig:dyna_4p_phase}(g)--(i)),
and rises in an oblique direction toward the open surface
of water (Figures \ref{fig:dyna_4p_phase}(j)--(l)).
It can be observed that
the pool of F1 fluid experiences significant
deformations in this process (Figures \ref{fig:dyna_4p_phase}(g)--(l)).

%% file: Summary.tex
\section{Concluding Remarks}
\label{sec:summary}


This paper focuses on the formulation and
simulation of multiphase flows consisting of $N$ ($N\geqslant 2$)
immiscible incompressible fluids with different densities,
dynamic viscosities and pair-wise surface tensions.
In particular, we have explored how to formulate
the N-phase system in a reduction-consistent and
thermodynamically consistent manner.
Thermodynamic consistency is achieved by adopting 
a thermodynamically consistent phase field model
for the N-phase system derived based on
the mass/momentum conservations and the second law of
thermodynamics.
Reduction consistency is achieved by the construction of
the mobility matrix and the free energy density function
with appropriate individual reduction properties,
which collectively guarantee the full reduction consistency
of the N-phase governing equations.

We have made two contributions in this work.
The main contribution lies in the method (Theorem \ref{thm:thm_1})
provided herein,
which allows for the systematic construction of reduction-consistent
 N-phase formulations. 
This method is quite general,
and it suggests  many possible forms for reduction-consistent
and thermodynamically consistent N-phase formulations.
Based on this method, we have developed and
presented a specific reduction-consistent
and thermodynamically consistent formulation for
incompressible N-phase ($N\geqslant 2$) flows, which is the second
contribution of this work. 
This specific N-phase formulation,
together with the boundary conditions presented in
Section \ref{sec:alg}, fully satisfies the
reduction consistency conditions ($\mathscr{C}1$),
($\mathscr{C}2$) and ($\mathscr{C}3$).


We can compare the approach to reduction consistency
in this work with that of \cite{Dong2017},
and also perhaps with that of \cite{BoyerM2014} for
a Cahn-Hilliard model (without hydrodynamic
interactions).
The approach of this paper allows us to
treat the reduction properties of the mobility matrix and the free energy
density function separately and individually. 
Therefore, it is considerably easier in the construction
to satisfy these reduction properties.
In contrast, in \cite{Dong2017} (and also \cite{BoyerM2014})
the reduction properties for the free energy density
function are entangled with that of the mobility matrix.
This poses an enormous challenge in the construction
of the free energy density function to satisfy
these reduction properties for full reduction consistency.
Consequently, only a partial reduction consistency
(between $N$ phases and two phases) has been achieved
in \cite{Dong2017,BoyerM2014} for an arbitrary
set of given pair-wise surface tensions.

For the specific reduction-consistent
and thermodynamically consistent N-phase formulation
presented herein, we have also presented
an efficient numerical algorithm for solving
the N-phase governing equations.
This is a semi-implicit splitting type scheme,
which de-couples the solution of different flow variables.
Despite the variable mobility matrix and
the variable mixture density/viscosity,
our algorithm involves only the solution of
linear algebraic systems with {\em constant and
  time-independent} coefficient matrices within each
time step for all variables.
In particular, when solving the system of coupled fourth-order phase field
equations, our algorithm only requires the solution of
$2(N-1)$ individual Helmholtz type equations in a
de-coupled fashion.

Extensive numerical experiments have been presented
for flow problems
involving multiple fluid components, large density ratios
and large viscosity ratios
to test the performance of the method presented herein,
and we have compared simulation results with
exact physical solutions or physical
theories from the literature.
In particular, we have conducted simulations
of three- or four-phase systems
in which one or two fluid components
are absent, and demonstrated that the simulation results
indeed match the results obtained on the smaller systems.
Comparisons with the Prosperetti's theory and 
the Langmuir-de Gennes theory show that
our method produces physically accurate results.


Reduction consistency and thermodynamic consistency are
important physical consistency requirements, and
can have a profound impact on the simulation results.
The lack of such consistency properties may lead
to un-physical results or result in gross errors.
For example, in \cite{Dong2017} we have documented
the detrimental effect of the lack of reduction
consistency in the formulation or boundary
condition on the simulation of wall-bounded
N-phase flows and N-phase contact angles.
It is observed that, due to the lack of reduction
consistency, a ``third'' fluid 
can be artificially generated at the base of a liquid drop
near the wall, causing a gross deviation of
the contact angle from the expected value \cite{Dong2017}.


%% file: Appendix.tex
\section*{Appendix A. Development of Thermodynamically
  Consistent N-phase Model}
\label{sec:develop}

%

In this Appendix we summarize the development of a thermodynamically consistent
N-phase model based on the mass conservation,
momentum conservation, and the second law of thermodynamics.
The development follows a similar procedure to that for
the model of \cite{Dong2014}, but we arrive at an N-phase model
that is different from those of \cite{Dong2014,Dong2017},
due to difference in the representation of mass balances and
the specification of constitutive relations to
satisfy the second law of thermodynamics.
The basic steps in the following development mirror those
of \cite{Dong2014}.
We refer the reader to the appendix of \cite{Dong2014} for 
the derivation of that model therein.

\paragraph{Definitions and Settings}

Consider an iso-thermal mixture of $N$ ($N\geqslant 2$) immiscible
incompressible fluids in some flow domain in two
or three dimensions.
Let $\tilde{\rho}_i$ ($1\leqslant i\leqslant N$) and
$\tilde{\mu}_i$ ($1\leqslant i\leqslant N$) respectively
denote the constant densities and constant dynamic
viscosities of these $N$ fluids.
Consider an arbitrary control volume $V$ taken from the mixture.
Let $M_i$ ($1\leqslant i\leqslant N$)
denote the mass of fluid $i$ contained within $V$,
$\rho_i$ ($1\leqslant i\leqslant N$) denote
the average density of fluid $i$ within $V$,
and $M$ and $\rho$ respectively denote the total mass and the average density
of the mixture within $V$.
Then the following relations hold,
\begin{equation}
  \rho_i = \frac{M_i}{V}, \quad
  \rho = \frac{M}{V} = \frac{M_1+\dots +M_N}{V} = \rho_1 + \rho_2 + \dots + \rho_N.
  \label{equ:rho_relation}
\end{equation}

We make the following assumption:
\begin{enumerate}[($\mathcal{A}$1):]

\item
  There is no volume addition or volume loss if any of
  these $N$ fluids are mixed, in other words, the volume
  occupied by a given amount of mass of any single fluid
  $i$ does not change after mixing.

\end{enumerate}
Let $V_i$ ($1\leqslant i\leqslant N$) denote
the volume occupied by the pure fluid $i$ (before mixing)
of mass $M_i$. It follows from the above assumption that
\begin{equation}
V = V_1 + V_2 + \dots + V_N.
\end{equation}
Let $c_i$ ($1\leqslant i\leqslant N$) denote the  volume
fraction of fluid $i$ within $V$. Then
\begin{equation}
  c_i = \frac{V_i}{V} = \frac{M_i/\tilde{\rho}_i}{M_i/\rho_i}
  = \frac{\rho_i}{\tilde{\rho}_i},
  \quad
  c_1 + c_2 + \dots + c_N = 1.
  \label{equ:volfrac_relation}
\end{equation}
Let the control volume $V\rightarrow 0$, and the
average quantities defined above become
field variables $\rho_i(\mathbf{x},t)$,
$c_i(\mathbf{x},t)$, $\rho(\mathbf{x},t)$.
These definitions follow those given in \cite{Dong2014}.

\paragraph{Mass Balance}

Let $\hat{\mathbf{J}}_i$ ($1\leqslant i\leqslant N$)
denote the mass flux vector of fluid $i$ in the mixture.
Then the mass balance of fluid $i$ leads to
\begin{equation}
  \frac{\partial\rho_i}{\partial t} + \nabla\cdot\hat{\mathbf{J}}_i = 0,
  \quad
  1\leqslant i\leqslant N.
  \label{equ:mass_balance}
\end{equation}
We define the individual velocity of fluid $i$, $\mathbf{u}_i$ ($1\leqslant i\leqslant N$),
by
$
  \hat{\mathbf{J}}_i = \rho_i \mathbf{u}_i.
$
Following \cite{Dong2014}, we define the mixture (or bulk)
velocity as the volume-averaged velocities of the individual
fluids,
\begin{equation}
  \mathbf{u} = \sum_{i=1}^N c_i\mathbf{u}_i
  = \sum_{i=1}^N \frac{\rho_i}{\tilde{\rho}_i} \mathbf{u}_i
  = \sum_{i=1}^N \frac{\hat{\mathbf{J}}_i}{\tilde{\rho}_i},
  \label{equ:mixture_vel}
\end{equation}
where we have used \eqref{equ:volfrac_relation}.
This mixture velocity is divergence free \cite{Dong2014},
\begin{equation}
  \nabla\cdot\mathbf{u} = \sum_{i=1}^N\frac{1}{\tilde{\rho}_i}\nabla\cdot\hat{\mathbf{J}}_i
  = -\frac{\partial}{\partial t}\left(\sum_{i=1}^N \frac{\rho_i}{\tilde{\rho}_i}\right)
  = -\frac{\partial}{\partial t}\left(\sum_{i=1}^N c_i\right)
  = -\frac{\partial }{\partial t} 1 = 0.
  \label{equ:continuity}
\end{equation}

Introduce the differential flux, 
$
\mathbf{J}_i = \hat{\mathbf{J}}_i - \rho_i\mathbf{u},
$
which is the flux of fluid $i$ relative to 
the bulk motion characterized by $\mathbf{u}$.  
$\mathbf{J}_i$ satisfies the relation
\begin{equation}
  \sum_{i=1}^N \frac{1}{\tilde{\rho}_i}\mathbf{J}_i
  = \sum_{i=1}^N \frac{\hat{J}_i}{\tilde{\rho}_i} - \sum_{i=1}^N \frac{\rho_i}{\tilde{\rho}_i}\mathbf{u}
  = \mathbf{u} - \left(\sum_{i=1}^N c_i  \right)\mathbf{u} = 0.
  \label{equ:diff_flux_relation}
\end{equation}
The mass balance equation \eqref{equ:mass_balance}
is then transformed into
\begin{equation}
  \frac{\partial\rho_i}{\partial t} + \mathbf{u}\cdot\nabla\rho_i
  = - \nabla\cdot\mathbf{J}_i,
  \quad 1\leqslant i\leqslant N
  \label{equ:mass_balance_1}
\end{equation}
where we have used \eqref{equ:continuity}.
Sum up the $N$ equations in \eqref{equ:mass_balance_1}
and we have
\begin{equation}
  \frac{\partial\rho}{\partial t} + \mathbf{u}\cdot\nabla\rho
  = -\nabla\cdot\tilde{\mathbf{J}}, \quad
  \text{where} \
  \tilde{\mathbf{J}} = \mathbf{J}_1 + \mathbf{J}_2 + \dots + \mathbf{J}_N.
  \label{equ:mixture_mass_balance}
\end{equation}
Equation \eqref{equ:mass_balance_1} can be written
in terms of the volume fractions as
\begin{equation}
  \frac{\partial c_i}{\partial t} + \mathbf{u}\cdot\nabla c_i
  = - \nabla\cdot\left(\frac{\mathbf{J}_i}{\tilde{\rho}_i}  \right),
  \quad 1\leqslant i\leqslant N.
  \label{equ:mass_balance_2}
\end{equation}

The mass balance of the $N$ individual fluids
 in the system
is represented by the equations \eqref{equ:mass_balance_1}
or \eqref{equ:mass_balance_2}.
The forms of the differential fluxes $\mathbf{J}_i$ ($1\leqslant i\leqslant N$)
in these equations
are to be specified by considering the second law
of thermodynamics, and must satisfy
the constraint \eqref{equ:diff_flux_relation}.

\paragraph{Momentum Balance}

Following \cite{AbelsGG2012,Dong2014}, we make the following
assumption:
\begin{enumerate}[($\mathcal{A}$1):]
  \addtocounter{enumi}{1}
  
\item
  The inertia and kinetic energy of the relative motion of
  any fluid with respect to the bulk motion is negligible, and 
  the mixture can be considered as a single fluid, which satisfies
  the linear-momentum conservation with respect to the
  volume-averaged velocity $\mathbf{u}$.
  
\end{enumerate}
Consider an arbitrary control volume $\Omega(t)$,
which moves with the bulk mixture velocity $\mathbf{u}$.
We assume that there is no external body force.
Then the momentum conservation on this control volume
is represented by
\begin{equation}
  \frac{d}{dt}\int_{\Omega(t)} \rho \mathbf{u}
  = \int_{\partial\Omega(t)}\mathbf{n}\cdot\mathbf{T}
  -\int_{\partial\Omega(t)} \sum_{i=1}^N (\mathbf{n}\cdot\mathbf{J}_i) \mathbf{u}
  \label{equ:momentum_balance}
\end{equation}
where $\partial\Omega(t)$ denotes the boundary of $\Omega(t)$,
$\mathbf{n}$ is the outward-pointing unit vector
normal to the boundary,
$\mathbf{T}$ denotes a stress tensor whose form is to be
specified by constitutive relations,
and the last term on the right hand side (RHS)
denotes the momentum transport due to
the relative motion of the fluids with respect to
the bulk motion.
Since the control volume is arbitrary,
by using the Reynolds transport theorem and the divergence
theorem we can transform this equation into
\begin{equation}
  \frac{\partial}{\partial t}(\rho\mathbf{u})
  + \nabla\cdot(\rho\mathbf{uu})
  + \nabla\cdot(\tilde{\mathbf{J}}\mathbf{u})
  =\nabla\cdot\mathbf{T}
  \label{equ:momentum_balance_1}
\end{equation}
where we have also used the
$\tilde{\mathbf{J}}$ expression in \eqref{equ:mixture_mass_balance}.
This equation can be further reduced to
\begin{equation}
  \rho\left(\frac{\partial\mathbf{u}}{\partial t} +
  \mathbf{u}\cdot\nabla\mathbf{u} \right)
  + \tilde{\mathbf{J}}\cdot\nabla\mathbf{u}
  = \nabla\cdot\mathbf{T}
  \label{equ:momentum_balance_2}
\end{equation}
where equation \eqref{equ:mixture_mass_balance}
has been used.

We assume that the stress tensor $\mathbf{T}$ is symmetric,
and re-write it as
\begin{equation}
  \mathbf{T} = \frac{1}{3}(\text{tr}\mathbf{T})\mathbf{I}
  + \mathbf{S} = -p\mathbf{I} + \mathbf{S}
  \label{equ:stress_tensor}
\end{equation}
where $\mathbf{I}$ is the identity tensor,
$\mathbf{S}$ is a trace-free symmetric tensor,
and $p=\frac{1}{3}\text{tr}\mathbf{T}$ will be
referred to as the pressure.
The momentum equation \eqref{equ:momentum_balance_2}
is then transformed into
\begin{equation}
  \rho\left(\frac{\partial\mathbf{u}}{\partial t} +
  \mathbf{u}\cdot\nabla\mathbf{u} \right)
  + \tilde{\mathbf{J}}\cdot\nabla\mathbf{u}
  = -\nabla p + \nabla\cdot\mathbf{S}.
  \label{equ:moment_balance}
\end{equation}
The form for the stress tensor $\mathbf{S}$
will be specified by considering
the second law of thermodynamics.

\paragraph{Constitutive Relations and Second Law of Thermodynamics}

We now consider how to specify the constitutive
relations for the tensor $\mathbf{S}$ and
the differential fluxes $\mathbf{J}_i$ ($1\leqslant i\leqslant N$)
based on the second law of thermodynamics.

In the spirit of the phase field approach
we introduce a free energy density function
$W(\vec{c},\nabla\vec{c})$,
where $\vec{c} = (c_1, c_2, \dots, c_N)$,
to account for the effect of the interfacial
energy (surface tensions) among the
$N$ fluids. The total energy density function
of the system is
$
e(\mathbf{u}, \vec{c}, \nabla\vec{c})
= \frac{1}{2}\rho |\mathbf{u}|^2
+ W(\vec{c},\nabla\vec{c}).
$

Consider an arbitrary control volume $\Omega(t)$ that
moves with the bulk velocity $\mathbf{u}$.
For an isothermal system,
the second law of
thermodynamics is represented by
the following inequality~\cite{GurtinFA2010},
\begin{equation}
  \frac{d}{dt}\int_{\Omega(t)}e(\mathbf{u},\vec{c},\nabla\vec{c})
  \leqslant P_c
  \label{equ:2nd_law}
\end{equation}
where $P_c$ denotes the total conventional
power (i.e.~excluding heat transfer)
expended on $\Omega(t)$.

The conventional powers expended on $\Omega(t)$
consist of several components:
\begin{itemize}

\item
  Work due to the stress tensor,
  $
  \int_{\partial\Omega(t)}\mathbf{n}\cdot\mathbf{T}\cdot\mathbf{u}.
  $

\item
  Kinetic energy transport due to the relative motion
  of the fluids with respect to the bulk motion,
  \begin{equation*}
    -\int_{\partial\Omega(t)}\sum_{i=1}^N (\mathbf{n}\cdot\mathbf{J}_i)
    \frac{1}{2}|\mathbf{u}|^2
    =-\int_{\partial\Omega(t)}(\mathbf{n}\cdot\tilde{\mathbf{J}})
    \frac{1}{2}|\mathbf{u}|^2.
  \end{equation*}
  Note that $-\mathbf{n}\cdot\mathbf{J}_i$ is the mass of
  fluid $i$ transported into $\Omega(t)$ due to the relative motion
  with respect to the bulk motion.

\item
  Free energy transport due to the relative motion of the
  fluids with respect to the bulk motion,
  \begin{equation*}
    -\sum_{i=1}^N \int_{\partial\Omega(t)} (\mathbf{n}\cdot\mathbf{J}_i)\mathcal{C}_i
  \end{equation*}
  where $\mathcal{C}_i$ ($1\leqslant i\leqslant N$) is the chemical potential
  of fluid $i$ (free energy per unit mass).

\item
  Work due to a surface microforce.
  Following \cite{Gurtin1996}, we assume the existence of
  a surface microforce $\bm{\xi}_i$ ($1\leqslant i\leqslant N$),
  whose power expended on the system is represented
  by (see \cite{Gurtin1996})
  \begin{equation*}
    \sum_{i=1}^N\int_{\partial\Omega(t)} \mathbf{n}\cdot\bm{\xi}_i
    \left(\frac{\partial c_i}{\partial t} + \mathbf{u}\cdot\nabla c_i  \right).
  \end{equation*}
  
\end{itemize}

By incorporating the above contributions, the inequality
\eqref{equ:2nd_law} becomes
\begin{equation}
  \begin{split}
  \frac{d}{dt}\int_{\Omega(t)} e(\mathbf{u},\vec{c},\nabla\vec{c})
  -\int_{\partial\Omega(t)}\mathbf{n}\cdot\mathbf{T}\cdot\mathbf{u}
  + &\int_{\partial\Omega(t)}(\mathbf{n}\cdot\tilde{\mathbf{J}})\frac{1}{2}|\mathbf{u}|^2
  \\
  &+ \sum_{i=1}^N \int_{\partial\Omega(t)}\mathbf{n}\cdot\mathbf{J}_i\mathcal{C}_i
  - \sum_{i=1}^N \int_{\partial\Omega(t)} (\mathbf{n}\cdot\bm{\xi}_i)\frac{Dc_i}{Dt}
  \leqslant 0
  \end{split}
  \label{equ:2nd_law_1}
\end{equation}
where
$
\frac{Dc_i}{Dt} = \frac{\partial c_i}{\partial t} + \mathbf{u}\cdot\nabla c_i
$
denotes the material derivative.
By invoking the Reynolds transport theorem
and the divergence theorem, and
noting that $\Omega(t)$ is arbitrary,
we transform the above inequality into
\begin{equation}
  -D_s \equiv \frac{\partial e}{\partial t}
  + \nabla\cdot(e\mathbf{u})
  -\nabla\cdot(\mathbf{T}\cdot\mathbf{u})
  + \nabla\cdot\left(\tilde{\mathbf{J}}\frac{1}{2}|\mathbf{u}|^2  \right)
  +\sum_{i=1}^N \nabla\cdot(\mathbf{J}_i\mathcal{C}_i)
  - \sum_{i=1}^N\nabla\cdot\left(\bm{\xi}_i\frac{Dc_i}{Dt}\right)
  \leqslant 0
  \label{equ:2nd_law_2}
\end{equation}
In light of equations \eqref{equ:continuity},
\eqref{equ:mixture_mass_balance}
and \eqref{equ:momentum_balance_2},
we can transform \eqref{equ:2nd_law_2} into
\begin{equation}
  \begin{split}
  -D_s=\frac{\partial W}{\partial t}
  + \mathbf{u}\cdot\nabla W
  - \mathbf{T}:\nabla\mathbf{u}
  + \sum_{i=1}^N(\nabla\cdot\mathbf{J}_i)\mathcal{C}_i
  +& \sum_{i=1}^N\mathbf{J}_i\cdot\nabla\mathcal{C}_i \\
  &- \sum_{i=1}^N(\nabla\cdot \bm{\xi}_i)\frac{Dc_i}{Dt}
  - \sum_{i=1}^N \bm{\xi}_i\cdot\nabla \frac{Dc_i}{Dt}
  \leqslant 0
  \end{split}
  \label{equ:2nd_law_3}
\end{equation}
where the symmetry assumption about $\mathbf{T}$
has been used.

In light of equation \eqref{equ:mass_balance_2} and the
relations
\begin{equation*}
  \left\{
  \begin{split}
    &
    \nabla\frac{Dc_i}{Dt} = \frac{D}{Dt}(\nabla c_i)
    + (\nabla\mathbf{u})\cdot\nabla c_i \\
    &
    \frac{DW}{Dt} = \sum_{i=1}^N \frac{\partial W}{\partial c_i}\frac{Dc_i}{Dt}
    + \sum_{i=1}^N \frac{\partial W}{\partial(\nabla c_i)}\cdot\frac{D}{Dt}(\nabla c_i)
  \end{split}
  \right.
\end{equation*}
we can transform equation \eqref{equ:2nd_law_3} into
\begin{equation}
  \begin{split}
  -D_s = \sum_{i=1}^N\left[\frac{\partial W}{\partial c_i}
    - \tilde{\rho}_i\mathcal{C}_i - \nabla\cdot\bm{\xi}_i  \right]
  &\frac{Dc_i}{Dt} 
  + \sum_{i=1}^N\left[\frac{\partial W}{\partial(\nabla c_i)}-\bm{\xi}_i  \right]\cdot\frac{D(\nabla c_i)}{Dt} \\
  &- \mathbf{T}:\nabla\mathbf{u}
  -\sum_{i=1}^N(\nabla c_i\otimes\bm{\xi}_i):\nabla\mathbf{u}
  + \sum_{i=1}^N\mathbf{J}_i\cdot\nabla \mathcal{C}_i
  \leqslant 0
  \end{split}
  \label{equ:2nd_law_4}
\end{equation}

We will make the following choices based on the inequality
\eqref{equ:2nd_law_4},
\begin{subequations}
  \begin{equation}
    \bm{\xi}_i = \frac{\partial W}{\partial(\nabla c_i)},
    \quad 1\leqslant i\leqslant N;
    \label{equ:const_relation_1}
  \end{equation}
  \begin{equation}
    \mathcal{C}_i = \frac{1}{\tilde{\rho}_i}\left[
    \frac{\partial W}{\partial c_i} - \nabla\cdot\frac{\partial W}{\partial(\nabla c_i)}
    \right],
    \quad 1\leqslant i\leqslant N.
    \label{equ:const_relation_2}
  \end{equation}
\end{subequations}
Note that these are specific choices made in this work to
guarantee the inequality \eqref{equ:2nd_law_4}.
They are not the most general possible
forms to satisfy \eqref{equ:2nd_law_4}.
Discussion of general constitutive relations is beyond
the scope of the current work.

Noting the choices \eqref{equ:const_relation_1}
and \eqref{equ:const_relation_2} and
the relation
\begin{equation*}
  (\nabla c_i\otimes\bm{\xi}_i):\nabla\mathbf{u}
  = \frac{1}{2}(\nabla c_i\otimes\bm{\xi}_i
  + \bm{\xi}_i\otimes\nabla c_i):\frac{1}{2}\mathbf{D}(\mathbf{u})
  + \frac{1}{2}(\nabla c_i\otimes\bm{\xi}_i
  - \bm{\xi}_i\otimes\nabla c_i):\frac{1}{2}(\nabla\mathbf{u}-\nabla\mathbf{u}^T),
\end{equation*}
where
$
\mathbf{D}(\mathbf{u}) = \nabla\mathbf{u}+\nabla\mathbf{u}^T,
$
we can transform \eqref{equ:2nd_law_4} into
\begin{equation}
  \begin{split}
  -D_s =& -\left[
    \mathbf{S} + \sum_{i=1}^N \frac{1}{2}\left(\nabla c_i\otimes\frac{\partial W}{\partial\nabla c_i}
  + \frac{\partial W}{\partial\nabla c_i}\otimes\nabla c_i\right)
  \right] : \frac{1}{2}\mathbf{D}(\mathbf{u}) \\
  & -\sum_{i=1}^N \frac{1}{2}\left( \nabla c_i\otimes\frac{\partial W}{\partial\nabla c_i}
  - \frac{\partial W}{\partial\nabla c_i}\otimes\nabla c_i
  \right):\frac{1}{2}(\nabla\mathbf{u}-
  \nabla\mathbf{u}^T) 
   + \sum_{i=1}^N \mathbf{J}_i\cdot\nabla \mathcal{C}_i \\
  & \leqslant 0
  \end{split}
  \label{equ:2nd_law_5}
\end{equation}
where we have used \eqref{equ:stress_tensor},
\eqref{equ:continuity}, and the symmetry of $\mathbf{S}$.
Since $\frac{1}{2}(\nabla\mathbf{u}-\nabla\mathbf{u}^T)$ is
independent of
$c_i$ ($1\leqslant i\leqslant N$) and
$W(\vec{c},\nabla\vec{c})$, and can attain
arbitrary values, we conclude that
\begin{equation}
  \sum_{i=1}^N\nabla c_i\otimes\frac{\partial W}{\partial(\nabla c_i)}
  = \sum_{i=1}^N \frac{\partial W}{\partial(\nabla c_i)}\otimes\nabla c_i,
  \label{equ:cond_energy}
\end{equation}
which is a condition the free energy density function
$W(\vec{c},\nabla\vec{c})$ must satisfy.

The inequality \eqref{equ:2nd_law_5} is then
reduced to
\begin{equation}
  -D_s = -\left[\mathbf{S} + \sum_{i=1}^N \nabla c_i\otimes
    \frac{\partial W}{\partial(\nabla c_i)} \right]
  : \frac{1}{2}\mathbf{D}(\mathbf{u})
  + \sum_{i=1}^N \left(\frac{1}{\tilde{\rho}_i}\mathbf{J}_i\right)\cdot\nabla(\tilde{\rho}_i\mathcal{C}_i)
  \leqslant 0.
  \label{equ:2nd_law_6}
\end{equation}
To ensure the above inequality we assume the
following constitutive relations
\begin{subequations}
  \begin{equation}
    \mathbf{S} + \sum_{i=1}^N \nabla c_i\otimes
    \frac{\partial W}{\partial(\nabla c_i)} = \mu(\vec{c})\mathbf{D}(\mathbf{u}),
    \label{equ:const_relation_3}
  \end{equation}
  \begin{equation}
    \frac{1}{\tilde{\rho}_i}\mathbf{J}_i
    = -\sum_{j=1}^N m_{ij}(\vec{c})\nabla(\tilde{\rho}_j\mathcal{C}_j)
    = -\sum_{j=1}^N m_{ij}(\vec{c})\nabla\left[
      \frac{\partial W}{\partial c_j} - \nabla\cdot\frac{\partial W}{\partial(\nabla c_j)}
    \right]
    \quad
    1\leqslant i\leqslant N,
    \label{equ:const_relation_4}
  \end{equation}
\end{subequations}
where $\mu(\vec{c})\geqslant 0$ plays the role of dynamic viscosity,
and the matrix formed by the coefficients $m_{ij}(\vec{c})$ ($1\leqslant i,j\leqslant N$)
\begin{equation}
  \mathbf{m} = \begin{bmatrix} m_{ij} \end{bmatrix}_{N\times N}
  \label{equ:mob_mat_def}
\end{equation}
is referred to as the mobility matrix.
$\mathbf{m}$ is required to be symmetric based on
the Onsager's reciprocal
relation and to be positive semi-definite in order
to ensure non-positivity of the second term in the inequality \eqref{equ:2nd_law_6}.
To ensure the relation \eqref{equ:diff_flux_relation}
for arbitrary $W(\vec{c},\nabla\vec{c})$,
we further require that
\begin{equation}
  \sum_{j=1}^N m_{ij}(\vec{c}) = \sum_{j=1}^Nm_{ji}(\vec{c}) = 0,
  \quad 1\leqslant i\leqslant N.
  \label{equ:mij_cond}
\end{equation}

In light of the condition \eqref{equ:mij_cond}, the constitutive
relation \eqref{equ:const_relation_4} can be re-written as
\begin{equation}
  \begin{split}
  \frac{1}{\tilde{\rho}_i}\mathbf{J}_i
  &= -m_{ii}\nabla(\tilde{\rho}_i\mathcal{C}_i)
  - \sum_{\substack{j=1\\ j\neq i}}^Nm_{ij}\nabla(\tilde{\rho}_j\mathcal{C}_j)
  =\sum_{\substack{j=1\\ j\neq i}}^Nm_{ij}\nabla(\tilde{\rho}_i\mathcal{C}_i)
  - \sum_{\substack{j=1\\ j\neq i}}^Nm_{ij}\nabla(\tilde{\rho}_j\mathcal{C}_j) \\
  &= \sum_{\substack{j=1\\ j\neq i}}^N m_{ij}\left[
    \nabla(\tilde{\rho}_i\mathcal{C}_i) - \nabla(\tilde{\rho}_j\mathcal{C}_j)
    \right]
  = \sum_{j=1}^N m_{ij}\left[
    \nabla(\tilde{\rho}_i\mathcal{C}_i) - \nabla(\tilde{\rho}_j\mathcal{C}_j)
    \right].
  \end{split}
\end{equation}
Consequently, the second term in the inequality \eqref{equ:2nd_law_6}
can be transformed into
\begin{equation*}
  \begin{split}
  \sum_{i=1}^N\frac{1}{\tilde{\rho}_i}\mathbf{J}_i\cdot\nabla(\tilde{\rho}_i\mathcal{C}_i)
  &= \sum_{i,j=1}^Nm_{ij}\left[
    \nabla(\tilde{\rho}_i\mathcal{C}_i) - \nabla(\tilde{\rho}_j\mathcal{C}_j)
    \right]\cdot\nabla(\tilde{\rho}_i\mathcal{C}_i) \\
  &=\sum_{i,j=1}^N\frac{1}{2}m_{ij}\left[
    \nabla(\tilde{\rho}_i\mathcal{C}_i) - \nabla(\tilde{\rho}_j\mathcal{C}_j)
    \right]\cdot \left[
    \nabla(\tilde{\rho}_i\mathcal{C}_i) - \nabla(\tilde{\rho}_j\mathcal{C}_j)
    \right] \\
  &= \sum_{\substack{i,j=1\\ j\neq i}}^N \frac{1}{2} m_{ij}\left|
  \nabla(\tilde{\rho}_i\mathcal{C}_i) - \nabla(\tilde{\rho}_j\mathcal{C}_j)
  \right|^2,
  \end{split}
\end{equation*}
where we have used the symmetry of $m_{ij}$.
Therefore, a sufficient condition to ensure the non-positivity
of above term, and the positive semi-definiteness of the mobility
matrix $\mathbf{m}$, is
\begin{equation}
  m_{ij}(\vec{c})\leqslant 0, \quad
  1\leqslant i\neq j\leqslant N.
  \label{equ:mij_cond_1}
\end{equation}

\paragraph{A Thermodynamically Consistent N-Phase Model}

Substituting the constitutive relations \eqref{equ:const_relation_3}
and \eqref{equ:const_relation_4} into
equations \eqref{equ:moment_balance} and \eqref{equ:mass_balance_2},
we obtain the following N-phase formulation
\begin{subequations}
  \begin{equation}
    \rho(\vec{c})\left(\frac{\partial\mathbf{u}}{\partial t} +
    \mathbf{u}\cdot\nabla\mathbf{u} \right)
    + \tilde{\mathbf{J}}\cdot\nabla\mathbf{u}
    = -\nabla p
    + \nabla\cdot\left[\mu(\vec{c})\mathbf{D}(\mathbf{u}) \right]
    - \sum_{i=1}^N \nabla\cdot\left[
    \nabla c_i \otimes \frac{\partial W}{\partial(\nabla c_i)}
    \right],
    \label{equ:nse_original}
  \end{equation}
  \begin{equation}
    \nabla\cdot\mathbf{u} = 0,
    \label{equ:continuity_original}
  \end{equation}
  \begin{equation}
    \frac{\partial c_i}{\partial t} + \mathbf{u}\cdot\nabla c_i
    = \sum_{j=1}^N \nabla\cdot\left[
      m_{ij}(\vec{c})\nabla\left(
      \frac{\partial W}{\partial c_j}
      - \nabla\cdot\frac{\partial W}{\partial\nabla c_j}
      \right)
      \right],
    \quad 1\leqslant i\leqslant N,
    \label{equ:CH_original}
  \end{equation}
\end{subequations}
where $W(\vec{c},\nabla\vec{c})$ is the free energy
density function whose form satisfies the condition
\eqref{equ:cond_energy},
and the coefficients $m_{ij}(\vec{c})$ ($1\leqslant i,j\leqslant N$)
form a symmetric positive semi-definite matrix and 
satisfy the condition \eqref{equ:mij_cond}.
Note that only $(N-1)$ equations among the
$N$ equations in \eqref{equ:CH_original}
are independent due to the conditions \eqref{equ:volfrac_relation}
and \eqref{equ:mij_cond}.
%
The mixture density is given by, in light of
\eqref{equ:rho_relation} and \eqref{equ:volfrac_relation},
\begin{equation}
  \rho(\vec{c}) = \sum_{i=1}^N\rho_i(\vec{c}) = \sum_{i=1}^N \tilde{\rho}_i c_i.
  \label{equ:rho_expr}
\end{equation}
$\tilde{\mathbf{J}}$ is given by, in light of \eqref{equ:mixture_mass_balance}
and \eqref{equ:const_relation_4},
\begin{equation}
  \tilde{\mathbf{J}}(\vec{c},\nabla\vec{c}) =
  -\sum_{i,j=1}^N\tilde{\rho}_im_{ij}(\vec{c})\nabla\left(
  \frac{\partial W}{\partial c_j}
  - \nabla\cdot\frac{\partial W}{\partial\nabla c_j}
  \right).
  \label{equ:J_tilde_expr}
\end{equation}
In the current work we assume that the mixture dynamic
viscosity $\mu(\vec{c})$ depends on $\vec{c}$ in a way
analogous to the mixture density $\rho(\vec{c})$,
\begin{equation}
  \mu(\vec{c}) = \sum_{i=1}^N\tilde{\mu}_i c_i,
  \label{equ:mu_expr}
\end{equation}
where $\tilde{\mu}_i$ ($1\leqslant i\leqslant N$) denote
the constant dynamic viscosities of these $N$ individual
fluids.
This model satisfies the mass conservation, momentum conservation,
and the second law of thermodynamics, and it is also
Galilean invariant.
This is a thermodynamically consistent N-phase model.


%% file: AppendB.tex
\section*{Appendix B. Proof of Properties About Reduction-Consistent
  Functions}

In this appendix we prove several properties of the reduction-consistent
and reduction-compatible functions listed in Section \ref{sec:formulation}.

\vspace{0.1in}
\noindent ($\mathscr{T}7$): If $v_{ij}(\vec{c})$ ($1\leqslant i,j\leqslant N$) are
  a reduction-consistent set of functions, and
  $w_i(\vec{c})$ ($1\leqslant i\leqslant N$) are a reduction-compatible set of
  functions, then $\sum_{j=1}^N v_{ij}w_j$ ($1\leqslant i\leqslant N$)
  and $\sum_{i=1}^N v_{ij}w_i$ ($1\leqslant j\leqslant N$)
  form two reduction-consistent sets of functions.

  \noindent\underline{Proof:} \ \
  Consider $N\geqslant 2$.
Let $z_i = \sum_{j=1}^N v_{ij} w_j$ ($1\leqslant i\leqslant N$). Suppose
fluid $k$ ($1\leqslant k\leqslant N$) 
is absent from the N-phase system.
Then $v_{ij}^{(N)}$ satisfy the reduction
relations in \eqref{equ:def_vij_compatible_1}
and \eqref{equ:def_vij_consis}.
$w_i^{(N)}$ satisfy the reduction relations
\begin{equation}
  w_i^{(N)} = \left\{
  \begin{array}{ll}
    w_i^{(N-1)}, & 1\leqslant i\leqslant k-1, \\
    w_{i-1}^{(N-1)}, & k+1\leqslant i\leqslant N.
  \end{array}
  \right.
  \label{equ:def_wi_compat}
\end{equation}
Consequently, we have
\begin{equation*}
  z_k^{(N)} = \sum_{j=1}^N v_{kj}^{(N)} w_j^{(N)}
  = 0.
\end{equation*}
For $1\leqslant i\leqslant k-1$,
\begin{equation*}
  \begin{split}
  z_i^{(N)} &= \sum_{j=1}^N v_{ij}^{(N)} w_j^{(N)}
  = \sum_{j=1}^{k-1} v_{ij}^{(N)} w_j^{(N)} + \sum_{j=k+1}^N v_{ij}^{(N)} w_j^{(N)}
  + v_{ik}^{(N)}w_k^{(N)} \\
  &
  = \sum_{j=1}^{k-1} v_{ij}^{(N-1)} w_j^{(N-1)}
  + \sum_{j=k+1}^N v_{ij-1}^{(N-1)} w_{j-1}^{(N-1)} 
  = \sum_{j=1}^{k-1} v_{ij}^{(N-1)} w_j^{(N-1)}
  + \sum_{j=k}^{N-1} v_{ij}^{(N-1)} w_{j}^{(N-1)} \\
  &
  = \sum_{j=1}^{N-1} v_{ij}^{(N-1)} w_j^{(N-1)} 
  = z_i^{(N-1)}.
  \end{split}
\end{equation*}
For $k+1\leqslant i\leqslant N$,
\begin{equation*}
  \begin{split}
  z_i^{(N)} &= \sum_{j=1}^N v_{ij}^{(N)} w_j^{(N)}
  = \sum_{j=1}^{k-1} v_{ij}^{(N)} w_j^{(N)} + \sum_{j=k+1}^N v_{ij}^{(N)} w_j^{(N)}
  + v_{ik}^{(N)}w_k^{(N)} \\
  &
  = \sum_{j=1}^{k-1} v_{i-1j}^{(N-1)} w_j^{(N-1)}
  + \sum_{j=k+1}^N v_{i-1j-1}^{(N-1)} w_{j-1}^{(N-1)} 
  = \sum_{j=1}^{k-1} v_{i-1j}^{(N-1)} w_j^{(N-1)}
  + \sum_{j=k}^{N-1} v_{i-1j}^{(N-1)} w_{j}^{(N-1)} \\
  &
  = \sum_{j=1}^{N-1} v_{i-1j}^{(N-1)} w_j^{(N-1)} 
  = z_{i-1}^{(N-1)}.
  \end{split}
\end{equation*}
We therefore conclude that $z_i=\sum_{j=1}^N v_{ij}w_j$ ($1\leqslant i\leqslant N$)
forms a reduction consistent set of functions.

One can show that $\sum_{i=1}^N v_{ij}w_i$ ($1\leqslant j\leqslant N$)
are a reduction-consistent set in a similar way.

\vspace{0.1in}
\noindent($\mathscr{T}8$): 
If $v_{ij}(\vec{c})$ ($1\leqslant i,j\leqslant N$) are a
reduction-compatible set of functions, and
$w_i(\vec{c})$ ($1\leqslant i\leqslant N$) are a reduction-consistent
set of functions, then $\sum_{j=1}^N v_{ij}w_j$ ($1\leqslant i\leqslant N$)
and $\sum_{i=1}^N v_{ij}w_i$ ($1\leqslant j\leqslant N$)
form two reduction-compatible sets of functions.

\noindent\underline{Proof:} \ \
Consider $N\geqslant 2$.
Let $z_i = \sum_{j=1}^N v_{ij} w_j$ ($1\leqslant i\leqslant N$).
Suppose
fluid $k$ ($1\leqslant k\leqslant N$) 
is absent from the system.
Then $v_{ij}^{(N)}$ satisfy the reduction
relations in \eqref{equ:def_vij_compatible_1}
and $w_i^{(N)}$ satisfy the reduction relations
\begin{equation}
  w_i^{(N)} = \left\{
  \begin{array}{ll}
    w_i^{(N-1)}, & 1\leqslant i\leqslant k-1, \\
    0, & i=k \\
    w_{i-1}^{(N-1)}, & k+1\leqslant i\leqslant N.
  \end{array}
  \right.
  \label{equ:def_wi_consis}
\end{equation}
Consequently, we have
for $1\leqslant i\leqslant k-1$,
\begin{equation*}
  \begin{split}
  z_i^{(N)} &= \sum_{j=1}^N v_{ij}^{(N)} w_j^{(N)}
  = \sum_{j=1}^{k-1} v_{ij}^{(N)} w_j^{(N)} + \sum_{j=k+1}^N v_{ij}^{(N)} w_j^{(N)} 
  = \sum_{j=1}^{k-1} v_{ij}^{(N-1)} w_j^{(N-1)}
  + \sum_{j=k+1}^N v_{ij-1}^{(N-1)} w_{j-1}^{(N-1)} \\
  &= \sum_{j=1}^{k-1} v_{ij}^{(N-1)} w_j^{(N-1)}
  + \sum_{j=k}^{N-1} v_{ij}^{(N-1)} w_{j}^{(N-1)} 
  = \sum_{j=1}^{N-1} v_{ij}^{(N-1)} w_j^{(N-1)} \\
  &= z_i^{(N-1)}.
  \end{split}
\end{equation*}
For $k+1\leqslant i\leqslant N$,
\begin{equation*}
  \begin{split}
  z_i^{(N)} &= \sum_{j=1}^N v_{ij}^{(N)} w_j^{(N)}
  = \sum_{j=1}^{k-1} v_{ij}^{(N)} w_j^{(N)} + \sum_{j=k+1}^N v_{ij}^{(N)} w_j^{(N)} 
  = \sum_{j=1}^{k-1} v_{i-1j}^{(N-1)} w_j^{(N-1)}
  + \sum_{j=k+1}^N v_{i-1j-1}^{(N-1)} w_{j-1}^{(N-1)} \\
  &= \sum_{j=1}^{k-1} v_{i-1j}^{(N-1)} w_j^{(N-1)}
  + \sum_{j=k}^{N-1} v_{i-1j}^{(N-1)} w_{j}^{(N-1)} 
  = \sum_{j=1}^{N-1} v_{i-1j}^{(N-1)} w_j^{(N-1)} \\
  &= z_{i-1}^{(N-1)}.
  \end{split}
\end{equation*}
We therefore conclude that $\sum_{j=1}^N v_{ij}w_j$ ($1\leqslant i\leqslant N$)
are a reduction-compatible set of variables.
$\sum_{i=1}^N v_{ij}w_i$ ($1\leqslant j\leqslant N$) can be shown
to be a reduction-compatible set in a similar way.

\vspace{0.1in}
\noindent ($\mathscr{T}9$):
If $v_i(\vec{c})$ ($1\leqslant i\leqslant N$) are a
reduction-consistent set of functions, then
$\sum_{i=1}^N v_i(\vec{c})$ is a reduction-consistent
function.

\noindent\underline{Proof:} \ \
Consider $N\geqslant 2$.
Let $z(\vec{c}) = \sum_{i=1}^N v_i(\vec{c})$.
Suppose fluid $k$ ($1\leqslant k\leqslant N$) is absent from
the system. Then $v_i^{(N)}$ satisfies the reduction relations given by
\eqref{equ:def_vi_compatible} and \eqref{equ:def_vi_consis}. 
Then
\begin{equation*}
  \begin{split}
  z^{(N)}(\vec{c}^{(N)}) &= \sum_{i=1}^N v_i^{(N)}
  = \sum_{i=1}^{k-1} v_i^{(N)} + \sum_{i=k+1}^N v_i^{(N)}
  = \sum_{i=1}^{k-1} v_i^{(N-1)} + \sum_{i=k+1}^N v_{i-1}^{(N-1)} \\
  &= \sum_{i=1}^{k-1} v_i^{(N-1)} + \sum_{i=k}^{N-1} v_{i}^{(N-1)}
  = \sum_{i=1}^{N-1} v_i^{(N-1)} \\
  &= z^{(N-1)}(\vec{c}^{(N-1)}).
  \end{split}
\end{equation*}
Therefore $\sum_{i=1}^Nv_i$ is a reduction-consistent function.

\vspace{0.1in}
\noindent ($\mathscr{T}10$):
If $v_{ij}(\vec{c})$ ($1\leqslant i,j\leqslant N$)
is a reduction-consistent set of functions,
then $\sum_{j=1}^{N}v_{ij}(\vec{c})$ ($1\leqslant i\leqslant N$)
and $\sum_{i=1}^N v_{ij}(\vec{c})$ ($1\leqslant j\leqslant N$)
are two reduction-consistent sets of functions,
and $\sum_{i,j=1}^Nv_{ij}(\vec{c})$ is a reduction-consistent function.

\noindent\underline{Proof:} \ \
Consider $N\geqslant 2$.
Let $z_i(\vec{c}) = \sum_{j=1}^N v_{ij}(\vec{c})$ ($1\leqslant i\leqslant N$)
and $z(\vec{c}) = \sum_{i,j=1}^N v_{ij}(\vec{c})$.
Suppose fluid $k$ ($1\leqslant k\leqslant N$) is absent from the
system. Then $v_{ij}$ satisfy the reduction relations
given by \eqref{equ:def_vij_compatible_1} and \eqref{equ:def_vij_consis}.
We have
\begin{equation*}
  z_k^{(N)} = \sum_{j=1}^N v_{kj}^{(N)}
  = 0.
\end{equation*}
For $1\leqslant i\leqslant k-1$,
\begin{equation*}
  \begin{split}
  z_i^{(N)} &= \sum_{j=1}^N v_{ij}^{(N)}
  = \sum_{j=1}^{k-1} v_{ij}^{(N)} + \sum_{j=k+1}^N v_{ij}^{(N)}
  = \sum_{j=1}^{k-1} v_{ij}^{(N-1)} + \sum_{j=k+1}^N v_{ij-1}^{(N-1)}
  = \sum_{j=1}^{k-1} v_{ij}^{(N-1)} + \sum_{j=k}^{N-1} v_{ij}^{(N-1)} \\
  &= \sum_{j=1}^{N-1} v_{ij}^{(N-1)}
  = z_i^{(N-1)}.
  \end{split}
\end{equation*}
For $k+1\leqslant i\leqslant N$,
\begin{equation*}
  \begin{split}
  z_i^{(N)} &= \sum_{j=1}^N v_{ij}^{(N)}
  = \sum_{j=1}^{k-1} v_{ij}^{(N)} + \sum_{j=k+1}^N v_{ij}^{(N)}
  = \sum_{j=1}^{k-1} v_{i-1j}^{(N-1)} + \sum_{j=k+1}^N v_{i-1j-1}^{(N-1)}
  = \sum_{j=1}^{k-1} v_{i-1j}^{(N-1)} + \sum_{j=k}^{N-1} v_{i-1j}^{(N-1)} \\
  &= \sum_{j=1}^{N-1} v_{i-1j}^{(N-1)}
  = z_{i-1}^{(N-1)}.
  \end{split}
\end{equation*}
We conclude that $\sum_{j=1}^N v_{ij}$ ($1\leqslant i\leqslant N$)
is a reduction-consistent set of functions.
One can show that $\sum_{i=1}^N v_{ij}$ ($1\leqslant j\leqslant N$)
is another reduction-consistent set of functions in a similar way.

Since 
$z_i(\vec{c})$ ($1\leqslant i\leqslant N$)
form a reduction-consistent set of functions and
$z(\vec{c}) = \sum_{i=1}^N z_i(\vec{c})$,
we conclude that $\sum_{i,j=1}^N v_{ij}$ is a reduction-consistent
function in light of the property ($\mathscr{T}9$).

\vspace{0.1in}
\noindent($\mathscr{T}11$): 
If $v_{ij}(\vec{c})$ ($1\leqslant i,j\leqslant N$) are
a reduction-compatible set of functions
and $w_i(\vec{c})$ ($1\leqslant i\leqslant N$) are
a reduction-consistent set of functions,
then $\sum_{i,j=1}^N v_{ij}w_iw_j$ is a reduction-consistent function.

\noindent\underline{Proof:} \ \
$\sum_{j=1}^Nv_{ij}w_j$ ($1\leqslant i\leqslant N$) are a reduction-compatible
set of functions according to property ($\mathscr{T}8$).
So $w_i\sum_{j=1}^Nv_{ij}w_j$ ($1\leqslant i\leqslant N$) form a
a reduction-consistent set of functions according to
property ($\mathscr{T}4$).
We then conclude that $\sum_{i,j=1}^Nv_{ij}w_iw_j$ is a reduction-consistent
function based on property ($\mathscr{T}9$).

\vspace{0.1in}
\noindent($\mathscr{T}12$): 
If $v_{ij}(\vec{c})$ ($1\leqslant i,j\leqslant N$) are
a reduction-consistent set of functions
and $w_i(\vec{c})$ ($1\leqslant i\leqslant N$) are
a reduction-compatible set of functions,
then $\sum_{i,j=1}^N v_{ij}w_iw_j$ is a reduction-consistent function.

\noindent\underline{Proof:} \ \
$\sum_{j=1}^Nv_{ij}w_j$ ($1\leqslant i\leqslant N$) are a reduction-consistent
set of functions according to property ($\mathscr{T}7$).
So $w_i\sum_{j=1}^Nv_{ij}w_j$ ($1\leqslant i\leqslant N$) form a
a reduction-consistent set of functions according to
property ($\mathscr{T}4$).
We then conclude that $\sum_{i,j=1}^Nv_{ij}w_iw_j$ is a reduction-consistent
function based on property ($\mathscr{T}9$).

%% file: AppendC.tex
\section*{Appendix C. Proof of Theorem \ref{thm:thm_1} }

It suffices to show that the $\mathscr{M}(\vec{c})$ defined
in \eqref{equ:def_nse} and $\mathscr{N}(\vec{c})$
defined in \eqref{equ:def_continuity} are each a
reduction-consistent function, and that
the $\mathscr{F}_i(\vec{c})$ ($1\leqslant i\leqslant N$) defined in
\eqref{equ:def_CH} form a reduction-consistent
set of functions.

Because $\mathscr{N}(\vec{c})=\nabla\cdot\mathbf{u}$ is independent
of $\vec{c}$, it is reduction-consistent based on the property
($\mathscr{T}14$) of Section \ref{sec:def_consis}.

Let us consider the reduction consistency of
$\mathscr{F}_i(\vec{c})$. Because
both $\frac{\partial c_i}{\partial t}$ ($1\leqslant i\leqslant N$)
and $\mathbf{u}\cdot\nabla c_i$ ($1\leqslant i\leqslant N$) are 
reduction-consistent sets of functions, it suffices
to show that
$
\sum_{j=1}^N m_{ij}\nabla\left(\frac{\partial W}{\partial c_j}
-\nabla\cdot\frac{\partial W}{\partial\nabla c_j} \right)
=\sum_{j=1}^N m_{ij}\nabla\left(\mathscr{H}_i - \mathscr{I}_i  \right)
$ for $1\leqslant i\leqslant N$
are a reduction-consistent set of functions.
That $\bm{\mathscr{G}}_i(\vec{c})$ ($1\leqslant i\leqslant N$) 
are a reduction-compatible set of functions implies that
$\mathscr{I}_i(\vec{c})$ ($1\leqslant i\leqslant N$) are a reduction-compatible
set according to property ($\mathscr{T}13$).
Because $\mathscr{H}_i$ and $\mathscr{I}_i$ are both reduction-compatible
sets of functions, $\nabla(\mathscr{H}_i-\mathscr{I}_i)$
is a reduction-compatible set of functions according to
properties ($\mathscr{T}3$) and ($\mathscr{T}13$) of Section \ref{sec:def_consis}.
Since $m_{ij}$ ($1\leqslant i,j\leqslant N$) are a reduction-consistent set,
$\sum_{j=1}^{N}m_{ij}\nabla\left(\mathscr{H}_i - \mathscr{I}_i  \right)$
($1\leqslant i\leqslant N$) are then a reduction-consistent set of functions according
to the property ($\mathscr{T}7$).
Therefore, all the three terms in \eqref{equ:def_CH}
are individually reduction-consistent sets of functions.
We can then conclude that $\mathscr{F}_i(\vec{c})$ ($1\leqslant i\leqslant N$)
are a reduction-consistent set of functions
based on the property ($\mathscr{T}2$).

Consider next the reduction consistency of $\mathscr{M}(\vec{c})$.
Note that both $\rho(\vec{c})$ and $\mu(\vec{c})$ are reduction-consistent functions
according to the property ($\mathscr{T}15$), and that
those terms independent of $\vec{c}$ are reduction-consistent
according to the property ($\mathscr{T}14$).
It suffices to show that the $\tilde{\mathbf{J}}$ term given by
\eqref{equ:J_tilde_expr} and
the term
$
\sum_{i=1}^N\nabla c_i\otimes\frac{\partial W}{\partial\nabla c_i}
=\sum_{i=1}^N\nabla c_i\otimes\bm{\mathscr{G}}_i
$
are both reduction-consistent functions.
According to \eqref{equ:J_tilde_expr},
$
\tilde{\mathbf{J}} = -\sum_{i=1}^N\tilde{\rho}_i\left[
  \sum_{j=1}^N m_{ij}\nabla\left(\mathscr{H}_j - \mathscr{I}_j \right)
\right ].
$
Since $\sum_{j=1}^N m_{ij}\nabla\left(\mathscr{H}_j - \mathscr{I}_j \right)$
($1\leqslant i\leqslant N$)
are a reduction-consistent set of functions and
$\tilde{\rho}_i$ ($1\leqslant i\leqslant N$) are a
reduction-compatible set of variables, we conclude based on
the properties ($\mathscr{T}4$) and ($\mathscr{T}9$) that
$\tilde{\mathbf{J}}$ is a reduction-consistent function.
Since $\nabla c_i$ ($1\leqslant i\leqslant N$) are
a reduction-consistent set and $\bm{\mathscr{G}}_i$ ($1\leqslant i\leqslant N$)
are a reduction-compatible set,
$\nabla c_i\otimes\bm{\mathscr{G}}_i$ ($1\leqslant i\leqslant N$)
form a reduction-consistent set of functions based on
property ($\mathscr{T}4$). Therefore, $\sum_{i=1}^N \nabla c_i\otimes\bm{\mathscr{G}}_i$
is a reduction-consistent function based on the property ($\mathscr{T}9$).
We can then conclude that $\mathscr{M}(\vec{c})$
is a reduction-consistent function.

%% file: AppendD.tex
\section*{Appendix D. Proof of Theorem \ref{thm:thm_2}}

\paragraph{Reduction Consistency of $m_{ij}(\vec{c})$}

We first show that the $m_{ij}(\vec{c})$ defined
by \eqref{equ:def_mij} form a reduction-consistent set
of functions. Consider an N-phase system ($N\geqslant 2$). 
Suppose that fluid $k$ ($1\leqslant k\leqslant N$)
is absent from the system, i.e.~the system is characterized by
\eqref{equ:special_nphase_system}, and the correspondence
relations in \eqref{equ:volfrac_correspondence} hold.
Then we have the following relations
\begin{equation}
  \tilde{m}_{ij}^{(N)} = \left\{
  \begin{array}{ll}
    \tilde{m}_{ij}^{(N-1)}, & 1\leqslant i\leqslant k-1, \ 1\leqslant j\leqslant k-1, \\
    \tilde{m}_{ij-1}^{(N-1)}, & 1\leqslant i\leqslant k-1, \ k+1\leqslant j\leqslant N, \\
    \tilde{m}_{i-1j}^{(N-1)}, & k+1\leqslant i\leqslant N, \ 1\leqslant j\leqslant k-1,\\
    \tilde{m}_{i-1j-1}^{(N-1)}, & k+1\leqslant i\leqslant N,\ k+1\leqslant j\leqslant N,
  \end{array}
  \right.
  \quad
  f(c_i^{(N)}) = \left\{
  \begin{array}{ll}
    f(c_i^{(N-1)}), & 1\leqslant i\leqslant k-1, \\
    0, & i=k \\
    f(c_{i-1}^{(N-1)}), & k+1\leqslant i\leqslant N,
  \end{array}
  \right.
\end{equation}
because of the properties of $f(c)$ given in \eqref{equ:func_f_property}
and the fact that $\tilde{m}_{ij}$
form a reduction-compatible set.

Let us look into the reduction for $m_{ij}(\vec{c})$.
We divide the problem into two cases:
(i) $i=k$ or $j=k$;
(ii) $i\neq k$ and $j\neq k$.
Consider the first case:  $i=k$ or $j=k$.
For $i=k$, if $j\neq k$, then
\begin{equation*}
  m_{kj}^{(N)} = -\tilde{m}_{kj}^{(N)} f(c_k^{(N)})f(c_j^{(N)}) = 0,
  \quad 1\leqslant j\leqslant N, \ j\neq k.
\end{equation*}
For $i=j=k$,
\begin{equation*}
  m_{kk}^{(N)} = -\sum_{\substack{j=1\\ j\neq k}}^N m_{kj}^{(N)} = 0.
\end{equation*}
For $j=k$ and $i\neq k$,
\begin{equation*}
  m_{ik}^{(N)} = -\tilde{m}_{ik}^{(N)} f(c_i^{(N)}) f(c_k^{(N)}) = 0,
  \quad 1\leqslant i\leqslant N, \ i\neq k.
\end{equation*}

Consider the second case: $i\neq k$ and $j\neq k$.
First consider the subcase $j\neq i$.
For $1\leqslant i\leqslant k-1$ and $1\leqslant j\leqslant k-1$
and $j\neq i$,
\begin{equation*}
  m_{ij}^{(N)} = -\tilde{m}_{ij}^{(N)} f(c_i^{(N)}) f(c_j^{(N)})
  = -\tilde{m}_{ij}^{(N-1)} f(c_i^{(N-1)}) f(c_j^{(N-1)})
  = m_{ij}^{(N-1)}.
\end{equation*}
For $1\leqslant i\leqslant k-1$ and $k+1\leqslant j\leqslant N$,
\begin{equation*}
  m_{ij}^{(N)} = -\tilde{m}_{ij}^{(N)} f(c_i^{(N)})f(c_j^{(N)})
  = -\tilde{m}_{ij-1}^{(N-1)} f(c_i^{(N-1)})f(c_{j-1}^{(N-1)})
  = m_{ij-1}^{(N-1)}.
\end{equation*}
For $k+1\leqslant i\leqslant N$ and $1\leqslant j\leqslant k-1$,
\begin{equation*}
  m_{ij}^{(N)} = -\tilde{m}_{ij}^{(N)} f(c_i^{(N)})f(c_j^{(N)})
  = -\tilde{m}_{i-1j}^{(N-1)} f(c_{i-1}^{(N-1)})f(c_j^{(N-1)})
  = m_{i-1j}^{(N-1)}.
\end{equation*}
For $k+1\leqslant i\leqslant N$ and $k+1\leqslant j\leqslant N$
and $j\neq i$,
\begin{equation*}
  m_{ij}^{(N)} = -\tilde{m}_{ij}^{(N)} f(c_i^{(N)})f(c_j^{(N)})
  = -\tilde{m}_{i-1j-1}^{(N-1)} f(c_{i-1}^{(N-1)})f(c_{j-1}^{(N-1)})
  = m_{i-1j-1}^{(N-1)}.
\end{equation*}
Now we consider the subcase $i=j\neq k$.
For $1\leqslant i\leqslant k-1$,
\begin{equation*}
  \begin{split}
  m_{ii}^{(N)} &= -\sum_{\substack{j=1\\ j\neq i}}^N m_{ij}^{(N)}
  = -\sum_{\substack{j=1\\ j\neq i}}^{k-1} m_{ij}^{(N)} - m_{ik}^{(N)}
  - \sum_{j=k+1}^N m_{ij}^{(N)}
  = -\sum_{\substack{j=1\\ j\neq i}}^{k-1} m_{ij}^{(N-1)}
  - \sum_{j=k+1}^N m_{ij-1}^{(N-1)} \\
  &= -\sum_{\substack{j=1\\ j\neq i}}^{k-1} m_{ij}^{(N-1)}
  - \sum_{j=k}^{N-1} m_{ij}^{(N-1)}
  = -\sum_{\substack{j=1\\ j\neq i}}^{N-1} m_{ij}^{(N-1)}
  = m_{ii}^{(N-1)}.
  \end{split}
\end{equation*}
For $k+1\leqslant i\leqslant N$,
\begin{equation*}
  \begin{split}
    m_{ii}^{(N)} &= -\sum_{\substack{j=1\\ j\neq i}}^N m_{ij}^{(N)}
    = -\sum_{j=1}^{k-1} m_{ij}^{(N)} - m_{ik}^{(N)}
    - \sum_{\substack{j=k+1\\ j\neq i}}^N m_{ij}^{(N)}
    = -\sum_{j=1}^{k-1} m_{i-1j}^{(N-1)}
    - \sum_{\substack{j=k+1\\ j\neq i}}^N m_{i-1j-1}^{(N-1)} \\
    &= -\sum_{j=1}^{k-1} m_{i-1j}^{(N-1)}
    - \sum_{\substack{j=k\\ j\neq i-1}}^{N-1} m_{i-1j}^{(N-1)}
    = -\sum_{\substack{j=1\\ j\neq i-1}}^{N-1} m_{i-1j}^{(N-1)}
    = m_{i-1i-1}^{(N-1)}.
  \end{split}
\end{equation*}

Combining the above, we have the following reduction relations
\begin{equation}
  m_{ij}^{(N)} = \left\{
  \begin{array}{ll}
    m_{ij}^{(N-1)}, & 1\leqslant i\leqslant k-1, \ 1\leqslant j\leqslant k-1 \\
    m_{ij-1}^{(N-1)}, & 1\leqslant i\leqslant k-1, \ k+1\leqslant j\leqslant N \\
    m_{i-1j}^{(N-1)}, & k+1\leqslant i\leqslant N, \ 1\leqslant j\leqslant k-1 \\
    m_{i-1j-1}^{(N-1)}, & k+1\leqslant i\leqslant N, \ k+1\leqslant j\leqslant N \\
    0, & i=k, \ 1\leqslant j\leqslant N \\
    0, & 1\leqslant i\leqslant N, \ j=k.
  \end{array}
  \right.
  \label{equ:mij_reduction_relation}
\end{equation}
Therefore, we conclude that $m_{ij}(\vec{c})$ ($1\leqslant i,j\leqslant N$)
defined by \eqref{equ:def_mij} are a reduction-consistent set of functions.

\paragraph{Reduction  Consistency of $W(\vec{c},\nabla\vec{c})$}

We will show that the two terms in the free energy density
function \eqref{equ:free_energy} are each a reduction-consistent function.
Consequently, $W(\vec{c},\nabla\vec{c})$
given by \eqref{equ:free_energy} is a reduction-consistent function.

The first term $\sum_{i,j=1}^N \frac{\lambda_{ij}}{2}\nabla c_i\cdot\nabla c_j$
is reduction-consistent. Since $\sigma_{ij}$ are a reduction-compatible set
(property ($\mathscr{T}16$)), $\lambda_{ij}$ ($1\leqslant i,j\leqslant N$)
given in \eqref{equ:free_energy_param} are a reduction-compatible set
according to property ($\mathscr{T}3$).
Noting that $\nabla c_i$ ($1\leqslant i\leqslant N$) are a
reduction-consistent set, we can conclude that
$\sum_{i,j=1}^N \frac{\lambda_{ij}}{2}\nabla c_i\cdot\nabla c_j$ is a
reduction-consistent function based on property ($\mathscr{T}11$).

Let us now show that the second term
$
H(\vec{c}) = \beta\sum_{i,j=1}^N \frac{\sigma_{ij}}{2}\left[
  g(c_i) + g(c_j) - g(c_i+c_j)
  \right]
$
is reduction-consistent.
Suppose fluid $k$ ($1\leqslant k\leqslant N$) is absent from
the system, i.e.~the system is characterized by \eqref{equ:special_nphase_system}
and the relations \eqref{equ:volfrac_correspondence}
hold.
Then we have the relations
\begin{equation}
  \sigma_{ij}^{(N)} = \left\{
  \begin{array}{ll}
    \sigma_{ij}^{(N-1)}, & 1\leqslant i\leqslant k-1, \ 1\leqslant j\leqslant k-1, \\
    \sigma_{ij-1}^{(N-1)}, & 1\leqslant i\leqslant k-1, \ k+1\leqslant j\leqslant N, \\
    \sigma_{i-1j}^{(N-1)}, & k+1\leqslant i\leqslant N, \ 1\leqslant j\leqslant k-1, \\
    \sigma_{i-1j-1}^{(N-1)}, & k+1\leqslant i\leqslant N, \ k+1\leqslant j\leqslant N,
  \end{array}
  \right.
  \ \
  g(c_i^{(N)}) = \left\{
  \begin{array}{ll}
    g(c_i^{(N-1)}), & 1\leqslant i\leqslant k-1, \\
    0, & i=k, \\
    g(c_{i-1}^{(N-1)}), & k+1\leqslant i\leqslant N.
  \end{array}
  \right.
  \label{equ:sigma_correspondence}
\end{equation}
Then we have
\begin{equation*}
  \begin{split}
  \frac{1}{\beta}H^{(N)} &= \sum_{i,j=1}^N \frac{\sigma_{ij}^{(N)}}{2}\left[
    g(c_i^{(N)}) + g(c_j^{(N)}) - g(c_i^{(N)}+c_j^{(N)})
    \right] 
  = \sum_{\substack{i,j=1\\ i,j\neq k}}^N \frac{\sigma_{ij}^{(N)}}{2}\left[
    g(c_i^{(N)}) + g(c_j^{(N)}) - g(c_i^{(N)}+c_j^{(N)})
    \right] \\
  &= \left(\sum_{i,j=1}^{k-1} + \sum_{i=1}^{k-1}\sum_{j=k+1}^N
  + \sum_{i=k+1}^N \sum_{j=1}^{k-1} + \sum_{i,j=k+1}^N\right)
  \frac{\sigma_{ij}^{(N)}}{2}\left[
    g(c_i^{(N)}) + g(c_j^{(N)}) - g(c_i^{(N)}+c_j^{(N)})
    \right] \\
  &= \sum_{i,j=1}^{k-1} \frac{\sigma_{ij}^{(N-1)}}{2}\left[
    g(c_i^{(N-1)}) + g(c_j^{(N-1)}) - g(c_i^{(N-1)}+c_j^{(N-1)})
    \right] \\
&\quad  +  \sum_{i=1}^{k-1}\sum_{j=k+1}^N \frac{\sigma_{ij-1}^{(N-1)}}{2}\left[
    g(c_i^{(N-1)}) + g(c_{j-1}^{(N-1)}) - g(c_i^{(N-1)}+c_{j-1}^{(N-1)})
    \right] \\
  &\quad  + \sum_{i=k+1}^N \sum_{j=1}^{k-1}
  \frac{\sigma_{i-1j}^{(N-1)}}{2}\left[
    g(c_{i-1}^{(N-1)}) + g(c_j^{(N-1)}) - g(c_{i-1}^{(N-1)}+c_j^{(N-1)})
    \right] \\
  &\quad + \sum_{i,j=k+1}^N \frac{\sigma_{i-1j-1}^{(N-1)}}{2}\left[
    g(c_{i-1}^{(N-1)}) + g(c_{j-1}^{(N-1)}) - g(c_{i-1}^{(N-1)}+c_{j-1}^{(N-1)})
    \right] \\
  &= \sum_{i,j=1}^{k-1} \frac{\sigma_{ij}^{(N-1)}}{2}\left[
    g(c_i^{(N-1)}) + g(c_j^{(N-1)}) - g(c_i^{(N-1)}+c_j^{(N-1)})
    \right] \\
&\quad  +  \sum_{i=1}^{k-1}\sum_{j=k}^{N-1} \frac{\sigma_{ij}^{(N-1)}}{2}\left[
    g(c_i^{(N-1)}) + g(c_{j}^{(N-1)}) - g(c_i^{(N-1)}+c_{j}^{(N-1)})
    \right] \\
  &\quad  + \sum_{i=k}^{N-1} \sum_{j=1}^{k-1}
  \frac{\sigma_{ij}^{(N-1)}}{2}\left[
    g(c_{i}^{(N-1)}) + g(c_j^{(N-1)}) - g(c_{i}^{(N-1)}+c_j^{(N-1)})
    \right] \\
  &\quad + \sum_{i,j=k}^{N-1} \frac{\sigma_{ij}^{(N-1)}}{2}\left[
    g(c_{i}^{(N-1)}) + g(c_{j}^{(N-1)}) - g(c_{i}^{(N-1)}+c_{j}^{(N-1)})
    \right] \\
  &= \sum_{i,j=1}^{N-1} \frac{\sigma_{ij}^{(N-1)}}{2}\left[
    g(c_{i}^{(N-1)}) + g(c_{j}^{(N-1)}) - g(c_{i}^{(N-1)}+c_{j}^{(N-1)})
    \right] \\
  &= \frac{1}{\beta}H^{(N-1)}.
  \end{split}
\end{equation*}
So we conclude that $H(\vec{c})$ is a reduction-consistent function.

\paragraph{Reduction Compatibility of $\mathscr{H}_i(\vec{c})$
  and $\bm{\mathscr{G}}_i(\vec{c})$ }

With $W(\vec{c},\nabla\vec{c})$ given by
\eqref{equ:free_energy},
\begin{equation}
  \mathscr{H}_i(\vec{c})=\frac{\partial W}{\partial c_i}
  = \beta\sum_{j=1}^N \sigma_{ij}\left[
    g^{\prime}(c_i) - g^{\prime}(c_i+c_j)
    \right],
  \quad 1\leqslant i\leqslant N;
  \label{equ:hi_expr}
\end{equation}
\begin{equation}
  \bm{\mathscr{G}}_i(\vec{c}) = \frac{\partial W}{\partial \nabla c_i}
  = \sum_{j=1}^N \lambda_{ij}\nabla c_j,
  \quad 1\leqslant i\leqslant N,
  \label{equ:qi_expr}
\end{equation}
where $g^{\prime}(c)$ is the derivative
of $g(c)$ defined in \eqref{equ:free_energy_param}.

$\bm{\mathscr{G}}_i(\vec{c})$ ($1\leqslant i\leqslant N$)
are evidently a reduction-compatible set of functions.
This is because $\lambda_{ij}$ ($1\leqslant i,j\leqslant N$)
are a reduction-compatible set and
$\nabla c_i$ ($1\leqslant i\leqslant N$) are a reduction-consistent set.
Based on the property ($\mathscr{T}8$) in Section \ref{sec:formulation},
we conclude that $\bm{\mathscr{G}}_i$ ($1\leqslant i\leqslant N$)
are a reduction-compatible set of functions.

We next show that $\mathscr{H}_i(\vec{c})$ ($1\leqslant i\leqslant N$)
given by \eqref{equ:hi_expr} are a reduction-compatible set of functions.
Suppose fluid $k$ ($1\leqslant k\leqslant N$) is absent from
the system, i.e.~the system is characterized by \eqref{equ:special_nphase_system}
and the reduction relations \eqref{equ:volfrac_correspondence}
hold. Then $\sigma_{ij}$ satisfy the reduction relations
given in \eqref{equ:sigma_correspondence}, and $g^{\prime}(c_i)$
(with $g(c)$ defined in \eqref{equ:free_energy_param})
satisfies the following relations
\begin{equation}
  g^{\prime}(c_i^{(N)}) = \left\{
  \begin{array}{ll}
    g^{\prime}(c_i^{(N-1)}), & 1\leqslant i\leqslant k-1, \\
    0, & i=k, \\
    g^{\prime}(c_{i-1}^{(N-1)}), & k+1 \leqslant i\leqslant N.
  \end{array}
  \right.
\end{equation}
For $1\leqslant i\leqslant k-1$,
\begin{equation*}
  \begin{split}
    \frac{1}{\beta}\mathscr{H}_i^{(N)} &=
    \sum_{j=1}^{k-1}\sigma_{ij}^{(N)}\left[
      g^{\prime}(c_i^{(N)}) - g^{\prime}(c_i^{(N)}+c_j^{(N)})
      \right]
    + \sum_{j=k+1}^{N}\sigma_{ij}^{(N)}\left[
      g^{\prime}(c_i^{(N)}) - g^{\prime}(c_i^{(N)}+c_j^{(N)})
      \right] \\
    &= \sum_{j=1}^{k-1}\sigma_{ij}^{(N-1)}\left[
      g^{\prime}(c_i^{(N-1)}) - g^{\prime}(c_i^{(N-1)}+c_j^{(N-1)})
      \right] \\
    &\quad + \sum_{j=k+1}^{N}\sigma_{ij-1}^{(N-1)}\left[
      g^{\prime}(c_i^{(N-1)}) - g^{\prime}(c_i^{(N-1)}+c_{j-1}^{(N-1)})
      \right] \\
    &= \sum_{j=1}^{k-1}\sigma_{ij}^{(N-1)}\left[
      g^{\prime}(c_i^{(N-1)}) - g^{\prime}(c_i^{(N-1)}+c_j^{(N-1)})
      \right] \\
    &\quad + \sum_{j=k}^{N-1}\sigma_{ij}^{(N-1)}\left[
      g^{\prime}(c_i^{(N-1)}) - g^{\prime}(c_i^{(N-1)}+c_{j}^{(N-1)})
      \right] \\
    &= \sum_{j=1}^{N-1} \sigma_{ij}^{(N-1)} \left[
      g^{\prime}(c_i^{(N-1)}) - g^{\prime}(c_i^{(N-1)}+c_{j}^{(N-1)})
      \right] \\
    &= \frac{1}{\beta}\mathscr{H}_i^{(N-1)}.
  \end{split}
\end{equation*}
For $k+1\leqslant i\leqslant N$,
\begin{equation*}
  \begin{split}
    \frac{1}{\beta}\mathscr{H}_i^{(N)} &=
    \sum_{j=1}^{k-1}\sigma_{ij}^{(N)}\left[
      g^{\prime}(c_i^{(N)}) - g^{\prime}(c_i^{(N)}+c_j^{(N)})
      \right]
    + \sum_{j=k+1}^{N}\sigma_{ij}^{(N)}\left[
      g^{\prime}(c_i^{(N)}) - g^{\prime}(c_i^{(N)}+c_j^{(N)})
      \right] \\
    &= \sum_{j=1}^{k-1}\sigma_{i-1j}^{(N-1)}\left[
      g^{\prime}(c_{i-1}^{(N-1)}) - g^{\prime}(c_{i-1}^{(N-1)}+c_j^{(N-1)})
      \right] \\
    &\quad + \sum_{j=k+1}^{N}\sigma_{i-1j-1}^{(N-1)}\left[
      g^{\prime}(c_{i-1}^{(N-1)}) - g^{\prime}(c_{i-1}^{(N-1)}+c_{j-1}^{(N-1)})
      \right] \\
    &= \sum_{j=1}^{k-1}\sigma_{i-1j}^{(N-1)}\left[
      g^{\prime}(c_{i-1}^{(N-1)}) - g^{\prime}(c_{i-1}^{(N-1)}+c_j^{(N-1)})
      \right] \\
    &\quad + \sum_{j=k}^{N-1}\sigma_{i-1j}^{(N-1)}\left[
      g^{\prime}(c_{i-1}^{(N-1)}) - g^{\prime}(c_{i-1}^{(N-1)}+c_{j}^{(N-1)})
      \right] \\
    &= \sum_{j=1}^{N-1} \sigma_{i-1j}^{(N-1)} \left[
      g^{\prime}(c_{i-1}^{(N-1)}) - g^{\prime}(c_{i-1}^{(N-1)}+c_{j}^{(N-1)})
      \right] \\
    &= \frac{1}{\beta}\mathscr{H}_{i-1}^{(N-1)}.
  \end{split}
\end{equation*}
Combining the above results, we conclude that
$\mathscr{H}_i(\vec{c})$ ($1\leqslant i\leqslant N$)
given by \eqref{equ:hi_expr} form a reduction-compatible
set of functions.

%% file: AppendE.tex
\section*{Appendix E. Algorithm for N-Phase Momentum Equations}

We summarize the algorithm developed in \cite{Dong2015}
for the N-phase momentum equations,
which is employed in the current work.
The algorithm is for the equations \eqref{equ:nse_trans_1}
and \eqref{equ:continuity_original}, together
with the boundary condition \eqref{equ:bc_vel}.
It is assumed that the volume fractions $c_i^{n+1}$ ($1\leqslant i\leqslant N$)
and the auxiliary variables $\psi_i^{n+1}$ ($1\leqslant i\leqslant N$)
have already been computed using the algorithm presented
in Section \ref{sec:alg}.
The goal here is to compute the velocity $\mathbf{u}^{n+1}$ 
and the pressure $P^{n+1}$
with given $\mathbf{u}^n$, $P^n$, $c_i^{n+1}$ and $\psi_i^{n+1}$.

The algorithm consists of two steps. The pressure and the velocity
are computed successively in a de-coupled fashion in the first and the second
steps, respectively. \\
\noindent\underline{For $P^{n+1}$:}
\begin{subequations}
\begin{equation}
\begin{split}
\frac{\gamma_0\tilde{\mathbf{u}}^{n+1}-\hat{\mathbf{u}}}{\Delta t}
+ \mathbf{u}^{*,n+1}\cdot\nabla\mathbf{u}^{*,n+1}
& + \frac{1}{\rho^{n+1}}\tilde{\mathbf{J}}^{n+1}\cdot\nabla\mathbf{u}^{*,n+1}
+ \frac{1}{\rho_0}\nabla P^{n+1}
= 
\left(\frac{1}{\rho_0}-\frac{1}{\rho^{n+1}}  \right)\nabla P^{*,n+1} \\
& - \frac{\mu^{n+1}}{\rho^{n+1}}\nabla\times\nabla\times\mathbf{u}^{*,n+1}
+ \frac{1}{\rho^{n+1}}\nabla\mu^{n+1}\cdot\mathbf{D}(\mathbf{u}^{*,n+1}) \\
& - \frac{1}{\rho^{n+1}}\sum_{i,j=1}^{N}\lambda_{ij}\left(\psi_j^{n+1}-\alpha c_j^{n+1}\right)\nabla c_i^{n+1}
+ \frac{1}{\rho^{n+1}}\mathbf{f}^{n+1},
\end{split}
\label{equ:pressure_1}
\end{equation}
\begin{equation}
\nabla\cdot\tilde{\mathbf{u}}^{n+1} = 0,
\label{equ:pressure_2}
\end{equation}
\begin{equation}
\left.\mathbf{n}\cdot\tilde{\mathbf{u}}^{n+1}\right|_{\partial\Omega}
= \mathbf{n}\cdot\mathbf{w}^{n+1}.
\label{equ:pressure_3}
\end{equation}
\end{subequations}
\\
\noindent\underline{For $\mathbf{u}^{n+1}$:}
\begin{subequations}
\begin{equation}
\frac{\gamma_0\mathbf{u}^{n+1}-\gamma_0\tilde{\mathbf{u}}^{n+1}}{\Delta t}
 - \nu_0 \nabla^2\mathbf{u}^{n+1}
= \nu_0 \nabla\times\nabla\times\mathbf{u}^{*,n+1},
\label{equ:velocity_1}
\end{equation}
\begin{equation}
\left.\mathbf{u}^{n+1}  \right|_{\partial\Omega} = \mathbf{w}^{n+1}.
\label{equ:velocity_2}
\end{equation}
\end{subequations}

In the above equations all the symbols follow
the same notation as outlined in Section \ref{sec:alg}.
$\mathbf{u}^{*,n+1}$ and $P^{*,n+1}$
are defined by \eqref{equ:def_var_star}.
$\hat{\mathbf{u}}$ and $\gamma_0$ are defined
by \eqref{equ:def_var_hat}.
$\tilde{\mathbf{J}}^{n+1}$ is
given by (see equation \eqref{equ:H_expr})
\begin{equation}
\tilde{\mathbf{J}}^{n+1} = -\sum_{i,j=1}^N \tilde{\rho}_i m_{ij}(\vec{c}^{n+1})
\nabla\left[
  -\sum_{k=1}^N\lambda_{jk}(\psi_k^{n+1}-\alpha c_k^{n+1})
  + \mathscr{H}_j(\vec{c}^{n+1})
\right].
\label{equ:J_expr_1}
\end{equation}
Note that in both the above equation and in equation
\eqref{equ:pressure_1} we have replaced
$\nabla^2c_i^{n+1}$ by
($\psi_i^{n+1}-\alpha c_i^{n+1}$) according to equations 
\eqref{equ:CH_phi} and \eqref{equ:def_psi_N}. 
$\mathbf{n}$ is the outward-pointing unit vector
normal to $\partial\Omega$.
$\tilde{\mathbf{u}}^{n+1}$ is an auxiliary
velocity that approximates $\mathbf{u}^{n+1}$.
$\rho^{n+1}$ and $\mu^{n+1}$ are
given by \eqref{equ:rho_expr} and \eqref{equ:mu_expr},
and in case of large density ratios we follow \cite{Dong2015}
and further clamp
the values of $\rho^{n+1}$ and $\mu^{n+1}$ as follows
(see \cite{Dong2015} for details)
\begin{equation}
    \rho^{n+1} = \left\{
    \begin{array}{ll}
      \rho^{n+1}, & \text{if} \ \rho^{n+1}\in[\tilde{\rho}_{\min},\tilde{\rho}_{\max}] \\
      \tilde{\rho}_{\max}, & \text{if} \ \rho^{n+1} > \tilde{\rho}_{\max} \\
      \tilde{\rho}_{\min}, & \text{if} \ \rho^{n+1} < \tilde{\rho}_{\min},
    \end{array}
    \right.
    \quad
    \mu^{n+1} = \left\{
    \begin{array}{ll}
      \mu^{n+1}, & \text{if} \ \mu^{n+1}\in[\tilde{\mu}_{\min},\tilde{\mu}_{\max}] \\
      \tilde{\mu}_{\max}, & \text{if} \ \mu^{n+1} > \tilde{\mu}_{\max} \\
      \tilde{\mu}_{\min}, & \text{if} \ \mu^{n+1} < \tilde{\mu}_{\min},
    \end{array}
    \right.
    \label{equ:rho_mu_clamp}
  \end{equation}
  where
  $\tilde{\rho}_{\max} = \max_{1\leqslant i\leqslant N}\{\tilde{\rho}_i\}$,
  $\tilde{\rho}_{\min} = \min_{1\leqslant i\leqslant N}\{\tilde{\rho}_i\}$,
  $\tilde{\mu}_{\max} = \max_{1\leqslant i\leqslant N}\{\tilde{\mu}_i\}$,
  and $\tilde{\mu}_{\min} = \min_{1\leqslant i\leqslant N}\{\tilde{\mu}_i\}$.
The constant $\rho_0$ is given by
$ 
\rho_0 = \tilde{\rho}_{\min} = \min_{1\leqslant i\leqslant N}\{\tilde{\rho}_i\}.
$ 
$\nu_0$ in \eqref{equ:velocity_1} is a chosen positive
constant that is sufficiently large.
We  employ an $\nu_0$ value
with the following property,
$ 
\nu_0 \geqslant \max\left(
\frac{\tilde{\mu}_1}{\tilde{\rho}_1},
\frac{\tilde{\mu}_2}{\tilde{\rho}_2},
\cdots,
\frac{\tilde{\mu}_N}{\tilde{\rho}_N}
\right).
$ 

The above algorithm employs a velocity
correction-type idea \cite{DongS2010,DongKC2014,Dong2015clesobc}  to de-couple
the computations for the pressure and the velocity.
With this algorithm the linear
algebraic systems resulting from the discretization
involve only {\em constant} and {\em time-independent} 
coefficient matrices, like
in two-phase flows \cite{DongS2012,Dong2012,Dong2014obc}. 

It is straightforward to derive
the weak forms for the pressure and the velocity
for the implementation of the algorithm using
$C^0$ spectral elements; see \cite{Dong2015} for details. 
We only provide the final weak forms here.
Let $q\in H^1(\Omega)$ denote the test function, and 
\begin{multline}
\mathbf{G}^{n+1} = 
\frac{1}{\rho^{n+1}}\mathbf{f}^{n+1}
- \left(
     \mathbf{u}^{*,n+1} 
     + \frac{1}{\rho^{n+1}} \tilde{\mathbf{J}}^{n+1}
  \right)\cdot\nabla\mathbf{u}^{*,n+1}
+ \frac{\hat{\mathbf{u}}}{\Delta t}
+ \left(\frac{1}{\rho_0} - \frac{1}{\rho^{n+1}}  \right)\nabla P^{*,n+1} \\
+ \frac{1}{\rho^{n+1}}\nabla\mu^{n+1}\cdot\mathbf{D}(\mathbf{u}^{*,n+1})
- \frac{1}{\rho^{n+1}}\sum_{i,j=1}^{N}\lambda_{ij}(\psi_j^{n+1}-\alpha c_j^{n+1})\nabla c_i^{n+1}
+ \nabla\left( \frac{\mu^{n+1}}{\rho^{n+1}} \right) \times \bm{\omega}^{*,n+1},
\label{equ:G_expr}
\end{multline}
where $\bm{\omega} = \nabla\times \mathbf{u}$ is
the vorticity.
The weak form for the pressure $P^{n+1}$ is
\begin{equation}
\int_{\Omega} \nabla P^{n+1} \cdot\nabla q
= \rho_0 \int_{\Omega} \mathbf{G}^{n+1}\cdot\nabla q
- \rho_0 \int_{\partial\Omega} \frac{\mu^{n+1}}{\rho^{n+1}} \mathbf{n}\times\bm{\omega}^{*,n+1}\cdot\nabla q
- \frac{\gamma_0\rho_0}{\Delta t}\int_{\partial\Omega}\mathbf{n}\cdot\mathbf{w}^{n+1} q,
\ \
\forall q\in H^1(\Omega).
\label{equ:p_weakform}
\end{equation}
Let 
$
H^1_0(\Omega) = \left\{ \
v \in H^1(\Omega) \ : \
v|_{\partial\Omega} = 0
\ \right\},
$
and $\varphi \in H_0^1(\Omega)$ denote
the test function.
The weak form about the velocity $\mathbf{u}^{n+1}$ is
\begin{multline}
\int_{\Omega}\nabla\varphi\cdot\nabla\mathbf{u}^{n+1}
+ \frac{\gamma_0}{\nu_0\Delta t}\int_{\Omega}\varphi\mathbf{u}^{n+1}
= \frac{1}{\nu_0}\int_{\Omega}\left(
    \mathbf{G}^{n+1} - \frac{1}{\rho_0}\nabla P^{n+1}
  \right) \varphi \\
- \frac{1}{\nu_0}\int_{\Omega} \left(
    \frac{\mu^{n+1}}{\rho^{n+1}} - \nu_0
  \right) 
  \bm{\omega}^{*,n+1} \times \nabla\varphi,
\qquad
\forall \varphi \in H_0^1(\Omega).
\label{equ:vel_weakform}
\end{multline}
These weak forms, 
\eqref{equ:p_weakform} and
\eqref{equ:vel_weakform}, can be
discretized in space using $C^0$ spectral 
elements 
in a straightforward
fashion~\cite{Dong2015}.

Solving the N-phase momentum equations
\eqref{equ:nse_trans_1} and \eqref{equ:continuity_original}
amounts to the following two successive operations. First,
solve equation \eqref{equ:p_weakform} for
pressure $P^{n+1}$. Then, solve
equation \eqref{equ:vel_weakform}, together with
the Dirichlet condition \eqref{equ:velocity_2} on $\partial\Omega$,
for $\mathbf{u}^{n+1}$.